\documentclass[twoside,12pt]{article}
\usepackage{epsfig}
\usepackage{graphicx}
\usepackage{amsmath,amssymb}
\usepackage{epstopdf}
\usepackage{pstricks}
\usepackage{multirow}
\usepackage{subfigure}

\newsavebox{\Staple}
\savebox{\Staple}{\begin{picture}(0,0)	
\thicklines
\put(0.0,0.1){\vector(0,1){0.9}}
\put(0.0,1.0){\vector(1,0){0.9}}
\put(0.9,1.0){\vector(0,-1){0.9}}
\end{picture}}

\newsavebox{\FiveStaple}
\savebox{\FiveStaple}{\begin{picture}(0,0)	
\thicklines
\put(0.0,0.1){\vector(0,1){0.9}}
\put(0.0,1.0){\vector(1,1){0.5}}
\put(0.5,1.5){\vector(1,0){0.9}}
\put(1.4,1.5){\vector(-1,-1){0.5}}
\put(0.9,1.0){\vector(0,-1){0.9}}
\end{picture}}

\newsavebox{\SevenStaple}
\savebox{\SevenStaple}{\begin{picture}(0,0)	
\thicklines
\put(0.0,0.1){\vector(0,1){0.9}}
\put(0.0,1.0){\vector(1,1){0.5}}
\put(0.5,1.5){\vector(1,2){0.3}}
\put(0.8,2.1){\vector(1,0){0.9}}
\put(1.7,2.1){\vector(-1,-2){0.3}}
\put(1.4,1.5){\vector(-1,-1){0.5}}
\put(0.9,1.0){\vector(0,-1){0.9}}
\end{picture}}

\newsavebox{\LepageStaple}
\savebox{\LepageStaple}{\begin{picture}(0,0)	
\thicklines
\put(0.0,0.1){\vector(0,1){0.9}}
\put(0.0,1.0){\vector(0,1){0.9}}
\put(0.0,1.9){\vector(1,0){0.9}}
\put(0.9,1.9){\vector(0,-1){0.9}}
\put(0.9,1.0){\vector(0,-1){0.9}}
\end{picture}}

\newsavebox{\Link}
\savebox{\Link}{\begin{picture}(0,0)	
\thicklines
\put(0.0,0.0){\vector(1,0){0.9}}
\end{picture}}

\newsavebox{\Naik}
\savebox{\Naik}{\begin{picture}(0,0)	
\thicklines
\put(0.0,0.0){\vector(1,0){0.9}}
\put(1.0,0.0){\vector(1,0){0.9}}
\put(2.0,0.0){\vector(1,0){0.9}}
\end{picture}}

\newsavebox{\OneFatNaik}
\savebox{\OneFatNaik}{\begin{picture}(0,0)	
\thicklines
\put(0,0){\usebox{\Link}\makebox(0,0)}
\put(1.1,0)+
\put(1.5,0){\usebox{\Naik}\makebox(0,0)}
\put(4.6,0)+
\put(5.1,0){\usebox{\Staple}\makebox(0,0)}
\end{picture}}

\def\p{\vec p}

\newcommand{\be}{\begin{equation}}
\newcommand{\ee}{\end{equation}}
\newcommand{\ba}{\begin{eqnarray}}
\newcommand{\ea}{\end{eqnarray}}
\newcommand{\nn}{\nonumber}
\newcommand{\Tr}{{\rm Tr}}
\newcommand{\hTr}{\hat{\rm T}{\rm r}}
\newcommand{\ho}{\hat{\omega}}
\newcommand{\ex}{{\rm e}}
\newcommand{\psib}{\bar{\psi}}
\newcommand{\re}{\text{Re}}
\newcommand{\csw}{\ensuremath{{c_\text{\tiny SW}}}}
\def\pbp{\langle\bar{\psi}\psi\rangle}
\newcommand{\vevsub}[1]{\ensuremath{\left<#1\right>_{\text{sub}}}}
\newcommand{\Real}{\ensuremath{\text{Re}}}
\newcommand{\Trace}{\ensuremath{\text{Tr}}}

\newcommand{\Oa}{\ensuremath{\mathcal{O}(a)}}
\newcommand{\Oas}{\ensuremath{\mathcal{O}(a^2)}}

\def\bfx{{\bf x}}

\def\bfz{{\bf z}}
\def\bfp{{\bf p}}

\def\lsi{\raise0.3ex\hbox{$<$\kern-0.75em\raise-1.1ex\hbox{$\sim$}}}
\def\gsi{\raise0.3ex\hbox{$>$\kern-0.75em\raise-1.1ex\hbox{$\sim$}}}
\newcommand{\lsim}{\mathop{\lsi}}
\newcommand{\gsim}{\mathop{\gsi}}
\newcommand{\eq}{Eq.~}
\newcommand{\eqs}{Eqs.~}
\newcommand{\fig}{Fig.~}

\newcommand{\link}[2]{  U_{ #1 , #2}  }
\newcommand{\linkdagger}[2]{  U_{ #1 , #2}^\dagger  }
\newcommand{\linkdaggerfat}[2]{  U_{ #1 , #2}^{{\rm fat} \dagger}  }
\newcommand{\kronecker}[2]{  \delta_{#1 \; #2}  }

\newcommand{\apart}[3]{      
\sum_{#3} \eta_{#3}(#1)\Big( \link{#1}{#3}^{\rm fat} \kronecker{#1}{#2-\hat{#3}} 
                -\linkdaggerfat{#1-\hat{#3}}{#3} 
                                        \kronecker{#1}{#2+\hat{#3}} \Big) }

\newcommand{\bparta}[3]{
\sum_{#3} \eta_{#3}(#1)\Big( \link{#1}{#3} \link{#1+\hat{#3}}{#3} \link{#1+2 \hat{#3}}{#3}
\kronecker{#1}{#2- 3 \hat{#3}}
- \linkdagger{#1-\hat{#3}}{#3} \linkdagger{#1-2\hat{#3}}{#3}
\linkdagger{#1-3 \hat{#3}}{#3} \kronecker{#1}{#2+ 3 \hat{#3}} \Big) }

\newcommand{\bpartb}[4]{
\sum_{#3} \eta_{#3}(#1)\sum_{#4\ne#3}\bigg[ \Big( \link{#1}{#3} \link{#1+\hat{#3}}{#4} \link{#1+\hat{#3} + \hat{#4}}{#4} \kronecker{#1}{#2-\hat{#3}-2\hat{#4}}
- \linkdagger{#1-\hat{#4}}{#4} \linkdagger{#1-2\hat{#4}}{#4} 
 \linkdagger{#1- \hat{#3} - 2\hat{#4}}{#3} \kronecker{#1}{#2+\hat{#3}+2\hat{#4}}\Big)\\
&&~~~+ \Big( \link{#1}{#4} \link{#1+\hat{#4}}{#4} \link{#1+
2\hat{#4}}{#3} \kronecker{#1}{#2-\hat{#3}-2\hat{#4}} 
 - \linkdagger{#1-\hat{#3}}{#3} \linkdagger{#1-\hat{#3}-\hat{#4}}{#4} 
     \linkdagger{#1- \hat{#3} - 2\hat{#4}}{#4}\kronecker{#1}{#2+\hat{#3}+2\hat{#4}}\Big)\\
&&~~~+ \Big( \linkdagger{#1-\hat{#4}}{#4} \linkdagger{#1-2\hat{#4}}{#4}
\link{#1-2\hat{#4}}{#3} \kronecker{#1}{#2-\hat{#3}+2\hat{#4}}
 - \linkdagger{#1-\hat{#3}}{#3} \link{#1-\hat{#3}}{#4} \link{#1- \hat{#3} +
\hat{#4}}{#4} \kronecker{#1}{#2+\hat{#3}-2\hat{#4}}\Big)\\
&&~~~+ \Big( \link{#1}{#3} \linkdagger{#1+\hat{#3}-\hat{#4}}{#4}
\linkdagger{#1+\hat{#3}-2\hat{#4}}{#3} \kronecker{#1}{#2-\hat{#3}+2\hat{#4}}
 - \link{#1}{#4} \link{#1+\hat{#4}}{#4} \linkdagger{#1- \hat{#3} +
2\hat{#4}}{#3} \kronecker{#1}{#2+\hat{#3}-2\hat{#4}}           
          \Big) \bigg] }

\begin{document}

\title{ \vspace{1cm} The QCD equation of state from the lattice}
\author{Owe Philipsen
\\
Institut f\"ur Theoretische Physik, Goethe-Universit\"at Frankfurt,\\
Max-von-Laue-Str. 1, 60438 Frankfurt am Main, Germany
}
\maketitle
\begin{abstract} 
The equation of state of QCD at finite temperatures and baryon densities has a wide range
of applications in many fields of modern particle and nuclear physics. It is the main 
ingredient to describe the dynamics of experimental heavy ion collisions,
the expansion of the early universe in the standard model era and the interior of compact stars.
On most scales of interest, QCD is strongly coupled and not amenable to perturbative
investigations. Over the past decade, first principles calculations 
using lattice QCD have reached maturity, in the sense that for particular discretisation schemes
simulations at the physical point have become possible, finite temperature 
results near the continuum limit are 
available and systematic errors begin to be controlled. 
This review summarises the current theoretical and numerical state of the art based on staggered and
Wilson fermions. 
\end{abstract}
\tableofcontents

\newpage

\section{Introduction}

Quantum Chromodynamics (QCD) is the fundamental quantum field theory describing the 
strong interactions within the Standard Model. As such it is also the fundamental theory of 
nuclear matter. One of its many well-tested features is asymptotic freedom, according to which
the coupling vanishes at asymptotically high energy scales revealing the nature of its
constituents, the quarks and gluons. Conversely, the coupling is 
strong on hadronic energy scales $\lsim  1$ GeV and we observe confinement of quarks and gluons
which is not amenable to weak coupling expansions. 
Asymptotic freedom implies fascinating changes of dynamics when QCD is considered under extreme
conditions, where either a high temperature sets the dominant momentum scale, as in the early
universe, or a high baryon density, as in compact stars. Heavy ion collision experiments
have succeeded in creating the hot quark gluon plasma in the laboratory and operate to further
understand its properties, while future heavy ion experiments and astronomical observations aim to
investigate cold and dense matter. Of particular phenomenological importance is the equation of 
state of the hot and/or dense system. As a fundamental property of the thermal system, it is in principle
accessible to experiment allowing for a direct comparison with theoretical predictions. Moreover, the equation of state constitutes important
input for further dynamical analyses of thermal systems, such as the hydrodynamical description
of heavy ion collisions~\cite{kh} or the search for new physics in the early universe~\cite{hp}.

Because of the inherently non-perturbative nature of the theory, numerical simulations of lattice QCD are the only tool allowing for predictions from first principles, for which it is known
how to remove the associated systematic errors. Investigations of the equation of state have been 
going on for about two decades. After rapid initial successes with the pure gauge plasma, the step to include
dynamical fermions has proved soberingly difficult. Systematic errors associated with fermion discretisations were initially underestimated, leading to apparent contradictions. However, after an impressive collective effort in man and machine power, these issues appear to be finally resolved. 
It was demonstrated
that, while it may take some time and effort, systematic errors eventually {\it can} be controlled and
removed, teaching us a lot about the underlying dynamics in the process. 
The equation of state at finite temperature and zero
baryon density is now known for $N_f=2+1$ quark flavours with physical masses 
very close to its continuum limit,
which should be taken within the next year.
This is based on the staggered fermion discretisation. Calculations with Wilson fermions are somewhat behind, but will soon serve as an independent cross check for remaining theoretical issues with 
the staggered formulation. Refinements to include the charm quark are also well on the way.
The situation at finite density is much less satisfactory. Because of the
sign problem of lattice QCD, direct simulations at finite baryon density are impossible and further
approximations have to be made which are valid for sufficiently small chemical potentials. 
Nevertheless, important first steps in this direction have been made and we also understand
the response of the equation of state to a small baryon chemical potential.

Rather than on a complete history, the
focus in this review is on the more recent calculations closest to the continuum or the physical point,
and hence on staggered and Wilson fermions.
The control of systematic errors being the main task of current lattice calculations, 
Secs.~\ref{sec:lat}-\ref{sec:meth} are devoted to a detailed discussion of
the different discretisation schemes and their associated cut-off effects along with improvement
schemes to minimise them. The  aim is to also give the non-practitioner
some insight into what is being done in different calculations. The numerical results on the equation of
state are then collected in Secs.\ref{sec:ym}-\ref{sec:mu}.  

\section{Statistical mechanics of QCD}

We wish to describe a system of particles in some volume $V$ which is in thermal
contact with a heat bath at temperature $T$. Associated with the particles  may be a set
of conserved charges $N_i, i=1,2,\ldots$ (such as particle number, electric charge, baryon number etc.).
In quantum field theory, the most direct description is in terms of the grand canonical ensemble.
Its density operator and partition function are given as
\be
\rho= \ex^{-\frac{1}{T}(H-\mu_iN_i)}\;,\qquad Z=\hTr \rho\;, \quad 
\hTr(\ldots)=\sum_n\langle n|(\ldots)|n\rangle\;,
\ee
where $\mu_i$ are chemical potentials for the conserved charges, and the quantum mechanical trace
is a sum over all energy eigenstates of the Hamiltonian. Thermodynamic averages
for an observable $O$ are then obtained as $ \langle O\rangle=Z^{-1}\hTr (\rho O)\;.$

From the partition function, all other thermodynamic equilibrium quantities follow by taking
appropriate derivatives. In particular, the (Helmholtz) free energy, pressure, entropy, 
mean values of charges and energy are obtained as

\begin{minipage}{0.5\textwidth}
\ba
F&=&-T\ln Z\;,\nn\\ 
p&=&\frac{\partial (T\ln Z)}{\partial V}\;,\nn\\
S&=&\frac{\partial (T\ln Z)}{\partial T}\;,\nn
\ea
\end{minipage}
 \begin{minipage}{0.45\textwidth}
 \ba \label{thermo}
\bar{N}_i&=&\frac{\partial (T\ln Z)}{\partial \mu_i}\;,\nn\\
 E&=&-pV+TS+\mu_i\bar{N}_i\;.\\ 
&&\mbox{}
\nn
\ea 
\end{minipage}
\vspace*{3mm}

\noindent
Since the free energy is known to be an extensive quantity, $F=fV$, and we are interested in the thermodynamic limit, it is often more convenient to consider the corresponding densities, 
\be
f=\frac{F}{V},\quad s=\frac{S}{V},\quad n_i=\frac{\bar{N}_i}{V},\quad \epsilon=\frac{E}{V}\;,
\ee 
for which we in addition have 
\be
p=-f,\quad \epsilon=\frac{T^2}{V}\frac{\partial \ln Z}{\partial T}=-\frac{1}{V}\frac{\partial \ln Z}{\partial T^{-1}}\;.
\label{thermo1}
\ee
In a relativistic system, another quantity of interest is the energy momentum tensor for a 
liquid or gas,
\be
T^{\mu\nu}=(p+\epsilon)u^\mu u^\nu-pg^{\mu\nu}\;,
\ee
where $u^\mu=(1,0,0,0)$ is the four velocity of a volume element in the rest frame of the medium.
Its so-called trace anomaly corresponds to
\be
I(T)\equiv T^{\mu\mu}(T)=T^5\frac{\partial}{\partial T}\frac{p(T)}{T^4}\;.
\label{eq:defi}
\ee
Various useful relations between these quantities are then
\be
I=\epsilon-3p\;,\quad s=\frac{\epsilon+p}{T}\;,\quad c_s^2=\frac{dp}{d\epsilon}\;,
\label{thermo2}
\ee
where the energy derivative of the pressure in the last equation corresponds to the speed of sound in the
thermal medium. The equation of state is the functional relationship among the thermodynamic 
parameters for a system in thermal equilibrium,
\be
f(p,V,T,\mu_i)=0\;,
\ee
and constitutes the input for many further analyses of a thermal system.

\subsection{QCD at finite temperature and quark density}

Let us now consider the grand canonical partition function of QCD. 
The derivation of its path integral representation is discussed in detail 
in the textbooks \cite{kapusta}, 
\be  
Z(V,{\mu_f},T;g,{m_f})=\hTr \left(\ex^{-(H-\mu_f {Q_f})/T}\right)=
\int DA \,D\bar{\psi}\,D\psi \; \ex^{-S_g[A_\mu]} \,\ex^{-S_f[\bar{\psi},\psi,A_\mu]},
\label{part}
\ee
with the Euclidean gauge and fermion actions
\ba
S_g[A_\mu]&=& \int\limits_0^{1/T} d\tau \int\limits_V d^3x \;
\frac{1}{2} {\rm Tr}\; F_{\mu\nu}(x) F_{\mu\nu}(x), \nn\\
S_f[\bar{\psi},\psi,A_\mu]&=& \int\limits_0^{1/T} d\tau \int\limits_V d^3x \;
 \sum_{f=1}^{N_f} \bar{\psi}_f(x) \left( \gamma_\mu
D_\mu+ m_f- \mu_f \gamma_0 \right) \psi_f(x) .
\label{lagrangian}
\ea
The index $f$ labels different quark flavours, and 
the covariant derivative contains the gauge coupling $g$,
\be
D_\mu=(\partial_\mu-igA_\mu),\quad A_\mu=T^aA_\mu^a(x), \quad a=1,\ldots N^2-1,\quad
F_{\mu\nu}(x)=\frac{i}{g}[D_\mu,D_\nu]\;.
\ee
The thermodynamic limit is obtained by sending the spatial three-volume $V\rightarrow \infty$.
The difference to the Euclidean path integral at $T=0$ describing vacuum physics is the compactified temporal direction
with a radius defining the inverse temperature, $1/T$.
To ensure Bose/Einstein statistics for bosons and the Pauli principle for fermions, the path integral is to be evaluated with periodic and anti-periodic boundary conditions in the temporal
direction for bosons and fermions, respectively,
\be
A_\mu(\tau,\bfx)=A_\mu(\tau+\frac{1}{T},\bfx),\qquad
\psi(\tau,\bfx)=-\psi(\tau+\frac{1}{T},\bfx)\;.
\label{bc}
\ee
The  path integral for vacuum QCD in infinite four-volume is  
smoothly recovered from this expression for $T\rightarrow 0$.

The partition function depends on 
the external macroscopic parameters $T,V,\mu_f$, as 
well as on the microscopic parameters like quark masses and the coupling constant. 
The conserved quark 
numbers corresponding to the chemical potentials $\mu_f$ are 
\be
Q_f=\bar{\psi}_f\gamma_0\psi_f\;.
\ee
We will consider mostly two and three flavours of quarks,
and always take $m_u=m_d$. The case $m_s=m_{u,d}$ is then denoted by $N_f=3$, while $N_f=2+1$
implies $m_s\neq m_{u,d}$. When the heavier charm quark is taken into account as well, we have $N_f=2+1+1$.

\subsection{Centre symmetry}\label{sec:zn}

Consider the partition function of pure gauge theory. The action is invariant under gauge transformations
$g(x)$ which obey the same periodic boundary conditions, $g(\tau,\bfx)=g(\tau+1/T,\bfx)$,
\be
A^g_\mu(x)=g(x)\left(A_\mu+\frac{i}{g}\partial_\mu\right) g^\dag(x)\;.
\ee
It is also invariant under gauge transformations $g'$, which are periodic up to a global $SU(N)$ matrix $h$,
$g'(\tau,\bfx)=hg(\tau+1/T,\bfx)$. The gauge transformed field satisfies the boundary condition,
\be
A^{g'}(\tau+1/T,\bfx)=hA^{g'}_\mu(\tau,\bfx)h^\dag\;,
\ee
which is the originally demanded one if $h\in Z(N)$ is in the centre of the group, 
$h=z{\bf 1}=\exp(i2\pi k/N)$, $k\in~\{0,1...N-1\}$. Hence pure gauge theory allows for gauge transformations
which are periodic up to a twist factor from the centre of the group.
For fermions one has instead
\be
\psi^{g'}(\tau+1/T,\bfx)=-z\psi^{g'}(\tau,\bfx)\;,
\ee
which is anti-periodic for $z=1$ only, i.e.~fermions break the centre symmetry of the pure gauge
action.  
The order parameter for centre symmetry is the Polyakov loop, i.e.~a traced temporal Wilson line,
\be
L(\bfx)=\Tr W(\bfx)= {\cal P} \exp\left(ig\int_0^{1/T} d\tau A_4(x)\right)\;,
\ee
which corresponds to the propagator of a static quark in euclidean time wrapping around the torus. 
Under gauge transformations, $W^g(\bfx)=g(x)W(\bfx)g^{-1}(x)$ it picks up a centre element when transformed with a winding transformation,
\be
L^{g}= L,\quad  L^{g'}=z^* L\;.
\label{polyz}
\ee
 The expectation value of the Polyakov loop gives the free energy difference between a 
Yang-Mills plasma
with and without the static quark. 
\be
\langle  L\rangle = \frac{1}{Z}\int DU \;\Tr W \; \ex^{-S_g}=\frac{Z_Q}{Z}=\ex^{-(F_Q-F_0)/T}\;,
\ee
For $T=0$ pure gauge theory is confining  and it costs infinite energy to remove the quark
to infinity, i.e.~$F_Q=\infty$ and therefore $\langle L \rangle =0$. For $T\rightarrow \infty$ one has
$\langle L \rangle \neq 0$, which
is no longer invariant under centre transformations and  signals the
spontaneous breaking of centre symmetry.
Therefore, QCD in the quenched limit has a true (non-analytic) deconfinement phase transition corresponding to the breaking of the global centre symmetry. When finite mass fermions are added,
the centre symmetry is broken explicitly, i.e.~$\langle L\rangle\neq 0$ always and there is no need
for a non-analytic phase transition.

\subsection{Chiral symmetry}

The QCD action is invariant under $U(1)$ transformations on the quark fields
\be
 \psi'_f=\exp(i\alpha)\;\psi_f\;,
 \ee
the corresponding conserved charge is baryon number. 
For $n_f$ degenerate quark flavours, the corresponding spinors can be written
as an $n_f$-plet $\psi=(\psi_f, \ldots \psi_{n_f})$, and the QCD action is symmetric 
under the global $SU(n_f)$ vector rotations
\be
\psi'_{f'}=\exp(i\theta^aT^a)_{f'f}\;\psi_f\;,\quad a=1,\ldots n_f^2-1, \quad\;.
\ee
In nature, $m_u\approx m_d$ and this is $SU(2)$ isospin symmetry.
For $n_f$ zero mass flavours QCD is furthermore
invariant under the corresponding $U_A(1)$ and $SU_A(n_f)$ axial transformations,
\be
\label{eq:axial}
\psi'_f=\exp(i\alpha \gamma_5)\;\psi_f\;,\quad \psi'_{f'}=\exp(i\theta^aT^a\gamma_5)_{f'f}\;\psi_f\;,\quad 
a=1,\ldots n_f^2-1\;.
\ee
Hence, in the chiral limit the total symmetry is 
\be
U(1)_B\times U(1)_A\times SU(n_f)_L\times SU(n_f)_R\;,
\ee 
where the $L,R$ denote the left and right handed linear combinations of the vector and axial-vector rotations. The axial $U(1)_A$ is anomalous, quantum corrections break it down to $Z(n_f)$. The remainder
gets spontaneously broken to the diagonal subgroup, $SU(N_f)_L\times SU(N_f)_R\rightarrow 
SU(N_f)_V$, giving rise to $n_f^2-1$ massless Goldstone bosons, the pions. 

At finite temperatures, chiral symmetry gets restored above some critical temperature $T_c$.
The order parameter
signalling chiral symmetry is the chiral condensate,
\be
\langle\bar{\psi}\psi_f\rangle =\frac{T}{V}\frac{\partial}{\partial m_f} \ln Z.
\ee
It is nonzero for $T<T_c$, when chiral symmetry is spontaneously broken, and zero for $T>T_c$.
Hence, in the chiral limit there is a non-analytic finite temperature phase transition corresponding to chiral symmetry restoration. This is of first order for $N_f\geq 3$ and either second order with $O(4)$ universality
or first order for $N_f§=2$, depending on the strength of the $U(1)_A$ anomaly  at $T_c$ \cite{pw}. 
For non-zero quark masses, chiral symmetry is broken explicitly and the chiral condensate 
$\langle\bar{\psi}\psi_f\rangle\neq 0$ for all temperatures. Again, in this case there is no need
for a non-analytic phase transition.

\subsection{The finite temperature phase transition of QCD}

\begin{figure}[t]
\centerline{
\includegraphics[width=0.4\textwidth]{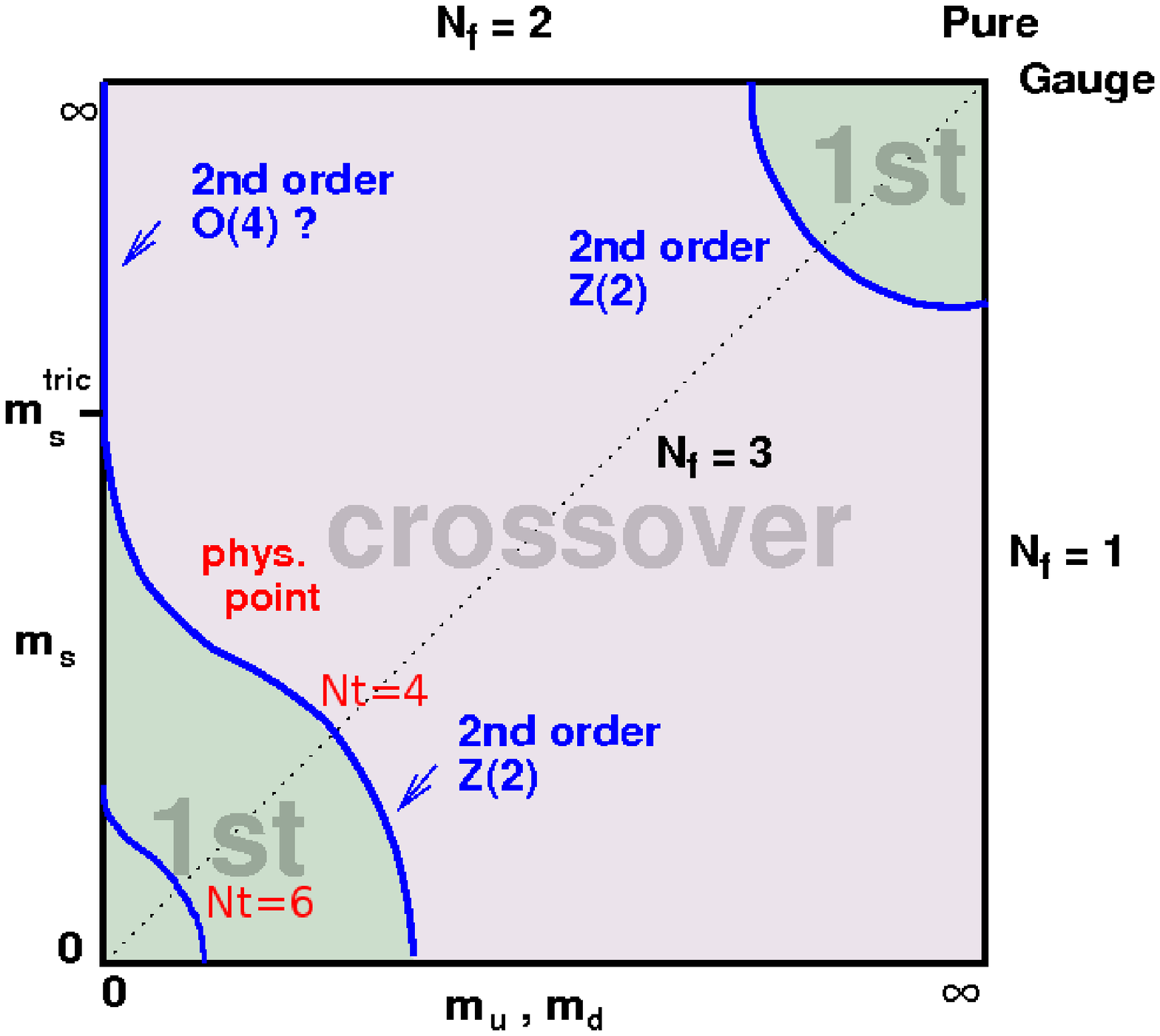}
\includegraphics[width=0.44\textwidth]{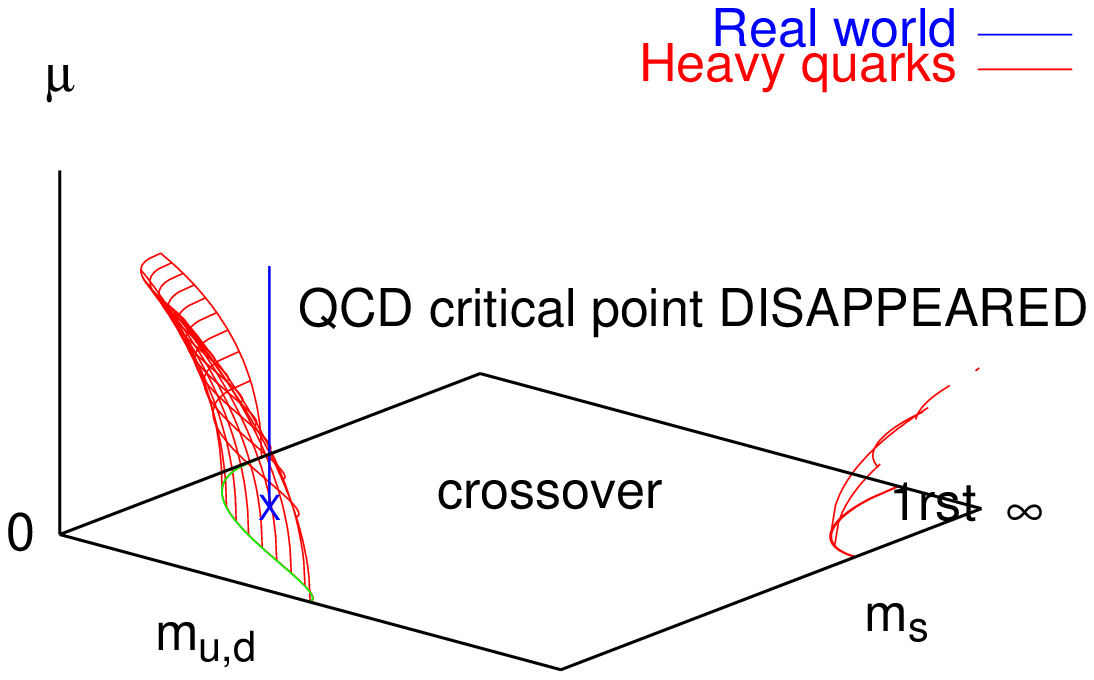}
}
\caption{\label{fig:schem} Left: Schematic phase transition behaviour of $N_f=2+1$
QCD for different choices of quark masses at
$\mu=0$. On finer lattices, the chiral critical line moves towards smaller quark masses. 
Right: Chiral and deconfinement critical surfaces swept by the 
critical lines as $\mu$ is turned on. Depending on its curvature, a QCD chiral critical
point is present or absent. 
}
\end{figure}
The order of the finite temperature phase transition at zero density depends on the quark masses and
is schematically shown in \fig\ref{fig:schem} (left).
In the limits of zero and infinite quark masses (lower left and upper 
right corners), order parameters corresponding to the breaking of a 
global symmetry exist, and for three degenerate quarks 
one numerically finds first order phase
transitions at small and large quark masses at some finite
temperatures $T_c(m)$. On the other hand, one observes an analytic crossover (without singularities
in the thermodynamic functions) at
intermediate quark masses, with second order boundary lines separating these
regions. Both lines have been shown to belong to the $Z(2)$ universality class
of the 3d Ising model \cite{kls,fp2}. 
The critical lines bound the quark mass regions featuring a chiral or deconfinement phase transition, 
and are called chiral and deconfinement critical lines, respectively.

QCD with physical values for the quark masses breaks both the chiral and the centre symmetries explicitly,
Hence, the finite temperature transition does not break or restore either symmetry and there is no
reason for it to be non-analytical. Indeed, physical QCD is in the crossover region as established
using the staggered fermion discretisation in two independent ways:
i) by showing that the finite temperature transition at the physical point is non-critical and 
analytic in the continuum limit \cite{nature}, ii) by explicitly mapping out the chiral critical line 
(so far on coarse lattices only) \cite{chirl}. 
At finite density the situation is much less clear, since straightforward Monte Carlo simulations are 
impossible due to the sign problem, cf.~Sec.~\ref{sec:mu}.  For larger chemical potentials
some techniques signal the presence of a critical point. However, these are not yet independently
confirmed. On the other hand, for small $\mu/T\lsim1$ all 
independent methods find that both the chiral and deconfinement transitions weaken with chemical 
potential, so there is no chiral critical point at small chemical potentials as in \fig\ref{fig:schem} (right). Reviews of the current status
can be found in \cite{murev,murev1}. 

Associated with the deconfinement and chiral phase transitions is a critical temperature $T_c(m_f,\mu_f)$,
which depends on the number of flavours and quark masses. The method to locate a transition in statistical mechanics is to look for inflection points in 
the changes of observables $O(\bfx)$, 
$O\in\{ L,\bar{\psi}\psi_f, \ldots\}$, or for the peak in the associated susceptibilities. 
Note that for an analytic crossover the pseudo-critical couplings defined
from different observables do not need to coincide. The partition function is analytic everywhere
and there is no uniquely specified ``transition''. This holds in particular for pseudo-critical couplings extracted from finite lattices. As the thermodynamic limit is approached, the 
couplings defined in different ways will merge where there is a non-analytic phase transition, 
and stay separate in the case of a crossover.

\subsection{Ideal gases: fermions, bosons and hadron resonances}

Much of our intuition for QCD at finite temperatures is based on the behaviour of free gases. 
For example, the qualitative expectation of a rise in the pressure as a signal for deconfinement is based
on the ideal hadron resonance gas in the low temperature phase and an ideal gas of quarks and gluons
in the deconfined phase.  This is also true on the lattice.
At high temperature perturbative investigations allow
for a controlled analytic understanding of the leading cut-off effects on the equation of state.
At low temperatures a strong coupling expansion provides new qualitative insights by showing the emergence of the hadron resonance gas
as a true effective theory of QCD in the strong coupling limit.  

The partition function for an ideal gas reads
\be
\ln Z_i^1(V,T)=\eta \nu_i V\int \frac{d^3p}{(2\pi)^3}\; \ln(1+\eta\,\ex^{-(\omega_i-\mu_i)/T})\;,
\label{freegas}
\ee
where $\eta=-1$ for bosons and $\eta=1$ for fermions, 
$\nu_i$ gives the number of degrees of freedom for the particle $i$ and $\mu_i$ is the chemical
potential for a conserved charge associated with the particle.
In the massless limit the momentum integration can be done in closed form and  gives the 
Stefan-Boltzmann pressure
\be
p=\nu_i\frac{\pi^2}{90}T^4\times \left\{\begin{array}{l}
1\quad \mbox{bosons} \\ \frac{7}{8}\quad \mbox{fermions}
\end{array} \right.\;.
\label{psb}
\ee
For an ideal gas of gluons and massless quarks 
we have two polarisation states for each of the $(N^2-1)$ gluons, 
two polarisation states per quark, $N$ colours per quark and the same for anti-quarks,
\be
\frac{p}{T^4}=\left(2(N^2-1)+4N N_f\frac{7}{8}\right)\frac{\pi^2}{90}\;.
\ee 
This is the Stefan-Boltzmann limit of the QCD pressure which is valid for vanishing coupling $g=0$, 
i.e.~in the limit $T\rightarrow \infty$.

There is another ideal gas system that is useful to model the low temperature behaviour
of QCD, assuming that at low temperatures, when we still have confinement, 
the QCD partition function is close to that of a free gas of hadrons. 
While strong coupling effects are responsible for  
confinement of the quarks and gluons, the colour forces are screened beyond the hadron radius and interactions between  colour neutral
hadrons are considerably weaker. Neglecting them, the QCD partition function factorises into
one-particle partition functions, $\ln Z(V,T,\mu_i)\approx \sum_i\ln Z_i^1(V,T,\mu_i)$,
and the pressure reads
\ba
p&=&\frac{T}{V}\sum_{i\in\{\text{mesons}\}}\ln Z_i^1(V,T,\mu_i)+\frac{T}{V}\sum_{i\in\{\text{baryons}\}}\ln Z_i^1(V,T,\mu_i)\;,
\ea
with $\eta=-1,1$ in the mesonic and baryonic terms, respectively.
The particles to be inserted are the known hadrons
and hadron resonances (for QCD purposes electroweak decays are to be neglected).
Taking experimental values for the energies from the particle data group \cite{pdg}, one can 
supply this formula with hundreds of hadron resonances, do the momentum integral and 
obtain a thermodynamic pressure that compares remarkably well with Monte Carlo simulations
of QCD \cite{hrg}.  It was later shown that this is no accident, but that 
the hadron resonance gas model can actually be derived from lattice QCD
as an effective theory for low temperatures, i.e.~strong coupling \cite{lppress}, see Sec.~\ref{sec:hrgsc}.

\section{Lattice QCD at finite temperature}\label{sec:lat}

Experience has shown that in order to avoid misleading conclusions, 
it is most important to understand and control systematic errors 
due to the lattice discretisation when attempting to extract continuum physics from results
of numerical simulations. For this reason it is worth discussing the discretisation
in some detail.

\subsection{Lattice gauge action}

We consider the lattice formulation of a $SU(N)$ pure gauge theory on a
hypercubic lattice, $N_s^3\times N_\tau$, with lattice spacing $a$ and spatial and temporal lengths $L=aN_s, T^{-1}=aN_\tau$, respectively.
The standard Wilson gauge action reads
\be
S_g[U]=\sum_{x}\sum_{1\leq\mu<\nu\leq 4}\beta\left(1-\frac{1}{3}{\rm Re}
\Tr U_{\mu\nu}(x)\right), 
\label{Sg}
\ee
where
$U_{\mu\nu}(x)=u_p(x)=U_\mu(x)U_\nu(x+a\hat{\mu})U^\dag_\mu(x+a\hat{\nu})U^\dag_\nu(x)$ is
the elementary plaquette, and the bare lattice and continuum gauge couplings 
are related by $\beta=2N/g^2$. 
For numerical simulations one  imposes periodic boundary conditions in all directions, 
\be
U_\mu(\tau,\bfx)=U_\mu(\tau+N_\tau,\bfx), \quad U_\mu(\tau,\bfx)=U_\mu(\tau,\bfx+N_s).
\ee
To see the relation to the continuum action, 
write the links as infinitesimal parallel transport between neighbouring lattice points. Inserting this
in the plaquette and expanding down the exponential yields,
\be
U_\mu(x)=\exp(-iagA_\mu(x))\;,\quad U_{\mu\nu}(x)=\exp\left(ia^2F_{\mu\nu}(x)+{\cal O}(a^3) \right)\;.
\ee
In the limit of small lattice spacings the action reproduces the classical continuum action,
\be
S_g[U]=\frac{a^4}{2g^2}\sum_x\sum_{\mu,\nu}\left(\Tr F_{\mu\nu}^2(x)+\Oas\right)\;.
\ee
However, there are infinitely many more terms suppressed by increasing powers of the lattice spacing.
As long as this is finite, the action and all results differ from the continuum action by these discretisation 
effects. As a consequence, the lattice action is not unique, only its classical continuum limit is. 
One may add or subtract any terms that vanish in the continuum limit without affecting the continuum 
physics. This is made use of in the construction of improved lattice actions, which attempt to reduce
discretisation effects.

The connection between zero and finite temperature physics is most easily 
exhibited by the transfer matrix, which evolves quantum mechanical states by 
one lattice spacing in Euclidean time.
It relates the 
path integral representation of a Euclidean lattice field theory to the 
Hamiltonian formulation and is completely specified by the lattice action \cite{mm}, to
put the partition
function in its quantum mechanical form
\be
Z=
\hTr(T^{N_\tau})=\hTr(\ex^{-N_\tau aH}), \quad \frac{1}{T}\equiv a N_\tau\;.
\label{temp}
\ee
In this form we can immediately see that $Z$ is equivalent to the partition function
of a thermal system if we identify the temporal lattice extent with inverse temperature.
The thermal expectation value of an observable is then
\be
\langle O \rangle = Z^{-1}\hTr(\ex^{-\frac{H}{T}} O)
=Z^{-1}\sum_n\langle n|T^{N_\tau}O|n\rangle
=\frac{\sum_n\langle n| O|n\rangle\, \ex^{-aN_\tau E_n}}{\sum_n \ex^{-aN_\tau E_n}}\;.
\label{boltz}
\ee 
As in the continuum, we are interested in the thermodynamic limit and hence $N_s\rightarrow \infty$, while
keeping $aN_\tau=1/T$ finite.
In this form we easily see the connection to $T=0$ physics:
projection on the vacuum expectation value is achieved by taking $N_\tau$ to infinity, with $E_0$ the
vacuum energy,
\be
\langle 0|O|0\rangle = \lim_{N_\tau\rightarrow\infty}
\frac{\sum_n\langle n| O|n\rangle\, \ex^{-aN_\tau (E_n-E_0)}}{\sum_n \ex^{-aN_\tau (E_n-E_0)}}\;.
\ee
In order to describe our gauge theory at finite temperatures, we simply need to dispense with this step.
The spectral decomposition of a thermal expectation value in \eq(\ref{boltz}) gives some important
qualitative insights. Its value is governed by the eigenvalues and eigenstates of the Hamiltonian, i.e.~the spectrum of the theory at {\it zero} temperature. The effect of temperature is the Boltzmann-weighted contribution of higher states.  This makes it plausible why for $T<T_c$ the hadron resonance
gas should give a good description of thermal expectation values: the $E_n$ in this case are
the masses, momentum and scattering eigenstates of hadrons, and the hadron resonance gas
description only neglects the latter. 

This argument is valid up to the phase transition. It can be extended
by defining a transfer
matrix and Hamiltonian in a spatial direction, $Z=\Tr(\ex^{-aN_zH_z})$. 
$N_\tau$ is now hidden in the definition of the Hamiltonian.
Thermal physics is thus equivalent to the ``zero temperature'' ($N_z\rightarrow\infty$) 
physics of the Hamiltonian $H_z$, which acts on states defined
on a space with two infinite and one finite, compactified direction. For large $N_\tau$ the
eigenstates of $H_z$ are equivalent to those of $H$, i.e.~the vacuum hadron states. 
Now finite temperature is nothing but a finite size effect:
 thermal effects become
noticeable once $N_\tau$ is small enough for the system to be sensitive to the boundary.
For $T>T_c$, the eigenstates of $H_z$ are the screening masses
$m\sim T$.
These general considerations remain the same when fermions are included.

\subsection{Including fermions}

Once a suitable fermion action 
\be
S_{f}= \sum_{x,y} \sum_f\bar{\psi}_f(x) D_{xy}(m_f)\, {\psi}_f(y)
\ee
has been selected and minding the appropriate anti-periodic boundary conditions in the temporal
direction,
the Gauss integral can be done and we end up with the partition function
\ba
Z(N_s,N_\tau;\beta,m_f)&=&\int DU\,\prod_f D\bar{\psi}_fD\psi_f\;\ex^{-S_g[U]-S_f[U,\bar{\psi}_f,\psi_f]}=\int DU\;\prod_f\det M(m_f)\; \ex^{-S_g[U]}\;,\nn\\
U_\mu(\tau,\bfx)&=&U_\mu(\tau+N_\tau,\bfx),\quad
 \psi(\tau,\bfx)=-\psi(\tau+N_\tau,\bfx)\;.
\ea
However, the selection of an appropriate lattice action is far more intricate than in the pure gauge case.
Even more difficult is the treatment of finite baryon chemical potential. Here we set $\mu_B=0$ 
and postpone the discussion of finite density to Sec.~\ref{sec:mu}.

\subsection{Naive lattice fermions}

In the naive discretisation, one simply replaces the partial derivative with a symmetrised 
finite difference,
\be
 \nabla_\mu f(x)=\frac{f(x+\hat{\mu})-f(x-\hat{\mu})}{2a}\quad
\stackrel{a\rightarrow 0}{\longrightarrow}\quad
\left(\partial_\mu+\frac{1}{6}a^2\partial_\mu^3+{\cal O}(a^4)\right)f(x)\;.
\label{eq:naivef}
\ee
which introduces leading order $\Oas$-corrections to the action.
Making the derivative covariant,
the Dirac operator reads (with $U_{-\mu}(x)=U_\mu^\dag(x-\hat{\mu})$)
\be
D_{xy}(m)=\left(\sum_{\mu=1}^4\gamma_\mu\nabla_\mu(U)+m\right)_{xy}=\sum_{\mu=1}^4\gamma_\mu\frac{1}{2a}\left(U_\mu(x)\delta_{x+\hat{\mu},y}
-U_{-\mu}(x)\delta_{x-\hat{\mu},y}\right)+m\;\delta_{xy}\;.
\label{naivferm}
\ee
The corresponding Fourier transform and quark propagator are 
\ba
D(p)&=&m+\frac{i}{a}\sum_{\mu=1}^4\gamma_\mu\sin(ap_\mu),\nn\\
D^{-1}(p)&=&\frac{m-ia^{-1}\sum_\mu\gamma_\mu\sin(ap_\mu)}{m^2+a^{-2}\sum_\mu\sin^2(ap_\mu)}
=\frac{m-ia^{-1}\sum_\mu\gamma_\mu\bar{p}_\mu}{m^2+a^{-2}\bar{p}^2}\;,
\ea
with the periodic lattice momentum $\bar{p}_\mu\equiv\sin (ap_\mu)$, which
is a consequence of the discrete Fourier transform. Consequently, in the chiral limit $m=0$,   
the fermion propagator has 16 poles ($ap_\mu=\{0,\pi\}$ per space-time direction $\mu$).
It thus describes 16 degenerate flavours of massless quarks where only one is desired. 
Since this is true for any lattice spacing however small, those 16 flavours survive an extrapolation
to zero lattice spacing, thus leading to an incorrect continuum limit with an enlarged particle content.
The same observation holds for $m\neq 0$ with poles shifted away from zero. There are several
ways of dealing with this problem, see e.g.~\cite{rothe,gl}. In the following we shall focus on 
Wilson fermions and staggered fermions, since they allow for the most realistic numerical simulations
presently. For all fermion species, at finite temperature the temporal component of momentum is represented by the Matsubara frequencies $\omega_n=(2n+1)\pi T$, 
reflecting the anti-periodic boundary conditions for the spinor fields in the temperature direction.

\subsection{Wilson fermions}
 
 A straightforward cure of the doubling problem is due to Wilson. 
His lattice Dirac operator is given by
\be
D^{W}_{xy}(m)=\left(m+\frac{4}{a}\right)\delta_{x,y}-\frac{1}{2a}\sum_{\mu=\pm 1}^{\pm 4}
(1-\gamma_\mu)U_\mu(x)\delta_{x+\hat{\mu},y}\;,
\ee
with the notation $\gamma_{-\mu}=-\gamma_\mu$. Note the additional
mass term $\sim 4/a$.
The corresponding Fourier transform 
and quark propagator read
\ba
D^{W}(p)&=&m+\frac{i}{a}\sum_{\mu=1}^4\gamma_\mu\sin(ap_\mu)
+\frac{1}{a}\sum_{\mu=1}^4(1-\cos(ap_\mu))\nn\\
 (D^W)^{-1}(p) &=& \frac{m-ia^{-1}\sum_\mu \gamma_\mu \bar{p}_\mu +\frac{1}{2}\hat{p}^2 + am}
{\bar{p}^2+\left(\frac{1}{2}\hat{p}^2 + am\right)^2}\;, 
\ea
with $\hat{p}_\mu=2\sin(ap_\mu/2)$. The 
third term in the Fourier transform is the Wilson term, which vanishes for $p=0$ but contributes $2/a$ whenever a 
momentum component $p_\mu=\pi/a$. Hence, the mass degeneracy of the doublers is lifted. 
In the massless limit there is only one pole at $p_\mu=0$ while the 15 others have poles $\sim a^{-1}$, with 
appropriate shifts for $m\neq 0$. So for $a\rightarrow 0$ the doublers become infinitely heavy and
decouple from the theory. Therefore, on sufficiently fine lattices Wilson fermions have no doubling problem.

The prize to pay is a complete loss of chiral symmetry, since the Wilson term
is not invariant under the transformations \eq(\ref{eq:axial}). 
That is, even in the massless limit there is no chiral symmetry and hence no spontaneous breaking
of it either. As a consequence, the quark mass receives both additive and multiplicative renormalisation.
One still may obtain massless pions by an appropriate tuning of the bare parameters,
but they do not correspond to Goldstone bosons for any finite lattice spacing. 
Consequently, there is no true chiral symmetry restoration at finite temperatures. 
Acting as a mass and hence symmetry breaking term, the Wilson term should soften any chiral
phase transition, similar to the quark masses in \fig\ref{fig:schem}.
Fortunately, since the Wilson term decouples in the continuum limit, all continuum symmetries get gradually restored as the lattice spacing vanishes. One would then expect
that Wilson fermions are more cumbersome when it comes to physics related to chiral symmetry, but
that eventually the correct continuum physics is obtained.

Finally, for numerical simulations it is often more convenient to express the 
Wilson Dirac operator in a slightly different form by a rescaling of fields and introducing the hopping
parameter $\kappa$,
\be
D^W_{xy}=(1-\kappa H_{xy}),\quad \kappa=\frac{1}{2am+8},  \quad H_{xy}=\sum_{\mu}\delta_{y,x+\hat{\mu}}(1+\gamma_{\mu})U_{\mu}(x)\;.
\label{eq:wilhop}
\ee

\subsection{Staggered fermions}\label{sec:stag}

The staggered action is obtained from the naive action \eq(\ref{naivferm}) by a series of 
field transformations mixing lattice and Dirac indices of the original spinor fields, for detailed formulae see \cite{rothe,gl}. 
The result is diagonal in the spinor indices, i.e.~corresponds to four identical terms in the action for each value of  the spinor index.  The fermion doubling problem may then be reduced by simply discarding three of those terms. The resulting action can be written as 
\be
S^\text{st}_{f}= \sum_{x,y} \bar{\chi}(x) D_{xy}(m)\, {\chi}(y)\;,
\label{stag}
\ee
with the staggered Dirac operator
\be
D^\text{st}_{x y}(m) =
{1\over 2} \sum_{\mu =1}^4 \eta_\mu
( \delta_{x+\hat{\mu},y} U_\mu(x) - \delta_{x,y+\hat{\mu}}
  U_\mu(y)^{\dagger}) 
+ \; \delta_{x y} am \;,
\ee
and the staggered sign function
\be
\eta_\mu(x)=(-1)^{x_1+\dots+x_{\mu-1}}, (\mu>1), \quad \eta_1(x)=1\;.
\ee
Note that this Dirac operator contains no $\gamma$-matrices and the fields $\chi(x)$ in the action
have no Dirac indices, i.e.~are one-component fields. Hence, this action is expected to describe only four fermion species compared to the 16 of the naive fermion action. 

As a by-product of this reduction, 
the one-component fermion field is significantly cheaper in numerical simulations than four-component spinors. A further advantage of the staggered discretisation is that it retains a part of the continuum chiral symmetry.
Defining $\eta_5(x)$ to be a staggered phase playing the role of the $\gamma_5$ matrix, 
\be
\eta_5(x)=(-1)^{x_1+x_2+x_3+x_4}\;,
\ee
one verifies that in the chiral limit, $m=0$, the action
\eq(\ref{stag}) is invariant under continuous transformations with $\alpha\in \mathbb{R}$,
\be
\chi(x)\rightarrow \ex^{i\alpha\eta_5(x)}\chi(x),\quad 
\bar{\chi}(x)\rightarrow \bar{\chi}(x)\ex^{i\alpha\eta_5}(x)\;.
\ee

While \eq(\ref{stag}) thus offers a few advantages,  one must also make sure that its reduced degrees of freedom describe Dirac fermions as desired. 
In order to see this, let us simplify without loss of generality to the non-interacting case, i.e.~$U_\mu(x)=1$.
The action can be reformulated once more 
by transforming the fields back into Dirac spinors $\psi^{(t)}_\alpha$  \cite{rothe,gl}, 
with a spinor index $\alpha=1,\ldots 4$ and $t=1,\dots 4$ labelling different
fermion species,
\ba
S_f&=&b^4\sum_z\left[\sum_{t=1}^4\left(m\,\bar{\psi}^{(t)}(z)\psi^{(t)}(z)+
\sum_{\mu=1}^4\bar{\psi}^{(t)}(z)\gamma_\mu\nabla_\mu\psi^{(t)}(z)\right)\right.\nn\\
&&\left. \quad\qquad -\frac{b}{2}\sum_{t,t'=1}^4\sum_{\mu=1}^4\bar{\psi}^{(t)}(z)\gamma_5(\tau_5\tau_\mu)_{tt'}
\Delta_\mu\psi^{(t')}(z)\right]\;.
\label{eq:taste}
\ea
These Dirac spinors live on a blocked lattice with twice the original lattice spacing,
$b=2a$, implying a new coordinate $z_\mu=0,1,\ldots N_s/2-1$, and finite differences over $b$,
\be
\nabla_\mu f(z)=\frac{f(z+\hat{\mu})-f(z-\hat{\mu})}{2b}, \quad 
\Delta_\mu f(z)=\frac{f(z+\hat{\mu})-2f(z)+f(z-\hat{\mu})}{b^2}\;,
\ee
where we have introduced matrices $\tau_\mu=\gamma^T_\mu, \mu=1,2,\ldots 5$.
The Fourier transform and quark propagator now read
\ba
D(p)&=&m+\frac{i}{b}\sum_{\mu=1}^4\gamma_\mu\sin(bp_\mu)
-\frac{1}{b}(1-\cos(bp_\mu))\gamma_5\tau_\mu\tau_5\nn\\
D^{-1}(p)&=&\frac{m-ib^{-1}\sum_{\mu=1}^4\gamma_\mu\sin(bp_\mu)
+\frac{2}{b}\gamma_5\tau_\mu\tau_5\sin^2(bp_\mu/2)}{\frac{4}{b^2}\sum_{\mu=1}^4\sin^2(bp_\mu/2)+m^2}\;.
\ea
Indeed, the poles at $p_\mu=\pi/a$ have disappeared thanks to the factor $1/2$ inside the $\sin$-function 
of the denominator, and we are left with only one pole at $p_\mu=0$ for each $\psi^{(t)}$.
The staggered action  thus indeed describes $N_t=4$ species of fermions. Since these are remnant
doublers, they are usually called ``tastes'' in order to  
distinguish lattice artefacts from the physical fermion species called flavours.

The kinetic and mass terms in the first line of \eq(\ref{eq:taste}) are diagonal and hence display a $SU(4)$ taste symmetry.
The third term breaks this symmetry by mixing different tastes, thus lifting the degeneracy of their masses. 
However, for $a\rightarrow 0$
the last term is suppressed compared to the other two and vanishes. Hence, in the continuum limit the 
four tastes become mass-degenerate and identical. 
This motivates the final step:  doing the Gaussian fermionic
path integral, the staggered action gives a determinant representing four tastes.  
In order to describe one single flavour it is then customary to take the fourth root of the staggered determinant, whereas two mass-degenerate flavours are approximated by taking the square root.
Hence, for $N_f=2+1$ staggered fermions,
\be
\int D\bar{\psi}D\psi \; \ex^{-S_f^\text{st}[\bar{\psi},\psi,U]}=\det[D^\text{st}(m_{ud})]^{1/2}
\det[D^\text{st}(m_s)]^{1/4}\;.
\ee
It must be stressed that this last step is only well-motivated in the continuum limit, when the tastes are degenerate. In a numerical simulation at finite lattice spacing, this is not the case and one has to 
worry about the systematic effect that this introduces. 
With four tastes of quarks, there are 16 meson states which fall into eight multiplets \cite{golter}, 
characterised by different masses and degeneracies. These can be ordered according to the
generator matrices that distinguish between the tastes, $\Gamma^F$, which can be chosen to be 
products of $\gamma$-matrices \cite{golter,morel}, as in Table \ref{tab:tastes0}.
Of the 16 pions listed in the table, only the one corresponding to $\Gamma^F=\gamma_5$ is a Goldstone
boson, i.e.~its mass $m_G$ vanishes for vanishing quark mass. Of course, this is consistent with the remnant pattern of chiral symmetry breaking for staggered fermions, where only one generator gets broken.
The other pseudo-scalar mesons are non-Goldstone bosons and have masses 
\be
m_i^2=m_G^2+\delta m_i^2,\quad \mbox{with}\quad \delta m^2_i\sim \alpha a^2\;.
\label{eq:tsplit} 
\ee
On coarse lattices with lattice spacing $a\gsim 0.25$ fm
some can be as heavy as several hundred MeV, as we shall see in detailed numerical 
analyses in Sec.~\ref{sec:taste}.
\begin{table}
\begin{center}
\begin{tabular}{c|c|c}
index&$\Gamma^F$&multiplicity $n_i$\\
\hline
0&$\gamma_5$&1\\
1&$\gamma_0\gamma_5$&1\\
2&$\gamma_i\gamma_5$&3\\
3&$\gamma_i\gamma_j$&3\\
4&$\gamma_i\gamma_0$&3\\
5&$\gamma_i$&3\\
6&$\gamma_0$&1\\
7&1&1\\
\hline
\end{tabular}
\end{center}
\caption{Pion taste multiplets labeled by taste matrices $\Gamma^F$ have 
multiplicity $n_i$ within the multiplets. 
}
\label{tab:tastes0}
\end{table}

Aside from the taste splitting, there are several other conceptual questions concerning staggered fermions which are still under investigation and debate. 
It is not entirely clear whether the rooted determinants can be represented by a local lattice field theory when re-exponentiated. This has repercussions on whether the continuum limit of the staggered theory
falls into the correct universality class. 
Another question concerns the coupling between topological sectors and zero modes of the staggered Dirac operator, since the coordinate and spinor degrees of freedom are staggered on different lattice sites. This is important in the small mass region close to the chiral limit, when the $U(1)_A$ anomaly plays a role. For references discussing these issues, see \cite{stag1,stag2,stag3,stag4}. Since so far there are no conflicting results between staggered and Wilson fermions, the working assumption for the purposes of this review is that these latter questions can be satisfactorily answered one day,
leaving cut-off effects and taste breaking as systematic errors to monitor in numerical simulations.

\subsection{Observables related to centre and chiral symmetry on the lattice}

In Sec.~\ref{sec:zn} we have seen that the centre symmetry of the partition function 
corresponds to a subgroup of the gauge symmetry of Yang-Mills theory. Since gauge symmetry
is fully realised in the lattice formulation, so is centre symmetry. The Polyakov loop 
is represented by a product of temporal links,
\be
L({\bfx})=\prod_{x_0}^{N_\tau}U_{0}(x)\;,
\label{eq:poly}
\ee 
with the same transformation behaviour as in the continuum. Also in analogy to the continuum, centre symmetry on the lattice is broken by including finite mass fermions. While the Polyakov loop 
is no longer an order
parameter for confinement in this situation, it nevertheless signals a change in the dynamics by
measuring the free energy of a static quark in the plasma.

The self energy of the static quarks cause a UV-divergence,
and for comparisons between different actions one needs to consider renormalised quantities.
Since the Polyakov loop is not directly related to an experimentally observable quantity, there is
some freedom in its renormalisation,
\be
L_R(T) = Z(\beta)^{N_\tau}L(\beta)\;.
\ee
One possible choice to fix the renormalisation factor $Z(\beta)$ is to relate it to 
the free energy of a static quark anti-quark pair at infinite separation, $L_R(T)=\exp{(-F_\infty(T)/2T)}$
\cite{kac}. A more practical procedure is to normalise the zero temperature static potential, obtained
from Wilson loops, at a given distance to a prescribed value, e.g.~$V_R(r_0)=0$, and then use the same additive shift for the free energy obtained from Polyakov loops at the same lattice spacing \cite{bmw_tc1}.
This translates into $Z(\beta)=\exp(V(r_0,\beta)/2)$, with the unrenormalised zero temperature static potential.

Similarly, the continuum chiral symmetry is
explicitly broken by the quark masses and the finite temperature transition is a smooth crossover instead
of a phase transition. 
While the chiral condensate for a light flavour is no longer a true order parameter, it still signals changes
in the dynamics.  
It is usually expressed by the inverse propagator,
\be
\langle \bar \psi \psi_f \rangle=\frac{1}{4} \frac{1}{N_{\sigma}^3 N_{\tau}} 
{\rm Tr} \langle D_f^{-1} \rangle,~~f=l,s \ ,
\ee
for the light and strange flavours, respectively.
Other interesting fermionic quantities are the corresponding quark number susceptibilities,
\begin{equation}
\frac{\chi_{f}}{T^2} = \frac{1}{VT^3}
\left. \frac{\partial^2\ln Z}{\partial(\mu_{f}/T)^2}\right|_{\mu_f=0} \; ,\;\; f=l,\; s \ . 
\label{chi_q}
\end{equation}
As we have already discussed, the lattice situation for chiral symmetry differs significantly from the
continuum. For Wilson fermions, chiral
symmetry is broken explicitly for any finite lattice spacing, even for massless quarks, 
whereas for $N_f=2$ staggered fermions  it is reduced to a $O(2)$ subgroup. 
Due to the breaking of chiral
symmetry on the lattice, both additive and multiplicative renormalisation of the chiral condensate are 
necessary. Renormalisation of the chiral condensate for Wilson fermions 
is discussed in \cite{giusti}, the presentation here 
follows \cite{bmw_w1,bmw_w2}. 
The dimension three operator $\bar{\psi}\psi$ features a divergence $\sim a^{-3}$ which can be removed
by subtracting the corresponding vacuum value. The multiplicative divergence gets removed by 
multiplying appropriate factors of quark masses displaying the same divergence, ensured by axial
Ward identities. Near the continuum limit the following relations hold between renormalised 
and PCAC quark masses,
\ba
\label{final}
m_R \pbp_R(T) &=& m_{PCAC} Z_A \Delta_{\bar{\psi}\psi}(T) + O(a)\;, \\
m_R \pbp_R(T) &=& 2 N_f m_{PCAC}^2 Z_A^2 \Delta_{PP}(T) + O(a)\;, 
\label{eq:defmR}
\ea
where $Z_A$ is a finite renormalisation constant. The latter can be eliminated together with $m_{PCAC}$
to arrive at
\be
m_R \pbp_R(T) = \frac{ \Delta_{\bar{\psi} \psi}^2(T) }{ 2 N_f \Delta_{PP}(T) } + O(a)\;.
\label{eq:chirratio}
\ee
For the condensate one has
\ba
\label{diff1}
\Delta_{\bar{\psi}\psi}(T) &=& \pbp(T) - \pbp(T=0)\;, \nn\\
\Delta_{PP}(T) &=& \int d^4 x \langle P_0(x) P_0(0) \rangle(T) 
- \int d^4 x \langle P_0(x) P_0(0) \rangle(T=0)\;, 
\ea
where $\pbp$ and $P_0(x)$ are the bare chiral condensate and bare pseudo-scalar 
density, respectively \cite{giusti}. Note that \eqs(\ref{eq:defmR}, \ref{eq:chirratio}) have different corrections
and hence different scaling behaviour, which must be determined numerically at this point.
Finally, a definition for a finite condensate which is used in many analyses employing staggered 
fermions is \cite{cheng_08} (the indices $\tau,0$ denote finite or zero temperature expectation values,
respectively)
\begin{equation}
\Delta_{l,s}(T)=\frac{\langle \bar\psi \psi \rangle_{l,\tau}-\frac{m_l}{m_s} \langle \bar \psi \psi \rangle_{s,\tau}}
{\langle \bar \psi \psi \rangle_{l,0}-\frac{m_l}{m_s} \langle \bar \psi \psi \rangle_{s,0}}\;.
\label{eq:deltals}
\end{equation}

\subsection{Tuning temperature, continuum limit and lines of constant physics}

According to \eq(\ref{temp}), 
the simplest way of tuning temperature in a numerical simulation 
on the lattice is by choosing $N_\tau$ for a given lattice spacing $a$. This is known as the
fixed scale approach whose application to the equation of state is discussed in Sec.~\ref{sec:fixed}.
Often this is not satisfactory as this allows only discrete temperatures. Then it is more practical to keep 
$N_\tau$ fixed and vary the lattice spacing $a$ via the coupling, $\beta=2N/g^2(a)$, thus affecting 
temperature. This implies a few marked differences to simulations at zero temperature. In particular, simulation 
points at different temperature correspond to different lattice spacings and thus have different cut-off effects.
The continuum limit, $a\rightarrow 0$, via the running coupling corresponds to the limit 
$\beta\rightarrow \infty$ while keeping observables fixed in physical units. For simulations at fixed
temperature $T$, this also implies $N_\tau\rightarrow \infty$. 
A continuum extrapolation therefore requires simulations
on a sequence of lattices with different $N_\tau$. 

In order to take the continuum limit, we need to compute expectation values of 
our observables of interest, $\langle O\rangle(\beta,m_f)$, in the thermodynamic limit for various
lattice spacings $a$. Then we can extrapolate to $a\rightarrow 0$ while keeping temperature and 
physical masses fixed. Similarly, when simulating
different temperatures on a lattice with fixed $N_\tau$ the lattice spacing is
different for different temperatures. This means that also the bare quark masses in lattice units, $am_f$,
need to be tuned such that the masses in physical units stay constant. 
A line of constant physics is an implicitly defined curve in parameter space  along which the bare quark masses are tuned together with the lattice gauge coupling, 
such that the renormalised quark masses in physical units is kept constant,
\be
m^R_f(am_{u,d}(\beta), am_s(\beta), \beta)=\text{const}\;.
\ee
Renormalised quark masses are scheme dependent and complicated to compute. In practice it is better
to define the lines of constant physics by physical observables that can be straightforwardly computed
in lattice simulations, such as hadron masses or decay constants. 
These are renormalisation group invariants and the
corresponding lines of constant physics read
\ba
O^\text{phys}_i(am_{u,d}(\beta), am_s(\beta), \beta)&=&\text{const},\quad i=1,2,3,\ldots\;,
\ea 
where we need as many observables as parameters to tune.
In principle, all chosen observables must lead to the same result. However, in practice 
the presence of cut-off effects also in those quantities renders different choices inequivalent.
For finite lattices spacings, observables display lattice corrections,
\be
O^\text{phys}(a)=O^\text{phys}+c_1a+c_2a^2+\ldots\;,
\ee
with different coefficients for different observables (and actions). Keeping some observables fixed
in physical units  will have other observables move while changing the lattice spacing.
Thus different lines of constant physics become equivalent only in the continuum limit, and
careful choices have to be made in order to minimise cut-off effects during the extrapolation.

The relation between the lattice spacing $a$ and the bare parameters $\beta, m_f$ is given by the renormalisation group. 
E.g., in units of the Lambda-parameter in lattice regularisation we have to leading 
order in perturbation theory for $SU(3)$ pure gauge theory,
\ba
a\Lambda_L&=&\left(\frac{6b_0}{\beta}\right)^{-b_1/2b_0^2}\;\ex^{-\frac{\beta}{12b_0}},\nn\\
b_0&=&\frac{1}{16\pi^2}\left(11-\frac{2}{3}N_f\right), \quad b_1=\left(\frac{1}{16\pi^2}\right)^2
\left[102-\left(10+\frac{2}{3}\right)N_f\right]\;.
\ea

Similar relations exist for the running of the quark masses.
However, perturbation theory is 
not convergent for most accessible lattice spacings. 
The way out is to express the calculated 
observables in terms of known physical quantities of the same mass dimension.
For example, if we want to compute the pseudo-critical temperature of the QCD phase transition, $T_c$,
by keeping $N_\tau$ fixed and tuning $\beta$, the location of 
a phase transition will be given as a critical coupling $\beta_c$.
The critical temperature in units of a hadron mass is then
\be
\frac{T_c}{m_H}=\frac{1}{a_cm_H N_\tau}=\frac{1}{a(\beta_c)m_H N_\tau}\;.
\ee

In order to set the physical scale for $T_c$, we then have to calculate the zero temperature 
hadron mass in lattice units at the value of the critical coupling, $(am_H)(\beta_c)$. Equating
this number with the hadron mass in physical units times the lattice spacing provides the scale for the lattice, 
\be
a^{-1}=\frac{m_H[\text{MeV}]}{(am_H)(\beta_c)}\;.
\ee
This is easily translated back to the critical temperature in physical units. 
Of course, the same procedure applies for observables other than the temperature, and besides a choice
of hadron masses, also decay constants or other observables that are easy to measure on the lattice
may be used.

This procedure is good as long as simulations take place at physical quark masses. For many problems,
this is not yet the case, and furthermore there are many interesting theoretical questions
concerning regimes with unphysical quark masses, such as the quenched and the chiral limits.
In order to set a scale in those cases, one uses quantities that display only little sensitivity to the quark mass values. Frequently used is the Sommer scale $r_0$ and its variant $r_1$, which are 
based on the potential of a static quark anti-quark
pair evaluated to prescribed values \cite{sommer}, 
\be
\left(r^2\frac{dV_{\bar{Q}Q}(r)}{dr}\right)_{r=r_0}=1.65\;,\quad 
\left(r^2\frac{dV_{\bar{Q}Q}(r)}{dr}\right)_{r=r_1}=1.0\;.
\ee
The advantage of these quantities is that they can be straightforwardly computed in
lattice calculations using Wilson loops, 
and that their numerical values change indeed very little with variations of the sea 
quark masses over wide ranges. 
The disadvantage is that the signal to noise ratio for the potential is worse than for typical hadron
masses, and some care has to be taken concerning rotational symmetry when using improved 
estimators. 
Moreover, numerical determinations yield dimensionless numbers in lattice units, $r_0/a, r_1/a$,
but there are no direct experimental measurements for these quantities. Therefore, additional lattice 
simulations are necessary in order to establish a relation to experimental observables, introducing 
additional systematics. The closest
relation of the static potential is to spectral quantities of heavy quarkonia. Examples  
used in recent precision calculations \cite{follana} are the mass splitting of different radial excitations of bottomonium, 
the mass splitting between the $D_s$ and $\eta_c$  mesons or 
the decay constant of the fictitious $\eta_{\bar{s}s}$, which is in turn related to $f_\pi,f_K$.
Another possibility is to set the scale for the lattice spacing by a simulation of a light hadron mass
at physical quark mass values as described above, and then compute $r_0/a,r_1/a$ in order to 
set the scale also away from the physical point.  

It should be clear that determining the lines of constant physics as well as setting the
physical scale, albeit being performed at zero temperature, are an important part of a finite temperature
simulation and play a major role in controlling and eliminating the cut-off dependence of numerical 
results. 

\subsection{Constraints on lattice simulations}
\label{sec:syst}

Lattice simulations 
are beset by systematic 
errors and uncertainties, the two obvious being finite lattice spacing and finite volume. 
An important question before running a simulation or when interpreting
its results is then: how large and fine does a lattice need to be for a particular problem?
The Compton wave length of a hadron is proportional to its inverse mass $m_H^{-1}$, and the largest Compton wave length constitutes the correlation length of the statistical system. 
To keep both finite size and discretisation errors reasonably small, we need to require
\be
a\ll m_H^{-1}\ll aN_s.
\label{latconst}
\ee
For the low $T$ confined phase we thus need a lattice size of several inverse pion masses. 
The push to do physical quark masses is only just beginning to be a possibility on the most powerful machines and with the cheapest actions
(i.e.~staggered, with Wilson rapidly catching up). On the other hand, at high $T$
screening masses scale as $m_H\sim T$, thus
\be
N_\tau^{-1}\ll 1\ll N_sN_\tau^{-1}.
\ee 
While we desire large $N_\tau$ in order to have fine lattices for a given temperature, the spatial
lattice size should be significantly larger than the temporal one. 
This combination of requirements limits the directly accessible temperatures to a few times $T_c$.

To get some numbers, for a temperature $T=200$ MeV $\sim 1$ fm$^{-1}$,  \eq(\ref{temp}) with $N_\tau=4,8,12$ implies a lattice spacing $a\approx 0.25,0.125,0.083$ fm. We shall see later that
the scaling region for the equation of state for most actions is entered only for $a\lsim 0.1$ fm. 
Demanding only $m_\pi L\gsim 2$ means for the physical pion $L\gsim 2\cdot 1.5$ fm. 
From this the required lattice sizes $N_s$ follows for $N_\tau=10$ to be at least $N_s\gsim 30$.
On the other hand, renormalisation requires also zero temperature calculations, i.e.~$N_\tau$ as large
as possible. For $a=0.1$ fm, $N_\tau=30$ corresponds to $T\sim 7$ MeV.

\subsection{The ideal boson gas on the lattice}

Similar to the continuum, we can evaluate the ideal gas for a bosonic field on the lattice.
Starting point is the equation employing the free propagator, 
\be
\ln Z_0=-\frac{1}{2} \ln\det \Delta=\frac{1}{2}\Tr\ln \Delta^{-1}
=V\sum_{n=-N_\tau/2}^{N_\tau/2-1}
\int_{\frac{\pi}{a}}^{\frac{\pi}{a}}\frac{d^3p}{(2\pi)^3}\ln(\hat{p}^2+(am)^2)\;,
\ee 
where now the integration is over the Brillouin zone and we have the lattice momenta,
\be
\hat{p}^2=4\sin^2(\frac{a\omega_n}{2})+4\sum_{j=1}^3\sin^2(\frac{ap_j}{2}),\quad
4\ho^2=4\sum_{j=1}^3\sin^2(\frac{ap_j}{2})+(am)^2\;.
\ee
Due to the finite lattice spacing
the Matsubara sum is only from $-N_\tau/2,....N_\tau/2-1$.
After performing the Matsubara sum (for details, see \cite{oplh}),   
and a change of variables by $\ho=\sinh(aE/2)$  the partition function reads
\be
\ln Z_0=-V\int_{-\frac{\pi}{a}}^{\frac{\pi}{a}} \frac{d^3p}{(2\pi)^3}\ln(1-\ex^{-N_\tau aE})\;.
\ee
Note that, despite the fact that the details of the calculation are quite different, this formula could have 
been obtained from the continuum case, \eq(\ref{freegas}), by simply replacing momenta, energies
and integration ranges by their lattice analogues.

One can now study the approach to the continuum by 
expanding in small lattice spacing. The integration range of the coefficients then gets
extended to infinity and 
the lattice corrections to the partition function are due
to the lattice corrections to the dispersion relation $E(\bf{p})$. The latter is defined by the zeros of the 
denominator of the propagator.  Writing
\be
E(\bfp)=E^{(0)}(\bfp)+aE^{(1)}+a^2E^{(2)}(\bfp)+\ldots, \qquad
E^{(0)}(\bfp)=\sqrt{\bfp^2+m^2},
\ee 
and expanding the dispersion relation
\ba
\sinh^2(\frac{aE}{2})&=&\sum_{j=1}^3\sin^2(\frac{ap_j}{2})+\frac{(am)^2}{4},\nn\\
\frac{(aE)^2}{4}+\frac{(aE)^4}{48}&=&\sum_{j=1}^3\left(\frac{(ap_j)^2}{4}-\frac{(ap_j)^4}{48}\right)
+\frac{(am)^2}{4}+O(a^6),
\ea
one finds the leading $a^2$-correction to the dispersion relation
\be
E^{(2)}(\bfp)=-\frac{1}{24E^{(0)}(\bfp)}\left(\sum_{j=1}^3p_j^4+E^{(0)4}(\bfp)\right).
\ee
Note that this correction term breaks the continuum rotational invariance of the dispersion relation
at the $p_i^4$-level of momentum components. 
Changing to dimensionless variables, $x_i=p_i/T, \varepsilon=E/T$, we have for the expansion
of the pressure
\be
 \frac{p}{T^4}=\left(\frac{p}{T^4}\right)_{\rm cont}-a^2\int\frac{d^3x}{(2\pi)^3}
 \frac{\varepsilon^{(2)}(x)}{\ex^{\varepsilon^{(0)}(x)}-1}+\ldots
\ee
Hence the pressure of
a free gas of bosons has leading lattice corrections of $O(a^2)$. 
For the massless case the integral can be done in closed form and one arrives at 
(for details, see \cite{oplh})\footnote{In the first calculations \cite{ekr,bkl} the leading coefficient is incorrectly 
reported as $10\pi^2/21$.}
\be
\frac{p}{p_{\rm cont}}=1+\frac{8\pi^2}{21}\frac{1}{N_\tau^2}+O\left(\frac{1}{N_\tau^4}\right)\;,
\label{eq:p_a}
\ee
where $p_{\rm cont}$ is the continuum result \eq(\ref{psb}). 
A similar calculation including finite size effects due to finite volume can be found in \cite{eks}.

\subsection{The ideal fermion gas on the lattice}

This procedure can be generalised to the fermionic case \cite{hegde_08}
to obtain
\be
\frac{p}{T^4}=N_\tau^3\int_{-\frac{\pi}{a}}^{\frac{\pi}{a}}
 \frac{d^3p}{(2\pi)^3}\left[\ln(1+\ex^{a\mu N_\tau} \ex^{-N_\tau aE})+\ln(1+\ex^{-a\mu N_\tau}\ex^{-N_\tau aE})\right]\;.
\ee 
For Wilson fermions, the inverse propagator is of the form
\be
D(p_4,\p)=\bar{p}^2+\left(\frac{\hat{p}^2}{2}+am\right)^2,\quad \bar{p}_\mu=\sin(ap_\mu)
\ee
Introducing again the energy variable $\sin^2 (ap_4)=-\sinh^2(aE)$, the dispersion relation is
\be
(aE)^2+\frac{1}{12}(aE)^4-\sum_i\left((ap_i)^2-\frac{(ap_i)^4}{3}\right)-(am)^2-\frac{a^4\bfp^4}{4}
+\frac{1}{2}(aE)^2a^2\bfp^2+am\left((aE)^2-a^2\bfp^2\right)=0\;,
\ee
which implies for the leading lattice correction
\ba 
E^{(0)}&=&\sqrt{\bfp^2+m^2},\quad E^{(1)}=-\frac{m^3}{2E^{(0)}}\;.
\ea
\begin{figure}[t]
\centerline{
\includegraphics[width=0.4\textwidth,angle=-90]{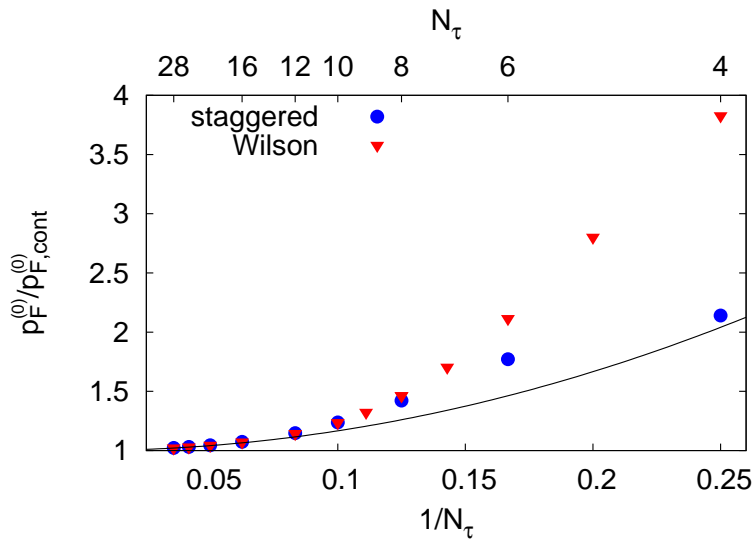}
}
\caption{\label{fig:free}
Stefan-Boltzmann limit for 
massless fermions, normalised to the continuum result. The solid line is the analytic 
$\mathcal{O}(1/N_\tau^2)$ prediction.
From \cite{pz}.
}
\end{figure}
The dispersion relation 
for staggered quarks to order $p_i^4$ is
\be
(aE)^2-a^2\bfp^2-(am)^2+\frac{1}{3}\sum_{j=1}^3(ap_j)^4+\frac{1}{3}(aE)^4=0
\ee
with the solution
\be
E^{(2)}(\bfp)=-\frac{1}{6}\left(\sum_{j=1}^3p_j^4+E^{(0)4}\right)\;.
\ee
As in the pure gauge case, the corrections to the continuum result start at $\Oas$, whereas for Wilson
fermions this is only true in the massless limit.
The approach to the continuum between free massless staggered and Wilson fermions
is shown in \fig\ref{fig:free}. Note that the scaling region, where leading lattice corrections 
are $\sim a^2=T^2/N_\tau^2$, sets in for $N_\tau\gsim 10$ only, for coarser lattices corrections
seem to be smaller in absolute size for staggered fermions.

\subsection{The equation of state from strong coupling expansions}\label{sec:sc}

Strong coupling expansions are well known from spin models and QCD at zero temperature. 
In contrast to the asymptotic series obtained by weak coupling expansions, they yield convergent 
series in the lattice gauge coupling $\beta=2N/g^2$
within a finite radius of convergence. The series approximate the true answer the better
the lower $\beta$. For a fixed $N_\tau$ this means low temperatures and hence the confined phase.
For pure gauge theory, $\beta_c$ of the deconfinement transition represents an upper bound
on the radius of convergence. 
A detailed introduction to strong coupling methods in the vacuum can be found 
in \cite{mm}.
Here we merely discuss modifications for finite temperature and the calculation
of the pressure \cite{lmp,lppress}. 

Using the formalism of moments and cumulants \cite{mm}, the pressure of pure gauge theory
on a lattice with temporal extent $N_\tau$ can be written as
\begin{eqnarray}
p(N_{\tau})=\frac{6}{a^4}\ln\,c_0(\beta)+\frac{1}{N_{\tau}V}\sum_{C=\left(X_i^{n_i}\right)}a(C)\prod_i\Phi(X_i)^{n_i},
\end{eqnarray} 
where $C$ specifies a cluster of
graphs and $a(C)$ is a combinatorial factor \cite{mm}.
$\Phi(X_i)$ is the contribution of a graph $X_i$, 
\be
\Phi(X_i)=\int DU \prod_{p\in X_i}d_r c_{r}\chi_{r}(U_p)\;,
\ee
with $\chi_r(U_p)$ the character of the $d_r$-dimensional 
representation $r$ of the plaquette $U_p$, and $c_r(\beta)$ are
expansion coefficients. The ones for the lowest-dimensional representations go as
\be
u=\frac{\beta}{18}+{\cal O}(\beta^2),\quad v=\frac{\beta^2}{432}+{\cal O}(\beta^4),\quad
w=\frac{\beta^2}{288}+{\cal O}(\beta^4)\;,
\ee
and the fundamental representation coefficient $c_f=u$ serves as an effective expansion
parameter.
The limit $N_\tau\rightarrow \infty$ corresponds to the zero temperature contribution.

\begin{figure}[t]
\centerline{
\includegraphics[height=2cm]{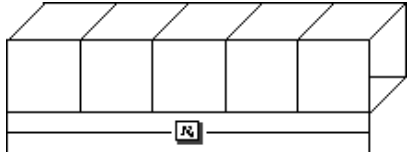}
}
\caption{Leading order tube contributing to the pressure for $N_{\tau}=4$. 
}
\label{fig:tubes}
\end{figure}

The physical pressure is given by the difference
\begin{eqnarray}
p_{\mathrm{phys}}=p(N_{\tau})-p(\infty)\;,
\end{eqnarray}
where the subtraction of vacuum contributions  
is necessary for renormalisation.
The lowest order graph $X_1$ contributing to the difference 
is a tube of length $N_\tau$ with a cross-section of one single plaquette, as shown in \fig\ref{fig:tubes}.
There are three spatial directions
for the cross section of the tube, giving a factor of 3. Translations in time take the graph into
itself and do not give a new contribution, while we get $V\Phi(X_1)$ from all spatial
translations, and a factor of two for the fundamental and anti-fundamental representation of $SU(3)$.
Thus, the leading order result for the pressure is
\be
p_{\mathrm{phys}}=\Phi(X_1)=\frac{3}{N_\tau}u^{4N_\tau}\;.
\ee
If we now consider fixed $N_\tau$, low $\beta$ and $u(\beta)$ corresponds to low temperature.
We thus recognise the exponential suppression of the pressure in the low temperature, strong coupling
regime. 

Corrections consist of plaquettes in higher dimensional representations inside the torus, additional
cubes on the tube and slits
by inserting two fundamental plaquettes at the same location. One finds for
all $N_\tau\geq5$ up to ${\cal{O}}(u^8)$ in the correction \cite{lppress} (with $c=1+3u+6v+8w-18u^2$)
\begin{eqnarray}
a^4p(N_{\tau},u)&=&\frac{3}{N_{\tau}}u^{4N_{\tau}}\left\lbrace c^{N_{\tau}}\left[1+12\,N_{\tau}{u}^{4}+42\,N_{\tau}{u}^{5}-{\frac {115343}{2048}}\,N_{\tau}{u}^{6}\right.\right.\nonumber\\
&&\left.\hspace{2cm}-{\frac {
597663}{2048}}\,N_{\tau}{u}^{7}+ \left(83\,{N_{\tau}}^
{2} +{\frac {72206061}{40960}}\,N_{\tau} \right) {u}^{8}\right]\nonumber\\
&&+b^{N_{\tau}}\left[1+12\,N_{\tau}{u}^{4}+30\,N_{\tau}{u}^{5}-{\frac {17191}{256}}\,N_{\tau}{u}^{6}\right.\nonumber\\
&&\left.\left.\hspace{2cm}-180\,N_{\tau}{u}^
{7}+ \left({83\,{N_{\tau}}^{2}+\frac {3819}{10}}\,N_{\tau} \right) {u}^{8}\right]\right\rbrace. 
\label{eq:scpress}
\end{eqnarray}

Similar techniques can be used to include dynamical fermions \cite{lppress}.
Wilson's fermion action in the form of \eq(\ref{eq:wilhop}) permits an expansion in powers
of the hopping matrix,
\be
-S_q^f=-\mathrm{tr}\,\ln\,(1-\kappa_fH)=\sum_{l}\,\frac{\kappa_f^l}{l}\,\mathrm{tr}\,H^l\;.
\ee
A truncation of this expansion is valid for heavy quarks, for which $\kappa_f$ is small.
For sufficiently strong coupling, the gluonic contribution to the partition function can be omitted and
\begin{eqnarray}
\label{hop}
Z(\kappa_f)=\int\left[dU\right] \exp\left[\sum_f\sum_{l=1}^{\infty}\frac{\kappa_f^l}{l}\mathrm{tr}\,M_f[U]^l\right].
\end{eqnarray}
In principle the sum extends over all closed loops $l$ on the lattice, which for 
finite temperature lattices can have non-trivial winding numbers,
resulting in Polyakov loops. 
%
Expanding all terms up to ${\cal{O}}(\kappa^{3N_{\tau}})$ and doing the group integrals 
one finds for the pressure for two flavours up and down \cite{lppress}\;, 
\begin{eqnarray}
p&=&\frac{1}{N_{\tau}a^3}\bigg\lbrace4(2\kappa_u)^{2N_{\tau}}+8(2\kappa_u2\kappa_d)^{N_{\tau}}+4(2\kappa_d)^{2N_{\tau}}\bigg\rbrace\nonumber\\
&+&\frac{1}{N_{\tau}a^3}\bigg\lbrace4(2\kappa_u)^{3N_{\tau}}+6\big[(2\kappa_u)^22\kappa_d\big]^{N_{\tau}}\nonumber\\
&+&6\big[2\kappa_u(2\kappa_d)^2\big]^{N_{\tau}}+4(2\kappa_d)^{3N_{\tau}}\bigg\rbrace\Big(\mathrm{e}^{3a\mu N_{\tau}}+\mathrm{e}^{-3a\mu N_{\tau}}\Big)\;.
\label{eq:hpress}
\end{eqnarray}

\subsection{The hadron resonance gas from strong coupling lattice QCD}\label{sec:hrgsc}


It is now interesting to ask how the QCD pressure can be interpreted in the strong
coupling regime.
It can be shown from first principles that it corresponds to a 
hadron resonance gas, 
without recourse to experimental numbers \cite{lppress}. 
Since the hopping expansion is valid for heavy quarks, 
it is convenient to use the non-relativistic approximations $E\simeq m+\bfp^2/(2m)$.
Inserting this in the expression for the ideal gas,
the momentum integration gets trivial and the ideal gas pressure takes the form
\begin{eqnarray}
p_i=\frac{\nu_i}{N_{\tau}a^3}
\eta\ln\left[1+\eta\mathrm{e}^{-\left(m_i-\mu_i\right)N_{\tau}}\right]\;.
\end{eqnarray}
For non-relativistic particles it is also justified to consider the limit
$m\gg T=1/N_{\tau}$, which yields
\begin{eqnarray}
p=\sum_ip_i=\frac{1}{N_{\tau}a^3}\sum_i\nu_i\mathrm{e}^{-(m_i-\mu_i)N_{\tau}}.
\end{eqnarray}
Using the fact that to each baryon with $\mu_B$ there is an anti-baryon with $-\mu_B$, we obtain
\begin{eqnarray}
p=\frac{1}{N_{\tau}a^3}\sum_M\nu_M\mathrm{e}^{-m_MN_{\tau}}+\frac{1}{N_{\tau}a^3}\sum_B\nu_B\mathrm{e}^{-m_BN_{\tau}}\cosh(\mu_BN_{\tau}),
\label{eq_hrglatt}
\end{eqnarray}
where we have split the sum into a mesonic and a baryonic part.
It is this form of the pressure which can be derived from a first principles strong coupling calculation.

In the case of $SU(3)$ pure gauge theory, the three lowest lying glueball states, which can be 
extracted from the strong coupling series of 
plaquette correlators, have masses of \cite{mu,seo}
\begin{eqnarray}
am_1(A)&=&-4\ln\,u-\ln\left(1+3u+6v_1+8w-18u^2\right)+{\cal{O}}(u^4)\nn\\
am_2(E)&=&-4\ln\,u-\ln\left(1+3u+6v_1+8w-18u^2\right)+{\cal{O}}(u^4)\nn\\
am_3(T)&=&-4\ln\,u-\ln\left(1-3u-6v_1+8w\right)+{\cal{O}}(u^4).
\end{eqnarray}
The arguments of the masses 
denote the spin representation of the associated point group for lattice rotations, and the subscript the corresponding spin degeneracy of the glueball. 
The strong coupling series for the pressure, \eq({\ref{eq:scpress}), can now be rewritten as
\begin{eqnarray}
p=\frac{1}{N_{\tau}a^3}\left[\mathrm{e}^{-m_1(A)N_{\tau}}+2\mathrm{e}^{-m_2(E)N_{\tau}}+3\mathrm{e}^{-m_3(T)N_{\tau}}\right]\left(1+{\cal{O}}\left(u^4\right)\right),
\end{eqnarray}
which has to be compared with Eq.~(\ref{eq_hrglatt}). 
We see that the hadron resonance gas result equals the one derived with 
strong coupling methods to the orders considered here. Corrections to the ideal hadron gas results appear in ${\cal{O}}(u^4)$.

These considerations can be extended to include dynamical fermions. 
Using
$am_f\approx-\ln2\kappa_f$, which is valid for heavy fermions  \cite{hasen,degrand},
one may derive the hadron masses to leading order hopping expansion
\begin{eqnarray}
\mbox{Mesons:}\qquad\quad am_{f\bar{f^\prime}}&=&-\ln2\kappa_f-\ln2\kappa_{f^\prime}\nonumber\\
\mbox{Baryons:}\qquad am_{ff^\prime f^{\prime\prime}}&=&-\ln2\kappa_f-\ln2\kappa_{f^\prime}-\ln2\kappa_{f^{\prime\prime}}\;.
\end{eqnarray}
The pressure, \eq(\ref{eq:hpress}), may then be rewritten as
\begin{eqnarray}
p&=&\frac{1}{N_\tau a^3}\left\lbrace\sum_{0^-}\mathrm{e}^{-m\left(0^-\right)N_\tau}+3\sum_{1^-}\mathrm{e}^{-m\left(1^-\right)N_\tau}\right\rbrace\nonumber\\
&+&\frac{1}{N_\tau a^3}\Bigg\lbrace4\sum_{\frac{1}{2}^+}\mathrm{e}^{-m\left(\frac{1}{2}^+\right)N_\tau}+8\sum_{\frac{3}{2}^+}\mathrm{e}^{-m\left(\frac{3}{2}^+\right)N_\tau}\Bigg\rbrace\cosh\big(\mu_BN_\tau\big),\label{eq_hrg}
\end{eqnarray}
where the sum over the pseudoscalar mesons includes the pions as well as the $\eta^0$, and the vector mesons include also the $\omega^0$, since the calculation is valid for heavy quarks. The prefactors for the baryons include the spin degeneracy as well as a factor $2$ counting the corresponding anti-baryons. 

We may thus conclude this section with the remarkable observation that, in the fully interacting theory, the hadron resonance gas 
emerges from a first principles calculation in the strong coupling limit of the equation of state for
lattice QCD. Interactions between hadrons are suppressed by high powers of the inverse coupling $1/g^2$
until very close to the critical temperature. Conversely, agreement of simulation data with 
hadron resonance gas predictions signals that the underlying dynamics is governed by strong coupling.
Furthermore, these results show from first principles that the pressure is exponentially small for temperatures not far below the quark hadron transition.

\section{Improved actions}\label{sec:impr}

\subsection{Generalities}

Since lattice actions are not unique, one may use the freedom in the choice of irrelevant terms to minimise
cut-off effects. This can be systematised by making all $a$-dependence explicit. We are interested in a lattice action $S_L$. Following 
Symanzik \cite{sym}, one may write down an effective continuum action with the same symmetries
as $S_L$,
\be
S_\text{eff}=\int d^4x\;\left\{{\cal L}_0(x)+a{\cal L}_1(x)+a^2{\cal L}_2(x)+\ldots\right\},
\ee
where the ${\cal L}_k$ are composite local operators with canonical dimension $4+k$ and ${\cal L}_0$ 
corresponds to the continuum theory. 
$S_\text{eff}$ thus produces the same vertex functions as $S_L$ up to a certain order in $a$. 
Similarly, one may define effective fields  
as a power series in the cut-off, $\phi_\text{eff}(x)=\phi_0(x)+a\phi_1(x)+\ldots$
The idea of improvement is to construct lattice representatives of the ${\cal L}_k,\phi_k$ and add them
to the lattice action $S_L$ and the fields in the observables of interest. 
If the coefficients are appropriately chosen,
lattice corrections to a certain power in $a$ can be made to disappear. The task of improvement lies in the choice of the coefficients. A straightforward way to determine them is lattice perturbation theory, but in this 
case the improvement is not complete as it holds to only a finite order in the coupling constant. Alternatively, one can determine the improvement coefficients non-perturbatively through simulations upon which the improvement is exact to a given order in the lattice spacing. However, this is often complicated or
associated with high computational cost. The only non-perturbatively complete improvements to date
are $\Oa$-improved Wilson fermion actions, see below.

Besides the Symanzik program, there is also a more ``empirical'' way to improve lattice actions and operators. These are based on the idea that the discretisation effects arise through short wave length fluctuations of
the gauge fields on the scale of the cut-off. Smoothing these fluctuations should decrease cut-off
effects. This can be achieved by replacing elementary links by more extended objects, such as
linear combinations of a link with the adjacent staples. These ``smeared''
or ``fattened'' objects have their dominant fluctuations on the scale of several lattice spacings instead of one. In some cases fat links are related to a perturbative improvement to some order in the coupling constant or
lattice spacing, but in most others the improvement is not complete to any definite order in $a$. Rather, the improvement effect has to be judged experimentally by comparing calculations in different schemes.   
Besides the theoretical performance of a particular improvement
scheme, its numerical efficiency is of course also an issue, but detailed comparative measurements
in the thermodynamic context are missing to date.
In the following sections, we discuss some schemes that have found applications in calculations of the equation of state. 

\subsection{Improved gauge actions} \label{sec:impr_gauge}

The Symanzik program for $SU(N)$ gauge theories has been discussed in \cite{weisz, weisz1}. In addition to 
the plaquette in the standard Wilson action \eq(\ref{Sg}), one may add larger planar loops of size 
$k\times l$ in such a way that the leading $O(a^2g^0)$ corrections are subtracted 
and deviations from the continuum start with $O(a^4g^0,a^2g^2)$. The improved gauge action then reads
\be
S_g=\beta\sum_{x}\sum_{1\leq\mu<\nu}^4 \left[c_0\left(1-\frac{1}{3}\re \Tr U_{\mu\nu}^{1\times1}(x)\right)+c_1\left(1
-\frac{1}{6}\re\Tr U_{\mu\nu}^{k\times l}\right)\right]\;,
\label{eq:wilson_impr}
\ee
where $U^{k\times l}_{\mu\nu}$ is the sum of the $k\times l$ and $l\times k$ rectangles.
The coefficients in general are functions of the gauge coupling, $c_i(g^2)$, and in the weak coupling limit
$g^2\rightarrow 0$ need to satisfy $c_0(0)=1-k^2l^2c_1(0)$ in order to reproduce the correct classical
continuum theory. For the choice 
\be
c_1(0)=-\frac{2}{k^2l^2(l^2+k^2-2)}\;,
\ee
the leading $O(a^2)$ corrections to the classical continuum limit are removed. Since this holds to leading
order in perturbation theory, the action is said to be tree-level improved. The standard Symanzik
improvement employs six-link rectangles, $k\times l=1\times 2$.

One can push this to higher orders in $g^2$ by adding further loops of larger length and non-planar loops. A non-perturbatively optimised expansion is obtained by multiplying the expansion coefficients with so called tadpole factors, which are to be determined self-consistently from plaquette expectation values \cite{tad},
\be
c_{1,\text{tad}}=u_0^{-2(k+l-2)}c_1(0),\quad u_0^4=
\frac{1}{6N_s^3N_\tau}\left\langle \sum_{x,\nu>\mu}\left(1-\frac{1}{3}\re\Tr U_{\mu\nu}(x)\right)\right\rangle\;.
\label{eq:tad}
\ee
This amounts to a resummation of higher order terms to create a better
expansion parameter, in the sense that convergence is faster. 
However, one should keep in mind that the resulting improvement is still not complete 
(i.e.~not to all orders in the coupling constant) and depends on the efficiency of the resummation.  
Finally, renormalisation group methods have been used to design an action which is close to the 
true effective action in the renormalisation group sense along a renormalised line of constant physics \cite{iwasaki}. This action can again be projected onto a two-parameter space as in \eq(\ref{eq:wilson_impr}).

Often combinations of these possibilities are taken, and the gauge action can be written as \cite{hotp4asq}
\begin{equation}
S_G = \sum_{x,\mu < \nu} \left[\beta_{\rm pl} (1-P_{\mu\nu}) +
\beta_{\rm rt} (1-R_{\mu\nu})\right] +
\beta_{\rm pg} \sum_{x,\mu < \nu<\sigma} (1-C_{\mu\nu\sigma}) 
\; .
\label{action_asqtad_G}
\end{equation}
It consists of the ordinary plaquette as well as of planar and non-planar six-link terms,
with $\mu,\; \nu,\; \sigma \in 1,\;2,\;3,\;4$ and $P_{\mu\nu}$, $R_{\mu\nu}$ and 
$C_{\mu\nu\sigma}$ denote 
the normalised trace of products of gauge field variables $U_{x,\mu}$:
\begin{eqnarray}
P_{\mu\nu} &=& \frac{1}{3}{\rm Re\; Tr}\; U_{x,\mu} U_{x+\hat{\mu},\nu} U^\dagger_{x+\hat{\nu},\mu}
U^\dagger_{x,\nu} \nonumber \\
R_{\mu\nu} &=& \frac{1}{6}{\rm Re\; Tr}\left( U_{x,\mu} U_{x+\hat{\mu},\mu} U_{x+2\hat{\mu},\nu}
U^\dagger_{x+\hat{\mu}+\hat{\nu},\mu} U^\dagger_{x+\hat{\nu},\mu} U^\dagger_{x,\nu} + (\mu \leftrightarrow \nu)
\right) \nonumber \\
C_{\mu\nu\sigma} &=& \frac{1}{12}
{\rm Re\; Tr}\left( 
U_{x,\mu} U_{x+\hat{\mu},\nu} U_{x+\hat{\mu}+\hat{\nu},\sigma}
U^\dagger_{x+\hat{\nu}+\hat{\sigma},\mu} U^\dagger_{x+\hat{\sigma},\nu}
U^\dagger_{x,\sigma} \right. \nonumber \\
&&\hspace*{1.55cm}\left. + (\mu \leftrightarrow \nu) + (\nu \leftrightarrow \sigma)
+(\mu \leftrightarrow \nu,\;\mu \leftrightarrow -\mu)
\right) \; .
\label{PRC}
\end{eqnarray}
The choice of parameters for different versions used in the context of the equation of state are summarised
in Table \ref{tab:impr_gauge}. 
\begin{table}[t]
\begin{center}
\begin{tabular}{|c|c|c|c|}
\hline
~&$\beta_{pl}$ & $\beta_{\rm rt}$ & $\beta_{\rm pg}$  \\[2mm]
\hline
& & &  \\[-4mm]
tree-level
Symanzik & $\displaystyle{\frac{5}{3}}$ & $\displaystyle{-\frac{1}{6}}$ & 0 \\[2mm]
\hline
& & &  \\[-4mm]
p4 & $\displaystyle{\frac{5}{3}}$ & $\displaystyle{-\frac{1}{6}}$ & 0 \\[2mm]
\hline
& & & \\[-4mm]
asqtad & 1 &
$\displaystyle{-\frac{1}{10u_0^2}(1+ 0.4805 \alpha_s)}$ &
$\displaystyle{-0.1330 \frac{1}{u_0^2} \alpha_s}$ \\[3mm]
\hline
& & &  \\[-4mm]
Iwasaki & $\displaystyle{\frac{5}{3}}$ & -0.662 & 0 \\[2mm]
\hline
\end{tabular}
\end{center}
\caption{
Couplings $\beta_x$, $x=$ pl, rt, pg defining the pure gauge part of the actions used in this review. 
Here $u_0$  is the tadpole coefficient, \eq(\ref{eq:tad}), and 
$\alpha_s = -4 \ln (u_0) / 3.0684$. 
}
\label{tab:impr_gauge}
\end{table}

\subsection{Clover-improved Wilson fermions}\label{sec:clover}

An improvement scheme for Wilson fermions which is frequently 
used in numerical simulations, 
both at zero and finite temperatures, are clover improved Wilson fermions \cite{sw},
\ba
S_\text{SW} &=& S_W+\csw\; iga^4\sum_{x,\mu,\nu}\frac{1}{4a}
\overline\psi(x)\sigma_{\mu\nu}F^\text{SW}_{\mu\nu}(x)\psi(x)\;,\nn\\
F^\text{SW}_{\mu\nu}(x)&=&\frac{1}{8iga^2}\sum_{\mu,\nu=\pm} 
\left(U_{\mu\nu}(x)-U^\dagger_{\mu\nu}(x)\right),\quad 
\sigma_{\mu\nu}=\frac{i}{2}[\gamma_\mu,\gamma_\nu]\;.
\ea
The additional term involves the clover of all four plaquettes sharing the lattice point $x$
in each plane. 
In this case one uses the propagator of standard Wilson  
fermions, but with rescaled bare quark mass and gauge coupling \cite{bs},
\be
m^\text{SW}=(m-m_c)(1+b_m\,am),\quad g^2_\text{SW}=g^2(1+b_g\,am)\;.
\label{clovscale}
\ee
Here $m_c(g^2)$ denotes the additive quark mass shift corresponding
to the chiral limit, which like the coefficients $\csw$, $b_m(g^2)$, $b_g(g^2)$ 
can be determined order by order in perturbation theory,
\be
\csw=1+{\cal O}(g^2),\quad b_m=-\frac{1}{2}+{\cal O}(g^2),\quad b_g={\cal O}(g^2)\;.
\ee
The tree-level dispersion relation at zero momentum reads
\be
E(0)=m(1-\frac{1}{2}am) + \Oas=m^\text{SW}+\Oas,
\ee
and is \Oa-improved after rescaling the bare parameters. The tuning of $\csw(\beta)$ has also been 
done non-perturbatively, resulting in an interpolation formula \cite{csw} which provides a completely $\Oa$-improved Wilson fermion action in the $\beta$-range covered.

\subsection{Twisted Mass Wilson fermions}

The twisted mass formulation 
is obtained by a chiral rotation in flavour space and 
requires an even number of pairwise degenerate flavours. 
For $N_f=2$, after a rescaling of the fermion 
fields its action takes the form
\be
S_{tm}=S_W+i\mu\;\psib\gamma_5\tau^3\psi.
\ee
Here $\mu$ denotes the twisted mass parameter
and the diagonal Pauli matrix $\tau^3$ acts in flavour space. 
The bare quark mass $m_q$ in this formulation is a combination of the bare mass $m$ in the standard
Wilson action and the twist parameter,
\be
m_q=\sqrt{m^2+\mu^2},\quad  m=m_q\cos(\omega), \quad \mu=m_q\sin(\omega)\;.
\label{massdef}
\ee
Twisted mass fermions provide an automatic $\mathcal{O}(a)$-improvement over Wilson fermions 
if the quark mass is solely determined by the twisted mass parameter $\mu$, 
which is the case at maximal twist, $\omega=\pi/2$.
For a detailed derivation and recent review of twisted mass QCD, see~\cite{shindler}. 

The free quark propagator for twisted mass fermions is 
\be
 D_\text{tm}(p) = \frac{-i\sum_\nu \gamma_\nu \overline{p}_\nu 
+ \frac{1}{2}\hat{p}^2 + am - ia\mu\gamma_5\tau^3}{\overline{p}^2
+\left(\frac{1}{2}\hat{p}^2 + am\right)^2 + (a\mu)^2}\;, 
\label{eq:propagator}
\ee 
with the usual dimensionless lattice momenta
\be
 \overline{p}_\mu = \sin(ap_\mu)\;, \quad
 \hat{p}_\mu      = 2\sin(ap_\mu/2)\;.
\ee
The standard Wilson propagator is recovered in the limit $\omega\rightarrow 0$.
The dispersion relation $E(\mathbf{p})$ can again be expanded 
in small lattice spacing and for 
vanishing spatial momentum we have 
\begin{equation}
E(0) = m_q\left(1-\frac{1}{2}am_q\cos(\omega)\right) + \mathcal{O}(a^2)\;.
\label{disp}
\end{equation}
Cut-off effects of $\mathcal{O}(a)$ set in with quark mass, and their removal at maximal twist, 
$\omega=\pi/2$, is apparent. 

\subsection{Naik and p4 improvement of staggered fermions}\label{sec:p4}

Let us recall that for the staggered fermion action \eq(\ref{stag}), the leading order corrections
are $\Oas$. Improvement of staggered fermions thus operates on the same level as 
that of the gauge action. Tree-level improvement for the staggered fermion action has been performed in \cite{naik}, by
replacing the lattice covariant derivative by a higher order discretised version, cf.~\eq(\ref{eq:naivef}),
\be
\nabla_\mu(U)\rightarrow \nabla_\mu(U)- \frac{a^2}{6}\nabla_\mu^3(U)\;.
\ee
The second term contains three links connecting fermion fields separated by one lattice spacing. 
A similar effect can 
be achieved when also including fermion fields separated by up to three lattice spacings. This
can be illustrated denoting the free and massless staggered Dirac operator by \cite{p4}
\be
D_{xy}=\sum_\mu\eta_\mu(x)\left(\sum_{i=1,3}[c_{i,0}\delta_{x+i\hat{\mu},y}-\delta_{x-i\hat{\mu},y}]
+\sum_{\nu\neq\mu}\sum_{j=\pm 2} c_{1,j}[\delta_{x+\hat{\mu}+j\hat{\nu},y}-
\delta_{x-\hat{\mu}+j\hat{\nu},y}]
\right)
\ee
The constraints
\ba
c_{1,0}+3c_{3,0}+6c_{1,2}&=&\frac{1}{2}\nn\\
c_{1,0}+3c_{3,0}+6c_{1,2}&=&24\,c_{1,2}
\ea
ensure that the dispersion relation is rotationally invariant through order $p^4$.
For the Naik action, this is achieved by the choice $c_{1,0}=9/16, c_{3,0}=-1/48$, with 
dispersion relation
\be
E=p+\frac{3}{40}\left(p^5-\frac{1}{p}\sum_ip_i^6\right)a^4+{\cal O}(a^6)\;,
\label{eq:naike}
\ee
which is duly improved through $\Oas$.
The so-called p4 action corresponds to the choice $c_{1,0}=3/8, c_{1,\pm 2}=1/48$. While this
particular choice still leaves $a^2$-corrections in the action, it has the same 
dispersion relation \eq(\ref{eq:naike}), but smaller coefficients ${\cal O}(a^6)$. 
To improve the thermodynamics of ideal gases improved dispersion relations
are sufficient, while higher order corrections 
are smaller than in the case of the Naik action. 
It should be stressed, however, that this construction 
relies on weak coupling which is realised only at very high temperatures. At lower temperature the perturbatively determined coefficients get increasingly out of tune with the renormalised
couplings in the true effective action. Moreover, both the Naik and the p4
improvement have little effect on the taste splitting caused by the staggered action. 

\subsection{Fat links for p4, asqtad and HISQ staggered actions} \label{sec:fat}

Taste splitting at order $a^2$ can be cancelled by adding four-quark operators to the staggered
action. However, a similar effect can be achieved by replacing the link in the lattice covariant
derivative with specific fat links \cite{tasteim1,tasteim2}.  
The simplest fat links are obtained by replacing the link with itself plus the surrounding staples \cite{fuzz},
such as the first term in the square brackets of \eq(\ref{smearing}).
Expanding down the gauge fields from the link variables, the fattened link adds a term \cite{fat}
\be
a^2w_3\bar{\psi}D_\nu F_{\nu\mu}\psi +{\cal{O}}(a^4)
\ee
to the fermionic action in continuum notation (as well as to the staggered action after field transformations). Clearly, it affects $\Oas$-improvement. From the pure gauge structure of this term, 
it will generally mix with similar terms appearing in improvement schemes for the pure gauge action,
and care has to be taken as to appropriate combinations of the two.
The staples used above are paths using three links. In a further extension also longer and non-planar paths are used \cite{fat7,fat71},
\begin{eqnarray}
U^\text{fat}_\mu(x)&=& c_1U_\mu(x)+\sum_\nu \Big[ w_3S^{(3)}_{\mu\nu}(x)
+ \sum_\rho \Big( w_5 S^{(5)}_{\mu\nu\rho}(x) + 
\sum_\sigma w_7 S^{(7)}_{\mu\nu\rho\sigma}(x)\Big)+ w_LS^{(L)}_{\mu\nu}(x)\Big] 
\label{smearing}
\end{eqnarray}
with the three-, five- and seven link terms, \fig\ref{fig:paths},
\ba
S^{(3)}_{\mu\nu}(x) &=& U_\nu(x)
         U_\mu(x+\hat{\nu})U^\dagger_\nu(x+\hat{\mu})\nonumber\\
S^{(5)}_{\mu\nu\rho}(x) &=& U_\nu(x)
 S^{(3)}_{\mu\rho}(x+\hat{\nu})U^\dagger_\nu(x+\hat{\mu})\nonumber\\
S^{(7)}_{\mu\nu\rho\sigma}(x) &=& U_\nu(x)
      S^{(5)}_{\mu\rho\sigma}(x+\hat{\nu})U^\dagger_\nu(x+\hat{\mu})\nn\\
      S^{(L)}_{\mu\nu}(x) &=& U_\nu(x)S^{(3)}_{\mu\nu}(x+\hat\nu)
         U^\dagger_\nu(x+\hat\mu)\;.         
\label{staples}
\ea
\begin{figure}[t]
\vspace*{-2cm}
\setlength{\unitlength}{1.0in}
\begin{center}\begin{picture}(6.0,2.0)
\put(0.5,0){\usebox{\Link}\makebox(0,0)}
\put(1.5,0){\usebox{\Staple}\makebox(0,0)}
\put(2.5,0){\usebox{\FiveStaple}\makebox(0,0)}
\put(3.5,0){\usebox{\SevenStaple}\makebox(0,0)}
\put(5.0,0){\usebox{\LepageStaple}\makebox(0,0)}
\end{picture}\end{center}
\caption{
   \label{fig:paths}
   The simple link, three link staple, five link staple, seven link
   staple and double staple used in suppressing taste symmetry breaking. From \cite{asqtad}.
}
\end{figure}
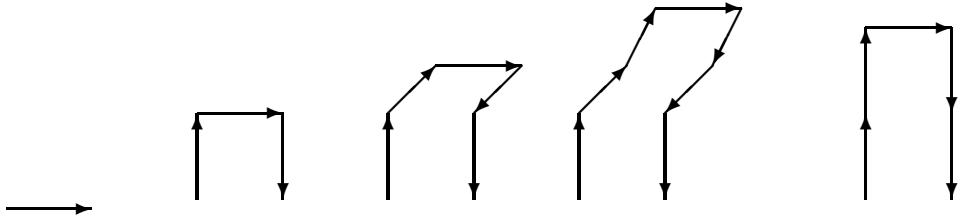
It can be shown that for appropriately tuned parameters with the first three terms the tree-level $\Oas$ taste splitting is removed, but an $a^2$-correction due to low momentum modes gets introduced by
the seven link path.
This is removed again by the double length staple $S^{(L)}$ \cite{lepage}. A staggered action
including the Naik term, fat-5, fat-7 links and the double staple removes all tree-level $a^2$ errors, resulting in leading corrections of ${\cal O}(a^4,g^2a^2)$ \cite{asqtad}. 
The corresponding action is hence called asq,  while the version with additionally tadpole improved
coefficients is called asqtad.
All these features can be summarised 
following the compact unified presentation (with changed normalisation of the coefficients) \cite{hotp4asq},
cf.~Table \ref{tab:pg_couplings},
\begin{eqnarray}
D[U]_{xy}&=& m~\delta_{xy} + \Big(c_{1} ~{ A[U]_{xy}} 
+ c_{3} ~{ B_1[U]_{xy}}+  c_{12} ~{ B_2[U]_{xy}} \Big)
\label{fmatrix2}
\\[2mm]
{A[U]_{xy}}
&=&\apart{x}{y}{\mu} \nonumber \\
{B_1[U]_{xy}}
&=&\bparta{x}{y}{\mu} \nonumber 
\end{eqnarray}
\vspace*{-0.5cm}
\begin{eqnarray*}
B_2[U]_{xy} &=&
\bpartb{x}{y}{\mu}{\nu}\; .\\
\end{eqnarray*}

\begin{table}[tb]
\begin{center}
\begin{tabular}{|c|c|c|c|}
\hline
~& $c_1$ & $c_3$ & $c_{12}$ \\[2mm]
\hline
&  &  & \\[-4mm]
p4 & $\displaystyle{\frac{3}{8}}$
& 0 & $\displaystyle{\frac{1}{96}}$\\[2mm]
\hline
&  &  & \\[-4mm]
asqtad &  $\displaystyle \frac{1}{2}$ &  $\displaystyle -\frac{1}{48 u_0^2}$ & $\displaystyle 0$
\\[3mm]
\hline
\end{tabular}
\end{center}
\caption[]{
Coefficients $c_i$ for the fermionic p4 and asqtad actions, \eq(\ref{fmatrix2}), as used in 
Sec.~\ref{sec:almost}.
}
\label{tab:pg_couplings}
\end{table}

A further significant improvement of the taste symmetry is achieved by the Highly Improved Staggered
Quark action. In its original design in  \cite{hisq} it includes additional one-loop taste changing operators
which are particularly useful for the spectroscopy of charmed mesons. For its use in thermodynamical simulations discussed here, it contains the same ingredients as the asqtad fermion action, 
but  the fat links are reunitarised. 
(Note that the sum of link matrices is no longer unitary). Its use in conjunction
with the tree-level Symanzik improved gauge action has been dubbed HISQ/tree action \cite{hot_hisq}. 

\subsection{Stout links}\label{sec:stout}

A more heuristic way to use smeared links in the Dirac operator is based on observations made
in spectrum calculations at zero temperature, where smeared observables have less fluctuations
on the scale on the cut-off.
This suggests to use similar methods also on the level of lattice actions. One would expect that  
fluctuations on the scale of a lattice spacing get already
damped in the generation of configurations, which 
is tantamount to reducing cut-off effects and hence improvement. 

\begin{figure}[t]
\setlength{\unitlength}{1.0in}
\begin{picture}(6.0,2.3)(-1.2,-1.15)
\thicklines
\linethickness{1.0mm}
\put(-1.0,0.0){\vector(1,0){0.9}}
\put(0.0,0.0){=}
\thinlines
\put(0.5,0.1){$\times 1$}
\put(0.5,0.9){$\times w$}
\put(0.5,-0.9){$\times w$}
\put(0.2,0.0){\vector(1,0){0.9}}
\put(0.2,0.1){\vector(0,1){0.9}}
\put(0.2,1.0){\vector(1,0){0.9}}
\put(1.1,1.0){\vector(0,-1){0.9}}
\put(0.2,-0.1){\vector(0,-1){0.9}}
\put(0.2,-1.0){\vector(1,0){0.9}}
\put(1.1,-1.0){\vector(0,1){0.9}}
\end{picture}\hspace*{-7.5cm}\includegraphics[width=0.6\textwidth,bb=0 0 598 199]{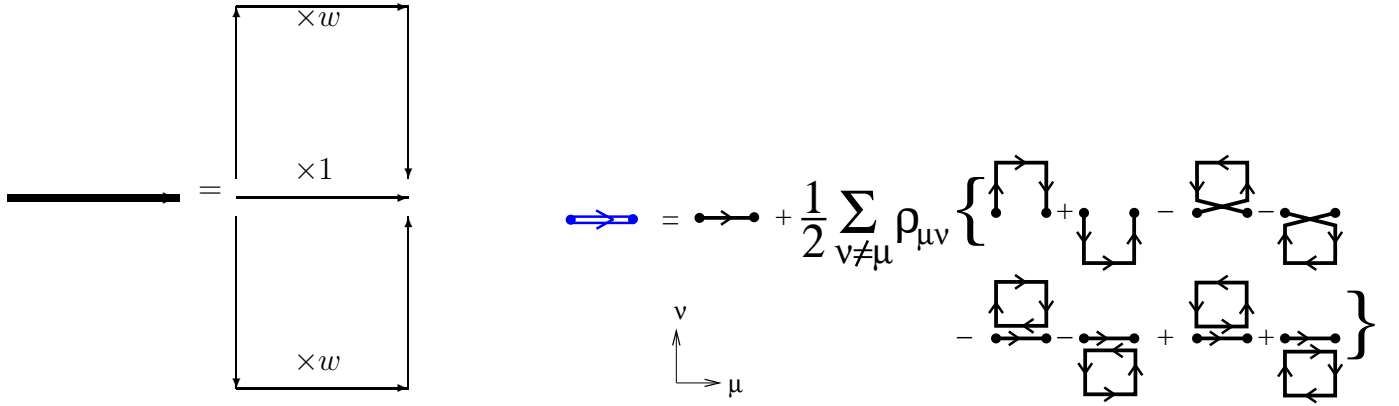}
\caption[]{Left: Fat links and APE smearing replacing a link by itself and the sum of surrounding staples. 
From \cite{fat}. Right: Expansion to first order in the $\rho_{\mu\nu}$
 of the stout link variable
 $U^{(1)}$ in terms of the original links.  Closed loops
 include a trace. From \cite{stout}.
\label{fig:paths1}
}
\end{figure}

The simplest possibility is called fuzzing or APE-smearing and consists of replacing
a spatial link with the sum of itself and a weight factor times the sum of the adjacent staples, as in 
\fig\ref{fig:paths1} (left) \cite{fuzz}.
To maintain all symmetries of 
the original link, a projection back to $SU(3)$ is applied.  
Empirical investigations show that the projection back to $SU(3)$ is very desirable for the
improvement of taste symmetry in staggered simulations.
On the other hand, the projection step
is non-analytic whereas 
for use in a Monte Carlo algorithm one would like to have differentiable objects such that the response
to infinitesimal changes can be computed. An analytic way to smear link variables 
goes under the name of stout links \cite{stout}. 
Consider the weighted sum of staples between neighbouring lattice sites,
\be
 C_\mu(x)=\sum_{\nu\neq \mu}\rho_{\mu\nu}\biggl(
 U_\nu(x) U_\mu(x\!+\!\hat{\nu}) U_\nu^\dagger(x\!+\!\hat{\mu})+ U^\dagger_\nu(x\!-\!\hat{\nu}) U_\mu(x\!-\!\hat{\nu})
  U_\nu(x\!-\!\hat{\nu}\!+\!\hat{\mu})\;.
\biggr), \label{eq:Cdef}
\ee
The weights $\rho_{\mu\nu}$ are tunable real
parameters, with a common choice $\rho\equiv \rho_{ik}, \rho_{\mu4}=\rho_{4\mu}=0$.
Then an $SU(3)$ matrix $Q_\mu$ can be constructed in the following way
\begin{eqnarray}
\Omega_\mu(x) &=& C_\mu(x)\ U_\mu^\dagger(x),
   \quad\mbox{(no summation)}\;,\nn\\
Q_\mu(x) &=& 
  \frac{i}{2}\Bigl(\!\Omega^\dagger_\mu(x)\!-\!\Omega_\mu(x)\!\Bigr)
   \!-\!\frac{i}{2N}\Tr\Bigl(\!\Omega^\dagger_\mu(x)
  \!-\!\Omega_\mu(x)\!\Bigr)\;.
 \label{eq:Qdef}
\end{eqnarray}
This is the basis for an iterative smearing procedure, with a step from level $n$ to $n+1$,
\begin{equation}
U^{(n+1)}_\mu(x)=\exp\Bigl(i Q_\mu^{(n)}(x)\Bigr)\ U_\mu^{(n)}(x)\;.
\label{eq:Ustout}
\end{equation}
One can show that this step maintains all the relevant symmetries of the original link.
The first such smearing step is shown schematically in \fig\ref{fig:paths1} (right). For a related
smearing procedure constructing hypercubic links, see \cite{hyp}.
It should be stressed, though, that these
methods do not implement a subtraction of terms of  
a specific power of the coupling or lattice spacing from the action, nor do we know a relation between
smearing steps and the renormalisation group. Rather, their improvement effect has to be monitored
in the results of simulations.

\subsection{Comparison of the weakly interacting gas for improved actions}

As the previous sections have illustrated, constructing and choosing improved actions is a complicated matter with many aspects to be taken into account. Let us now see how the different choices perform
under different conditions. First, let us discuss the high temperature behaviour, which is affected by
the perturbative improvements. 
The lattice pressure for free quarks can be expanded as a power series in the lattice spacing $a$, which 
at finite temperature is equivalent to expanding in $1/N_\tau$,
\be
\frac{p}{T^4} = \frac{(aN_\tau)^3}{2\pi^3}\int d^3p\,\ln\left(1+e^{-N_\tau aE(\mathbf{p},m)}\right)
           =\, \frac{p^{(a^0)}}{T^4} + \frac{p^{(a^1)}}{T^4}\frac{1}{N_\tau} +  \frac{p^{(a^2)}1}{T^4}\frac{1}{N_\tau^2} + \ldots
\ee
For massless quarks, the coefficients $p^{(a^2)}/T^4$ and $p^{(a^4)}/T^4$  can be calculated in closed form~\cite{hegde_08} and remarkably are the same for standard staggered and Wilson-type fermions. Differences are introduced only in higher orders of the lattice spacing.
In particular, for the Naik and p4 improved staggered actions, corrections now start as
\ba
\frac{f^\text{Naik}}{f_{SB}}&=&1-\frac{1143}{980}\left(\frac{\pi}{N_\tau}\right)^4
-\frac{365}{77}\left(\frac{\pi}{N_\tau}\right)^6+\ldots \nn\\
\frac{p^\text{p4}}{p_{SB}}&=&1-\frac{1143}{980}\left(\frac{\pi}{N_\tau}\right)^4
-\frac{73}{2079}\left(\frac{\pi}{N_\tau}\right)^6+\ldots
\ea

A comparison of the free energy density of a gas of free massless
staggered quarks is shown in \fig\ref{fig:pSB} (left), where the standard, Naik and p4 actions
are each normalised by the continuum result. Note that the leading cut-off effects of the 
Naik and p4 actions have the opposite sign compared to the standard action. It is apparent that 
the continuum value of the free energy density is approached more rapidly for the improved actions
than for the bare action. However, one also sees that the improvement effect becomes significant
only for sufficiently fine lattices, $N_\tau\gsim 6$. 
\fig\ref{fig:pSB} (right) shows the continuum approach
between the p4 action and various actions using fat and smeared links. 
By design the p4, asqtad and HISQ actions remove $\Oas$ discretisation errors in the dispersion relation and hence in the thermodynamic functions, whereas stout smearing still shows a remnant 
$a^2$-dependence in the free fermion gas. 

However, for a controlled continuum 
extrapolation, it is not necessarily  important that cut-off effects are small in absolute size, but
that their functional behaviour is controlled. The expected $a^2$-behaviour is also shown by the stout
action for $N_\tau\gsim 10$, allowing for a continuum extrapolation, whereas for coarser lattices $a^4$-corrections become important. 
\begin{figure}[t]
\includegraphics[width=0.5\textwidth]{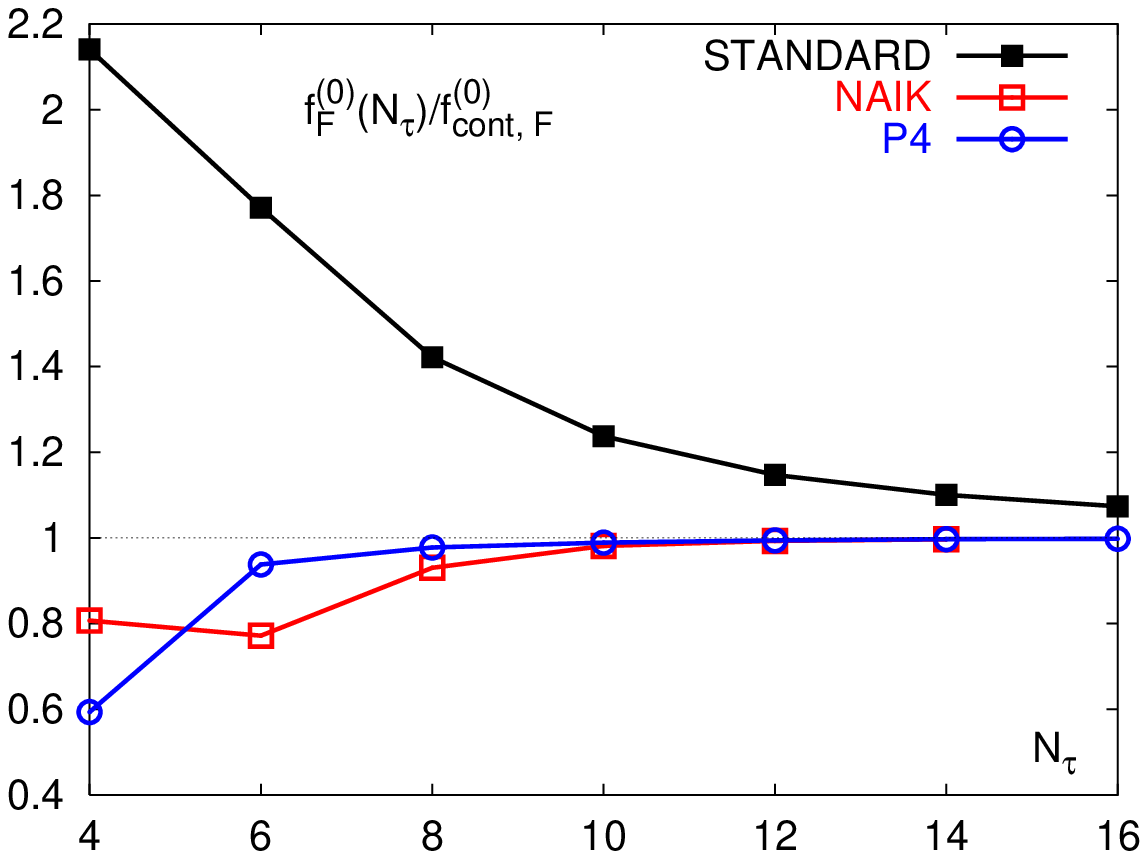}
\includegraphics[width=0.5\textwidth]{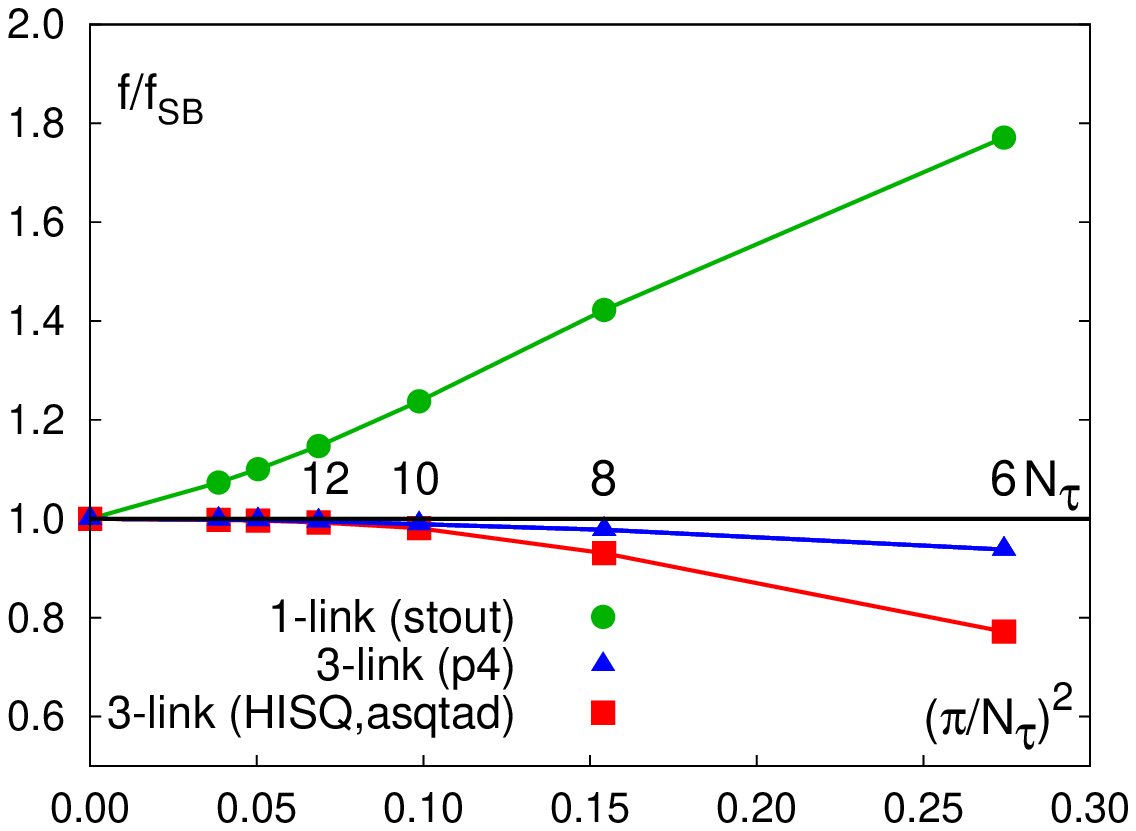}
\caption{The free energy density of an ideal quark gas calculated
for different values of the temporal extent $N_{\tau}$ 
divided by the corresponding value for $N_\tau = \infty$. From \cite{p4} (left), \cite{hot_hisq} (right).}
\label{fig:pSB}
\end{figure}

The ideal gas and massless limits are somewhat extreme simplifications and far from the cases of practical 
interest. In \cite{pz} the effects of mass and interactions were
taken into account. Note that in order to compare different discretisations in the massive case, the renormalised mass needs to 
be the same between different discretisation schemes.
For non-vanishing masses the lattice corrections of Wilson actions to the free fermion gas
start at $\Oa$, in contrast to the staggered actions. However, the linear coefficient is found to be 
exceedingly small compared to the  $\Oas$ contribution which dominates in \fig\ref{pmass} (left).
This also follows from the fact that the $\Oa$-improved Wilson actions do not scale better
than the unimproved one. Thus, for the ideal gas
Wilson actions scale comparably to the unimproved staggered action and a continuum extrapolation
using $a^2$-behaviour is possible for $N_\tau\gsim 10$.

The leading ${\cal O}(g^2)$ corrections to the fermionic pressure due to interactions are 
given by two diagrams,
\begin{equation}
p_{F}^{(2)}=-\frac{1}{2}\frac{1}{\beta V}\left(
\begin{minipage}{1.3cm}
\includegraphics[width=\textwidth]{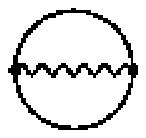}
\end{minipage}
+
\begin{minipage}{2cm}
\includegraphics[width=\textwidth]{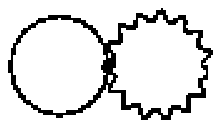}
\end{minipage}\right).
\end{equation}
The result for massive fermions is shown in \fig\ref{pmass} (right). 
Despite the rather large numerical uncertainties, one finds a qualitative correspondence to non-interacting case, especially the small dependence on the quark mass and the particular discretisation. 
As in the free case, on lattices with $N_\tau\geq 8$ Wilson fermions are found to be competitive with 
unimproved staggered fermions. 
Again, $a^2$ scaling is not observed for any of the discretisations before $N_\tau\gsim10$.
Note that for $N_\tau<8$ the absolute size of the cutoff effects for the unimproved Wilson action appears smaller than for the clover action. This indicates significant higher order contributions for the $N_\tau$ considered here.

The comparison of the scales between the free and the interacting situation shows 
that considerations based on the ideal gas limit alone may be insufficient
for the discussion of improvement schemes.
\begin{figure}[t]
\includegraphics[height=0.5\textwidth,angle=-90]{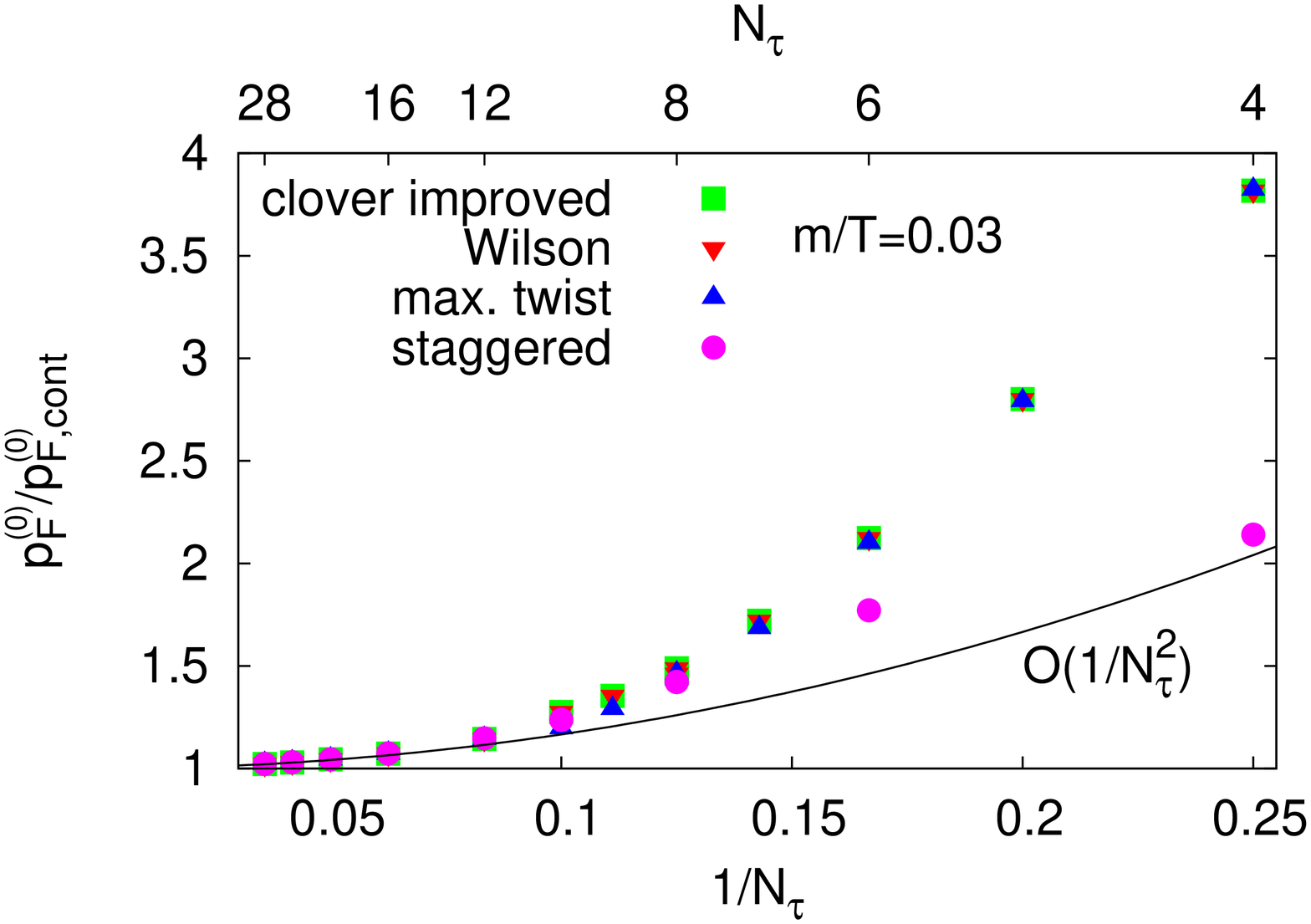}
\hfill\includegraphics[height=0.5\textwidth,angle=-90]{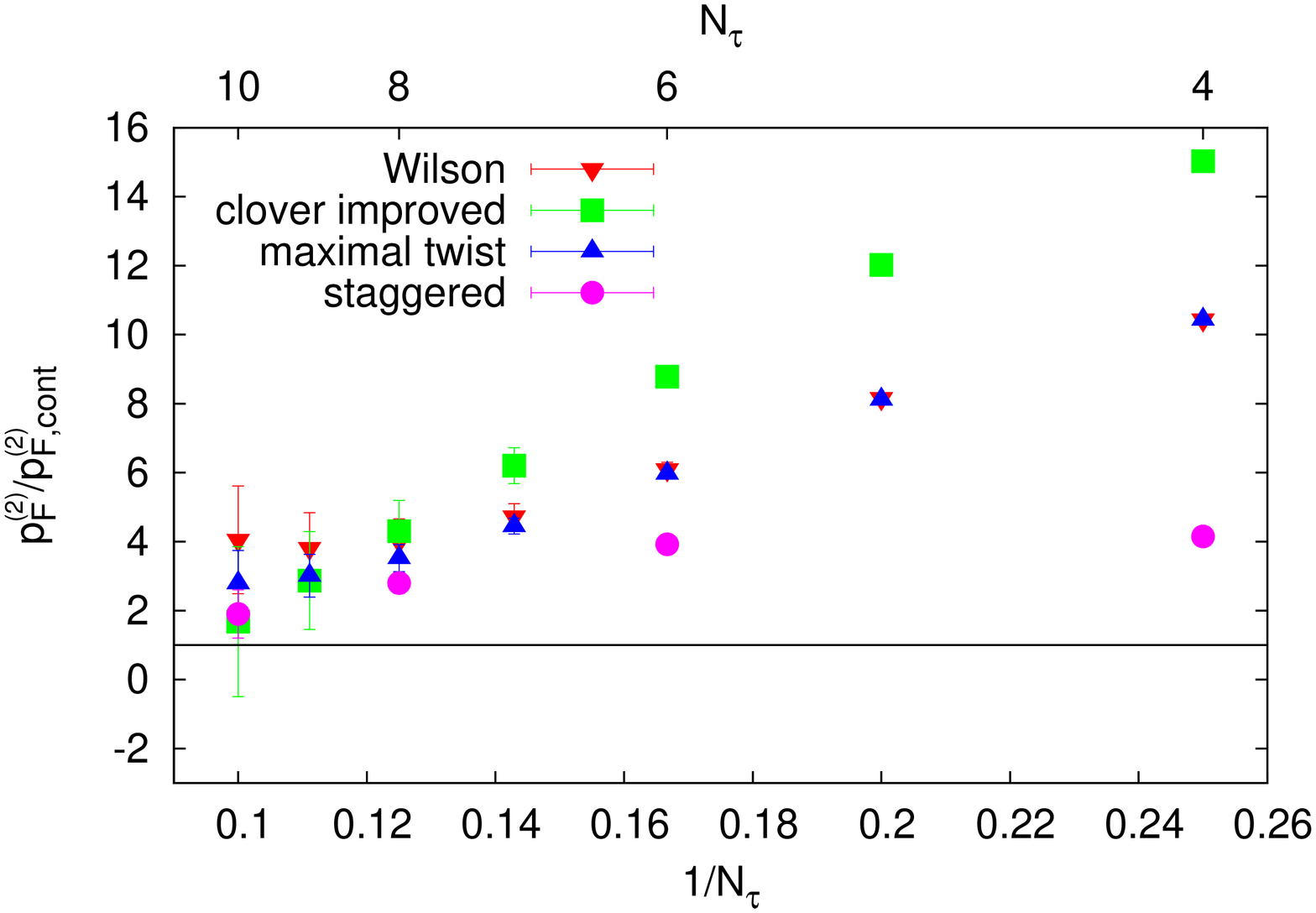}
\caption{Continuum approach of the pressure of massive fermions with a fixed renormalised quark mass $m_R/T=0.03$. 
Left: Ideal gas. Right: Two-loop contribution. 
From \cite{pz}.
\label{pmass}}
\end{figure}
In a regime, where perturbation theory is valid, we can consider the relative cutoff error for the pressure, 
\be
\left|\Delta p_F\right|/p_{F,\text{cont}}=\left|p_{F,\text{lat}}-p_{F,\text{cont}}\right|/p_{F,\text{cont}},
\label{series}
\ee
where $p_{F,\text{cont}}$ and $p_{F,\text{lat}}$ are each calculated to some specified order in the coupling.
In the weak coupling limit the denominator is dominated by the 
ideal gas limit, and the cutoff effects of the interactions get normalised to that number. 
However, one cannot draw any conclusions from this for the behaviour of the cut-off errors near $T_c$.
Since the sign of the leading order corrections is negative, the relative error quickly develops a pole 
with growing coupling, beyond which results cannot be extrapolated. For longer series the pole might disappear, but the relative cut-off error as a rational function will be non-monotonic in general.

\subsection{Comparison of taste splitting for staggered actions}\label{sec:taste}

A main source of systematic error for  simulations with staggered fermions  is the taste splitting, whose effect on thermodynamics has long been underestimated. Recall that the splittings are
\be
m_i^2=m_G^2+\delta m_i^2,\quad \delta m_i^2\sim(\alpha_sa^2)\;,
\label{m_ps}
\ee
to leading order in discretisation errors, where $m_G$ corresponds to the true Goldstone boson
with taste matrix $\Gamma^F=\gamma_5$, cf.~Table \ref{tab:tastes0}. In \cite{bmw_tc2} it was observed that the lattice spacing
needs to be $a\lsim 0.15$ fm in order for this leading behaviour to dominate, larger lattice spacings 
still show splittings of even higher order in the lattice spacing.
A detailed investigation of cut-off dependence of the splitting on finer lattices
is shown for the asqtad and stout actions in \fig\ref{fig:taste1}, with
the eight pion multiplets labelled as in Table \ref{tab:tastes0}. Both actions enter the $a^2$-scaling
region within the plot, with the stout action featuring smaller splittings and significantly lower
mass values for the heavier multiplets on coarser lattices.
A similar study comparing the scaling between HISQ/tree and stout actions is shown in \fig\ref{fig:taste2}, 
with even smaller mass splittings for the HISQ/tree action. 
The lines correspond to linear continuum extrapolations
in $\alpha_s^2a^2$, where the gauge coupling has been 
fixed in calculations of the static potential, $\alpha_s=\alpha_V$ \cite{aubin}. 
\begin{figure}
\scalebox{.71}{
\includegraphics{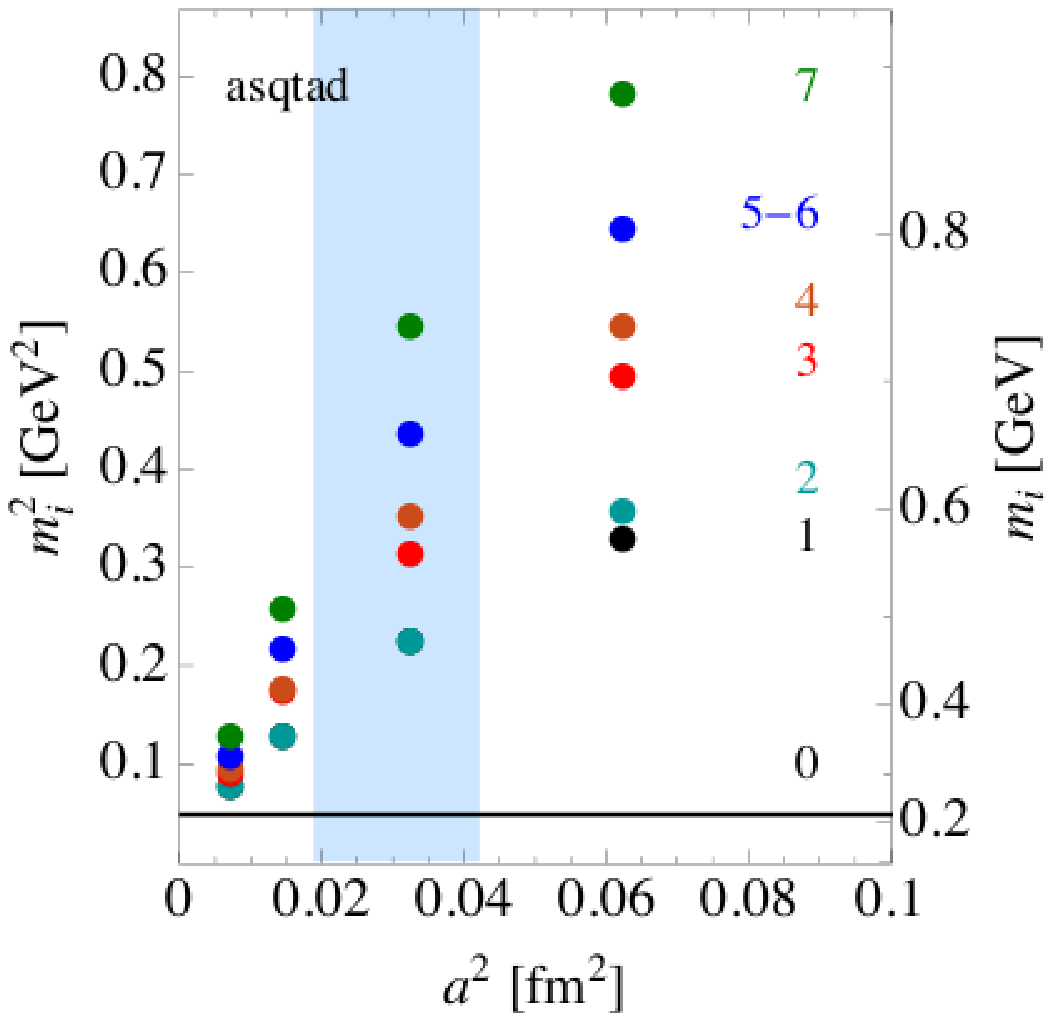}\\}\hspace*{1.5cm}
\scalebox{.71}{
\includegraphics{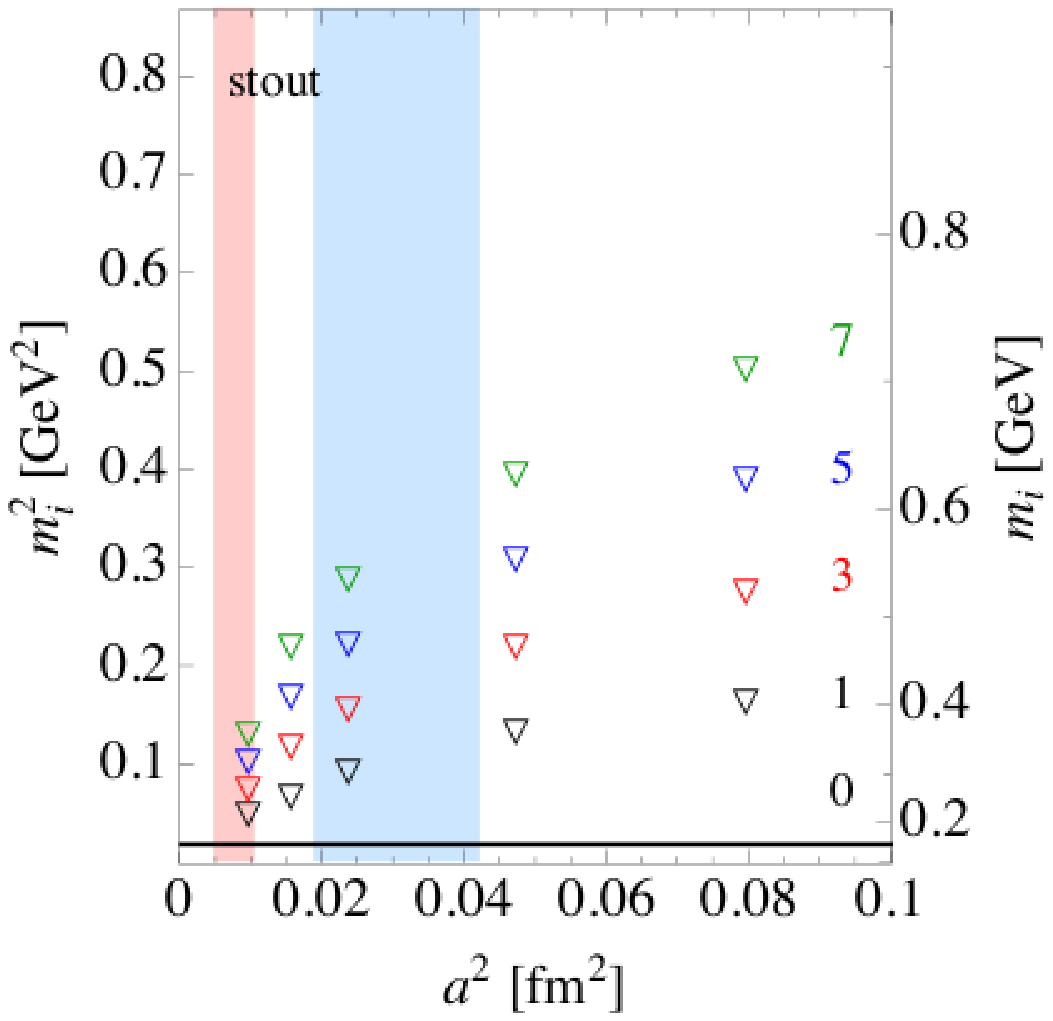}\\}
\caption{Pion multiplet masses as functions of the lattice spacing. 
Left: asqtad action \cite{hotQCD_rev}. Right: stout action \cite{bmw_tc3}. 
Numbers label the taste multiplets, Table \ref{tab:tastes0}. 
The broad band indicates the range of
lattice spacings for a thermodynamics study at $N_\tau=8$ between $T=$ 120 and 180
MeV. The narrow band in the right panel corresponds to the same temperature range
and $N_\tau=16$. The line labelled ``0" is the
pseudo-Goldstone boson with mass 220 MeV for the asqtad and
135 MeV for the stout results. From \cite{bmw_tc3}. 
}
\label{fig:taste1}
\end{figure}

\begin{figure}
\includegraphics[width=8cm]{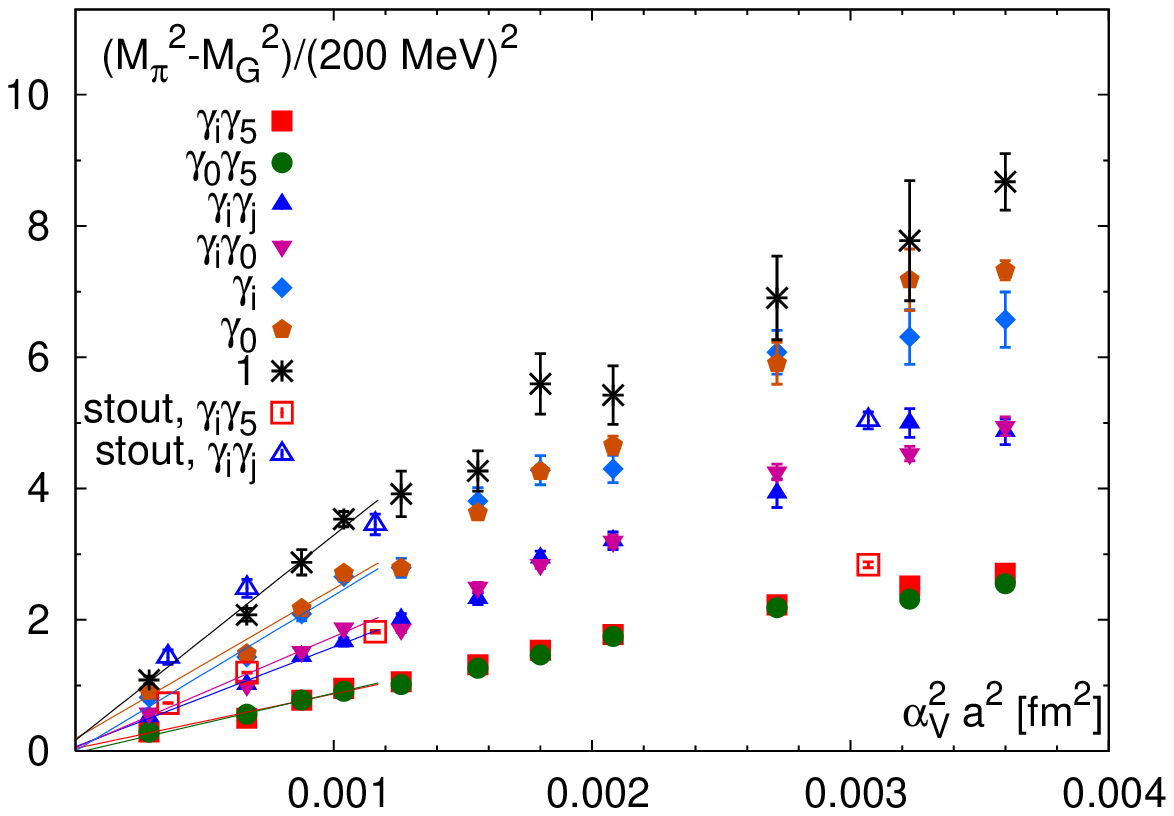}
\includegraphics[width=8cm]{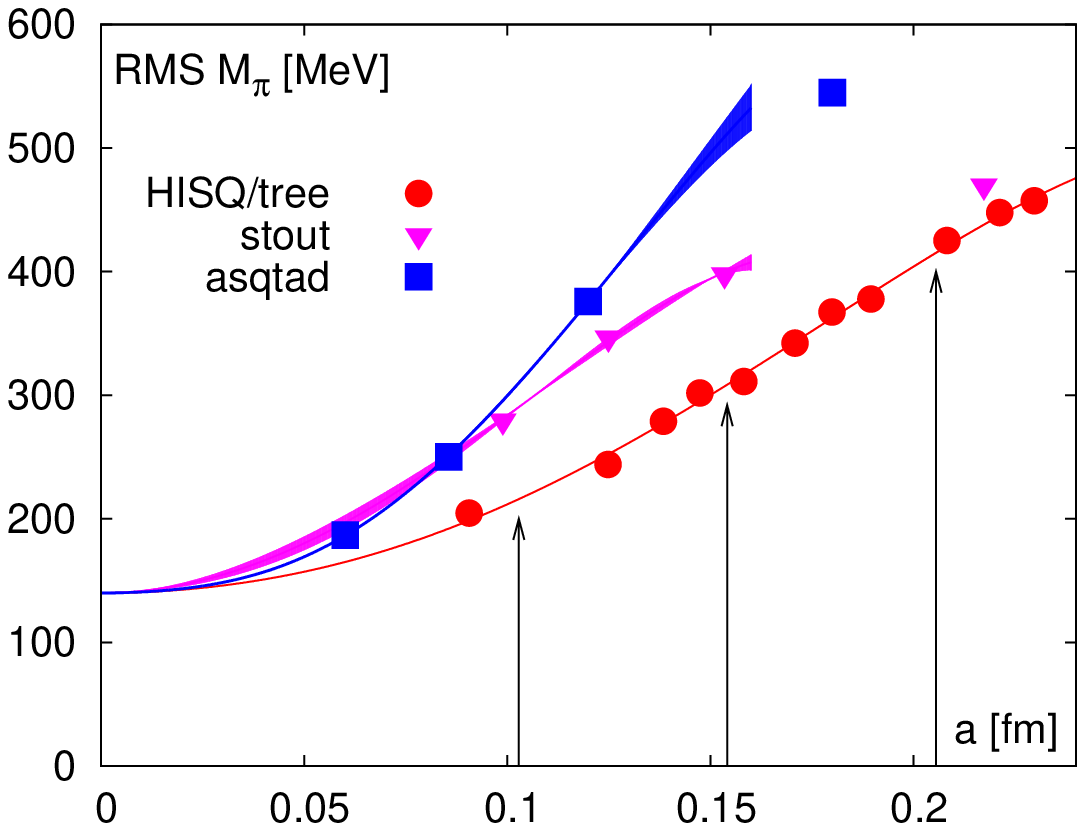}
\caption{Left: Pion multiplet masses calculated with the HISQ/tree \cite{hot_hisq} 
  and stout \cite{bmw_tc2} actions as a
  function of $\alpha_V^2 a^2$. Right: RMS
  pion mass with $M_G=140$ MeV as a function of the lattice spacing
  for the asqtad \cite{aubin}, stout \cite{bmw_tc3} and HISQ/tree \cite{hot_hisq} actions. 
  Vertical arrows
  indicate the lattice spacing corresponding to $T \approx 160$ MeV
  for $N_{\tau}=6$, $8$ and $12$. From \cite{hot_hisq}.}
\label{fig:taste2}
\end{figure}

Knowing that in the continuum limit the tastes become degenerate,
one may identify the particle mass with the root mean square average over the tastes, which is 
another measure for the taste splitting. For the 
example of the pion,
\be
  M_\pi^{RMS}= \sqrt{
  \frac{1}{16}\left(M_{\gamma_5}^2+M_{\gamma_0\gamma_5}^2
  +3M_{\gamma_i\gamma_5}^2+3M_{\gamma_i\gamma_j}^2
  +3M_{\gamma_i\gamma_0}^2+3M_{\gamma_i}^2
  +M_{\gamma_0}^2+M_{1}^2\right)} \; .   
\label{rmspi}
\ee
This quantity is shown in \fig\ref{fig:taste2} (right). According to the respective size of taste splittings, 
this quantity is largest for asqtad, whereas the HISQ/tree action features the smallest average mass. Note that the differences show up in a region where all actions have an averaged pion mass significantly away from the physical value, i.e.~the lattices are too coarse to reproduce pion dynamics. In order for the averaged mass to be close to the physical mass, lattices with $N_\tau\gsim 12$ are required. 

As discussed in Sec.~\ref{sec:hrgsc}, for low temperatures the hadron resonance gas should give
a good approximation to the equation of state and it is useful to compare its predictions against 
lattice data.  However, the taste splitting plots illustrate that for staggered fermions the hadron
spectrum is distorted by taste splitting and gets fully restored only  
in the continuum limit. 
For comparisons with staggered fermions at finite lattice spacings, 
it was therefore suggested \cite{pasi10} to use the lattice spectrum including cut-off effects instead
of the continuum spectrum in the hadron resonance gas. For this purpose the taste splittings are
parametrised in terms of the scale parameter $r_1$ as
\begin{equation}
r_1^2 \cdot \delta m_{ps_i}^2=\frac{a_{\rm ps}^i x+b_{\rm ps}^i x^2}{{(1+c_{\rm ps}^i x)}^{\beta_i}},
~x=(a/r_1)^2\;.
\label{delta_ps}
\end{equation} 
An excellent description of the scaling behaviour of the different pion tastes is achieved 
in \fig\ref{fig:pion}.
\begin{figure}
\centerline{
\includegraphics[width=0.5\textwidth]{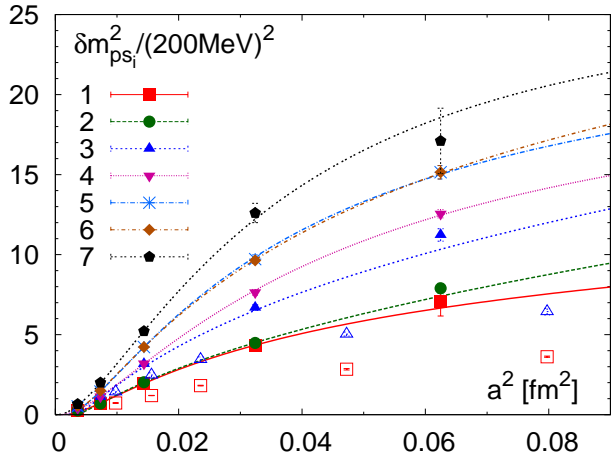}
}
\caption[]{The quadratic  splittings of non-Goldstone pseudo-scalar mesons in the seven different
multiplets calculated with asqtad action 
at different lattice spacings \cite{hotp4asq}. The lines show the parametrisation Eq.~(\ref{delta_ps}).
Open symbols refer to lattice data obtained with the stout action \cite{bmw_tc2}.
From \cite{pasi10}.}
\label{fig:pion}
\end{figure}

One may now take the non-degenerate tastes of pions and kaons into the hadron 
resonance gas separately with a contribution \cite{pasi10} 
\begin{equation}
p^{\pi,K}/T^4=\frac{1}{16} \frac{1}{VT^3} \sum_{i=0}^7 d_{\rm ps}^i \ln{\cal Z}^{M}(m_{{\rm ps}_i}),
\label{ps_contr}
\end{equation} 
 where $m_{{\rm ps}_i},~i=1-7$ is calculated according to Eq. (\ref{m_ps}) and $m_{{\rm ps}_0}$ is
equal to the pion or kaon mass used in the actual lattice calculations. We shall see in 
Sec.~\ref{sec:almost} that a good description of the asqtad equation of state is achieved in this way.
 
In summary, taste symmetry violation appears to be a systematic error with a large effect in 
staggered simulations. 
While formally of ${\cal O}(\alpha_s a^2)$ for p4, stout and ${\cal O}(\alpha_s^2a^2)$ for asqtad and 
HISQ/tree actions, its main effect on thermodynamical simulations is in the distortion of the pion spectrum. 
Identifying the Goldstone pion with the physical pion is not sufficient, since also the non-Goldstones contribute to the equation of state at finite lattice spacing. Hence, apart from an $\Oas$-errors in thermodynamic functions, the taste splitting effectively implies a pion mass out of tune with the physical 
value and distorted contributions of the tastes to the equation of state.

\section{Computing the equation of state by Monte Carlo simulation}\label{sec:meth}

Once a suitable lattice action is chosen, the task at hand
is to compute the free energy density from the full
QCD partition function. A technical obstacle here is that, in a Monte Carlo simulation, 
one cannot compute the partition function directly, 
since all expectation values are normalised to $Z$,
\be
\langle O\rangle=Z^{-1}\Tr(\rho O)\;.
\ee
So we need to measure the expectation value of some observable $O$ in order 
to access the partition function. This is achieved by calculating derivatives of the partition function 
with respect to some parameter, as in the derivative and integral methods in the next two sections.

\subsection{The derivative method for anisotropic lattices}\label{sec:deriv}

Theoretically the most direct way to obtain thermodynamic functions from lattice simulations is by the derivative method using anisotropic lattices with two lattice spacings $a_s,a_t$ in the spatial
and temporal directions, respectively. Temperature and spatial volume are then given by
\be
T=\frac{1}{a_\tau N_\tau},\quad V=(a_sN_s)^3\;,
\ee
and the anisotropic pure gauge action reads
\be
S_g[U]
 =\sum_{x}\sum_{1\leq i<j\leq 3}\beta_s\left(1-\frac{1}{3}{\rm Re}
\Tr U_{ij}(x)\right) +\sum_{x}\sum_{1\leq \mu< 4}\beta_\tau\left(1-\frac{1}{3}{\rm Re}
\Tr U_{\mu 4}(x)\right)\;, 
\ee
which is split into terms containing only spatial and temporal plaquettes, respectively.
There are now two bare lattice gauge couplings, 
\be
\beta_s=\frac{2N}{g_s^2\xi},\quad \beta_\tau=\frac{2N\xi}{g_\tau^2}\;,
\ee
which are related by the anisotropy parameter $\xi\equiv a_s/a_\tau$. Using the thermodynamic
relations \eq(\ref{thermo},\ref{thermo1}),
one derives the lattice versions of energy density and pressure as
\be
\frac{\epsilon}{T^4}=\left(\frac{N_\tau}{\xi N_s}\right)^3\left\langle a_\tau\frac{\partial S}{\partial a_\tau}\right\rangle,\quad 
\frac{p}{T^4}=-\frac{1}{3}\left(\frac{N_\tau}{\xi N_s}\right)^3\left\langle a_s\frac{\partial S}{\partial a_s}\right\rangle\;.
\ee  
Denoting the expectation values of the spatial and temporal plaquettes by 
\be
R_{sp}=\left\langle1-\frac{1}{N}{\rm Re}
\Tr U_{ij}(x)\right\rangle ,\quad R_{tp}=\left\langle1-\frac{1}{N}{\rm Re}\Tr U_{i4}(x)\right\rangle\;,
\ee
one gets for the derivatives 
\ba
\left\langle a_\tau\frac{\partial S}{\partial a_\tau}\right\rangle&=&
\left(\beta_s+a_\tau\frac{\partial \beta_s}{\partial a_\tau}\right)\sum_{sp}R_{sp}
+\left(-\beta_\tau+a_\tau\frac{\partial \beta_\tau}{\partial a_\tau}\right)\sum_{tp}R_{tp}\;,\nn\\
\left\langle a_s\frac{\partial S}{\partial a_s}\right\rangle&=&
\left(-\beta_s+a_s\frac{\partial \beta_s}{\partial a_s}\right)\sum_{sp}R_{sp}
+\left(\beta_\tau+a_s\frac{\partial \beta_\tau}{\partial a_s}\right)\sum_{tp}R_{tp}\;.
\ea
Thus, in order to compute the equation of state we need to know the beta-functions of the gauge
couplings and measure the expectation values of spatial and temporal plaquettes.
It is convenient to introduce a gauge coupling $g^2$ corresponding to an isotropic lattice with
lattice spacing $a_s$. Then the anisotropic couplings can be expressed in terms of $g$ and the 
anisotropy $\xi$,
\ba
g_s^{-2}&=&g^{-2}+C_s(g^2,\xi)=g^{-2}+c_s(\xi)+{\cal O}(g^2)\nn\\
g_t^{-2}&=&g^{-2}+C_t(g^2,\xi)=g^{-2}+c_t(\xi)+{\cal O}(g^2)\;,
\ea
where the right hand expressions represent a perturbative expansion for small coupling in lattice perturbation theory \cite{kaniso,2hasen}. 

Similar expressions can be derived when quarks are included, whereupon also the beta-functions
for the running of the quark masses appear.
Note that also the anisotropy parameter $\xi$ gets renormalised and has a corresponding beta-function. 
These various beta-functions are only known to some low order in 
perturbation theory, while their accurate
non-perturbative determination is complicated. Moreover, while anisotropic lattices allow for a finer
temperature resolution by choosing small $a_t$, there are clearly more parameters to tune than in the 
isotropic case, making
for example the determination of lines of constant physics more difficult. These are less important
when temperature is tuned via $N_\tau$, but that in turn implies relatively coarse temperature steps.
For these reasons, the most popular method in practice 
is the integral method described in the next section and we shall not further pursue the derivative method here. Recent numerical results and a discussion of the associated difficulties using standard staggered fermions on anisotropic lattices can be found in \cite{levaniso}. 

\subsection{The integral method}\label{sec:integ}

The most frequently used method in practice is the integral method \cite{integral}, in which a  
derivative of the free energy with respect to some parameter serves as observable, 
which then gets integrated again to yield the free energy density, e.g.~$x=T$, 
\be
\frac{f}{T^4} {\biggl |}_{T_o}^{T} \; = \; - {1\over V} \int_{T_o}^{T} 
{\rm d}x \; \frac{\partial}{\partial x }\ \;\;x^{-3} \ln Z(V,x) \;.
\ee
On the lattice, when doing simulations at fixed $N_\tau$,  
it is convenient to take derivatives with respect to the bare lattice parameters instead,
\ba
\frac{f}{T^4} {\biggl |}_{(\beta_o,m_{f0})}^{(\beta,m_f)} \; &=& \;  -\frac{N_\tau^3}{N_s^3} 
\int_{\beta_o,m_{f0}}^{\beta,m_f}
\left({\rm d}\beta ' \left[\left\langle \frac{\partial \ln Z}{\partial \beta'} \right\rangle - 
\left\langle \frac{\partial \ln Z}{\partial \beta'}\right \rangle_{T=0}\right] \right.\nn\\
 &&\hspace*{2cm}\left.+\sum_f{\rm d}m'_f \left[\left\langle\frac{\partial \ln Z}{\partial m'_f}\right\rangle-\left\langle\frac{\partial \ln Z}{\partial m'_f}\right \rangle_{T=0}\right] 
\right)\;. 
\label{freediv}
\ea
We have subtracted the corresponding vacuum expectation value in 
each line to renormalise the 
free energy density and pressure to zero for $T=0$.
The expectation values of the derivatives correspond to the average plaquette action and 
chiral condensate, respectively,
\ba
\frac{1}{N_\tau N_s^3}\frac{\partial \ln Z}{\partial \beta}&=&\frac{1}{N_\tau N_s^3}\left\langle \sum_p U_p\right\rangle=
\langle -s_g\rangle \;,\nn\\
\frac{1}{N_\tau N_s^3}\frac{\partial \ln Z}{\partial m_f}&=&\frac{1}{N_\tau N_s^3}
\left\langle \sum_{x}\bar{\psi}_f(x)\psi_f(x)\right\rangle\;.
\ea

The vacuum expectation values correspond to $N_\tau\rightarrow\infty$. In a simulation
a low temperature lattice with sufficiently large $N^0_\tau$ has to be chosen, which introduces an error
$\sim (N_\tau/N_\tau^0)^4$ in the subtracted quantity. The integration in \eq(\ref{freediv}) has to be performed along a line of constant physics, i.e.~a curve $\beta(m_f)$ such that the renormalised quark masses
stay constant. In this case the left hand side 
corresponds to the difference of the renormalised free energy density
evaluated at two different temperatures,
\be
\frac{f}{T^4} {\biggl |}_{(\beta_o,m_{f0})}^{(\beta,m_f)} =\frac{f(T)}{T^4}-\frac{f(T_0)}{T_0^4}\;.
\ee
The integration has thus introduced a lower integration constant, 
which needs to be fixed for the result to be
meaningful. Note, that for sufficiently small $\beta_0$ in principle this is possible with an analytic strong coupling calculation
as in Sec.~\ref{sec:sc}. However, these exist
only for Wilson's pure gauge action so far, whereas 
dynamical simulations require light fermions as well
as modifications for improved actions.
Many practitioners therefore simply choose
$\beta_0$ low enough so that the pressure is numerically consistent with zero and can be neglected.
However, this procedure is not practical for the chiral limit, when pions become massless
and stay relativistic down to very low temperatures.
Alternatively, below the quark hadron transition 
the free energy $f(T_0)$ can be matched to the hadron resonance gas, 
which is again validated
through the strong coupling expansion, cf.~Sec.~\ref{sec:hrgsc}. 

Finally, using \eq(\ref{eq:defi}), the derivatives appearing in the integrand of \eq(\ref{freediv}) 
are also related to the trace anomaly,
\ba
\frac{I(T)}{T^4} \frac{dT}{T}&=&
N_\tau^4\left(d\beta \langle -s_g \rangle^{\rm sub} +
\sum_f d m_f \langle \bar\psi_f\psi_f\rangle^{\rm sub}\right)\;,\nn\\
\frac{I(T)}{T^4} &=&-N_\tau^4\left(a\frac{d\beta}{da} \langle -s_g \rangle^{\rm sub} +
\sum_f a\frac{d m_f}{da} \langle \bar\psi_f\psi_f\rangle^{\rm sub}\right)\;,
\label{eq:tracealat}
\ea
where the superscript ``${\rm sub}$'' indicates that the difference with the corresponding vacuum quantity
is to be taken. We have also converted the temperature derivative into one with respect to the lattice 
spacing, such that we need the lattice beta-functions for the variation of the gauge coupling and the masses
along a line of constant physics. Hence, the quantity computed directly in a standard approach is  the trace anomaly, while the pressure and other quantities are obtained by numerical integration of those data.
Note that this introduces another systematic error, since the trace anomaly data have to be interpolated
numerically for this purpose.

\subsection{The fixed scale method}\label{sec:fixed}

In the last section the temporal lattice extent $N_\tau$ was held fixed and temperature was tuned via the lattice gauge coupling and the lattice spacing
$a(\beta)$ in the relation $T=(aN_\tau)^{-1}$. As a consequence, when calculating a given quantity like
the thermodynamic functions as a function of temperature, one has different lattice spacings and hence 
different cut-off errors for different temperatures. This commonly leads to rather coarse lattices   
in the low temperature region and to a different quality of continuum extrapolation for 
different temperatures. 
An alternative use of the integral method is the so-called fixed scale approach, where one keeps
the lattice gauge coupling, and hence the lattice spacing, constant, and varies $N_\tau$ instead to tune
the temperature. 
Obviously, this allows only discrete temperature steps and the method has only become
viable recently, since lattices are now fine enough such that reasonably interesting temperature
steps can be obtained. 

However, in this case integration over $\beta$ is inapplicable.  
Instead, one uses $T$ as integration variable \cite{whot_fix}.
Starting point is again the evaluation of the trace anomaly followed by integration over temperature,
\be
\frac{p}{T^4}=\int_{T_0}^T dT\;\frac{I(T)}{T^5}\;,\quad \frac{I(T)}{T^4}=\frac{1}{T^3V}\;a\frac{db_i}{da}
\left\langle\frac{\partial S}{\partial b_i}\right\rangle^\text{sub}\;,
\ee
where $\vec{b}=(\beta,m_f,\ldots)$ is a vector denoting all parameters of the lattice action with 
the corresponding beta-functions $d\vec{b}/da$ along a line of constant physics. The crucial difference
is that now integration is over temperature while holding $\beta$ fixed, i.e.~over $N_\tau$. Clearly, 
this requires an interpolation to real values for $N_\tau$, introducing a systematic error which 
has to be monitored. In \cite{whot_fix} it was found that for sufficiently 
fine lattices (corresponding to $N_\tau \gsim 8$ in the usual fixed $N_\tau$ approach) results are consistent
between the two, cf.~Sec.~\ref{sec:ym}.

The fixed scale approach is complementary to fixed $N_\tau$ in that it has large
discretisation effects at high $T$, when $T\gsim a^{-1}$ and typical thermal fluctuations occur on
the scale of the cut-off. On the other hand, at low and intermediate temperatures there are several
advantages: for a fixed lattice spacing all temperatures are immediately on the same line of constant physics,
without any tuning of parameters. Furthermore, only one zero temperature simulation at that same
$\beta$ is required in 
order to renormalise all observables. Finally, the continuum limit proceeds straightforwardly in complete
analogy to vacuum calculations, by a series of simulations at different fixed lattice spacings $a$.
Because of their complementarity, it will be extremely useful in the future to control systematic errors
by using them both. 

\subsection{Thermodynamic potentials from momentum distributions}\label{sec:mom}

Both, the fixed $N_\tau$ and the fixed scale methods discussed above, require the subtraction of 
vacuum expectation values and have to deal with a lower integration constant that is not precisely 
fixed. These circumstances render both methods cumbersome and expensive when either very high 
or low temperatures are considered. A new approach avoiding these technical difficulties has been
proposed in \cite{gm1}. Starting point is the observation that momentum
is a conserved quantity because of translation invariance. 
Hence quantum mechanical states with different momenta are orthogonal and
a projection operator $\hat{\rm P}^{({\bfp})}$ onto states with momenta $\bfp$ can be defined.
The momentum distribution in the grand canonical ensemble is then given by
\be
\frac{R(T,\mu,{\bfp})}{L^3} = \langle \hat{\rm P}^{({\bfp})}\rangle = 
\frac{{\rm Tr}\{e^{-(\hat H-\mu\hat N)/T} 
\hat{\rm P}^{({\bfp})}\}}{{\rm Tr}\{e^{-(\hat H-\mu\hat N)/T}\}}\;.
\ee
One may now define a generating function $K$ for the connected cumulants of the momentum distribution,
\ba
e^{-K(\beta,\mu,{\bfz})}& =& 
\frac{1}{L^3} \sum_{\bfp} e^{i{\bfp}\cdot{\bfz}} \, R(\beta,\mu,{\bfp})\;,\nn\\
\frac{K_{\{2 n_1, 2 n_2, 2 n_3\}}}{(-1)^{n_1+n_2+n_3+1}} &=& 
\frac{\partial^{2n_1}}{\partial \bfz_1^{2n_1}} \frac{\partial^{2n_2}}{\partial \bfz_2^{2n_2}}
\frac{\partial^{2n_3}}{\partial \bfz_3^{2n_3}}
\frac{K(\bfz)}{L^3}\Big|_{\bfz=0}\; .
\ea
Employing Ward identities for the conservation of energy
and momentum, one can show that for $\mu=0$ the entropy $s=(\epsilon+p)/T$ is connected to
the second order cumulant, while
the specific heat is related to a combination of
the second and fourth order cumulants \cite{gm1,gm2},
\be
K_{2,0,0}((T,\mu)=T(\epsilon+p)\;, \qquad
c_v = \frac{K_{4,0,0}}{3 T^4} - \frac{3\, K_{2,0,0}}{T^2}
=\frac{K_{2,2,0}}{T^4} - \frac{3\, K_{2,0,0}}{T^2}\;.
\ee
On the other hand, the generating functional for the momentum distribution can also be expressed
as a ratio of partition functions,
\be
e^{-K(T,\mu,{\bfz})} = \frac{Z(T,\mu,\bfz)}{Z(T,\mu)}\;,\quad 
Z(T,\mu,\bfz)=\Tr\{e^{-(\hat H-\mu\hat N)/T+i\hat\bfp\bfz}\}\;,
\ee
where $Z(T,\mu,\bfz)$ is a partition function in which states of momentum $\bfp$ are weighted
by a phase $e^{i\bfp\cdot\bfz}$.
It can be represented as a Euclidean path integral, where the weight factor is absorbed in the shifted boundary conditions for the generic field, 
\be
\phi(1/T,{\bfx})=\pm\phi(0,{\bfx}+{\bfz})\;.
\label{eq:bc}
\ee
Note that the cumulants represent connected correlation functions of conserved charges, i.e.~momentum.
As such they are physical observables and do not require renormalisation, even on the lattice \cite{gm2},
which therefore saves the vacuum subtraction required in the integral method.
Hence, calculation
of the ratio of the shifted and unshifted partition functions permits an evaluation of the entropy and
other thermodynamic functions of relativistic field theories. Applications to $SU(3)$ pure gauge theory
and scalar field theory have been presented in \cite{gm1,gm2}, respectively.

For pure gauge theory, the required ratio of partition functions is
split up in factors 
\be\label{eq:prod}
\frac{Z(T,\bfz)}{Z(T)} = \prod_{i=0}^{n-1}
\frac{Z(T,r_i)}{Z(T,r_{i+1})}\; ,
\ee
with partition functions $Z(T,r_i)$, ($r_i=i/n$, $i=0,1,\dots,n$)
and corresponding actions 
\be
\overline S(U,r_i)= r_i S(U) + (1-r_i) S(U^z)\;,
\ee 
where $U^z$ obeys shifted boundary conditions, see \eq(\ref{eq:bc}). 
The continuum value of the second cumulant
is obtained from a discrete lattice by a limiting procedure,
\be
\frac{s(T)}{T^3} = \frac{K_{2,0,0}(T)}{T^5} = 
\lim_{a\rightarrow 0} 
\frac{2 K(T, \bfz, a)}{{|{\bfz}|}^2 T^5 L^3}\; ,
\label{eq:sT3}
\ee
where $\bfz=(n_z a,0,0)$ and $n_z$ has to stay constant 
when $a \rightarrow 0$.   Numerical results obtained by this method are shown in Sec.~\ref{sec:hight}.
Note that, while no subtraction of vacuum lattices is necessary in this approach, it requires the 
computation of an entire series of products of partition functions, \eq(\ref{eq:prod}), which is again
computationally costly. It will be interesting to see in future investigations whether an optimised 
choice for the interpolating partition functions can render this method computationally advantageous.

\section{Numerical results for pure gauge theory}\label{sec:ym}

\subsection{$SU(3)$ Yang-Mills}\label{sec:ymlow}

The results for the equation of state of pure gauge theory \cite{boyd} are by now ``historical'' and well-known. Since
systematic errors in this case are relatively benign, it still represents the only case with a complete continuum extrapolation and sets 
the standard of what can be achieved with lattice simulations. 
It is based on Wilson's standard pure gauge
action using lattices with $N_\tau=4,6,8$ and aspect ratios $N_s/N_\tau=4-5.3$. 

$SU(3)$ pure gauge theory has a first order finite temperature deconfinement 
transition at unique critical couplings $\beta_c(N_\tau)$ associated with centre symmetry
breaking, Sec.~\ref{sec:zn}. In \cite{boyd} those were determined  
from the location of the peak of the Polyakov loop susceptibility and converted to 
units of the string tension by calculating the latter on $32^4$ vacuum lattices for the same $\beta$-values.
In the continuum limit,
\be
{T_c \over \sqrt{\sigma}} = 0.625 \pm 0.003~(+0.004)\;,
\label{Tcratio}
\ee
 where the second error corresponds to estimated finite size corrections to the critical couplings.
 Together with $\sqrt{\sigma}=420$ MeV this gives a critical temperature $T_c\approx 265$ MeV  
 in terms of which all other dimensionful quantities can be expressed. 
 
The equation of state was calculated by the integral method, Sec.~\ref{sec:integ}. 
The trace anomaly for the pure gauge case reduces to
\be
\frac{I(T)}{T^4}=N_\tau^4a\frac{d\beta}{da}\left(3\langle\Tr U_p^t+ \Tr U_p^s\rangle-6\langle \Tr U_p\rangle_{T=0}\right)\;,
\ee
and requires evaluation 
of temporal and spatial plaquettes as well as the lattice beta-function. 
For the latter the difference in critical couplings corresponding to a doubling of the lattice spacing is
computed, $\Delta\beta=\beta_c(2N_\tau)-\beta_c(N_\tau)$, and modelled by the perturbative two-loop
beta-function modified by an effective coupling and a scaling factor. 
The corresponding trace anomaly for the three
different values of $N_\tau$ is shown in \fig\ref{fig:ym} (left), including smooth spline functions as numerical
interpolation between the data points. Note that, for a first order phase transition, 
a discontinuity in the energy density is expected in the thermodynamic limit. However, on a finite size
lattice, there is tunnelling between the metastable phases resulting in a continuous average value for
all temperatures.  The pressure is then obtained by integrating the trace anomaly according 
to the first line of \eq(\ref{freediv}). The free energy density is neglected (statistically compatible with zero) 
for a lower integration limit below the critical coupling, $\beta_0\lsim \beta_c$.
\begin{figure}[t]
\includegraphics[width=0.35\textwidth]{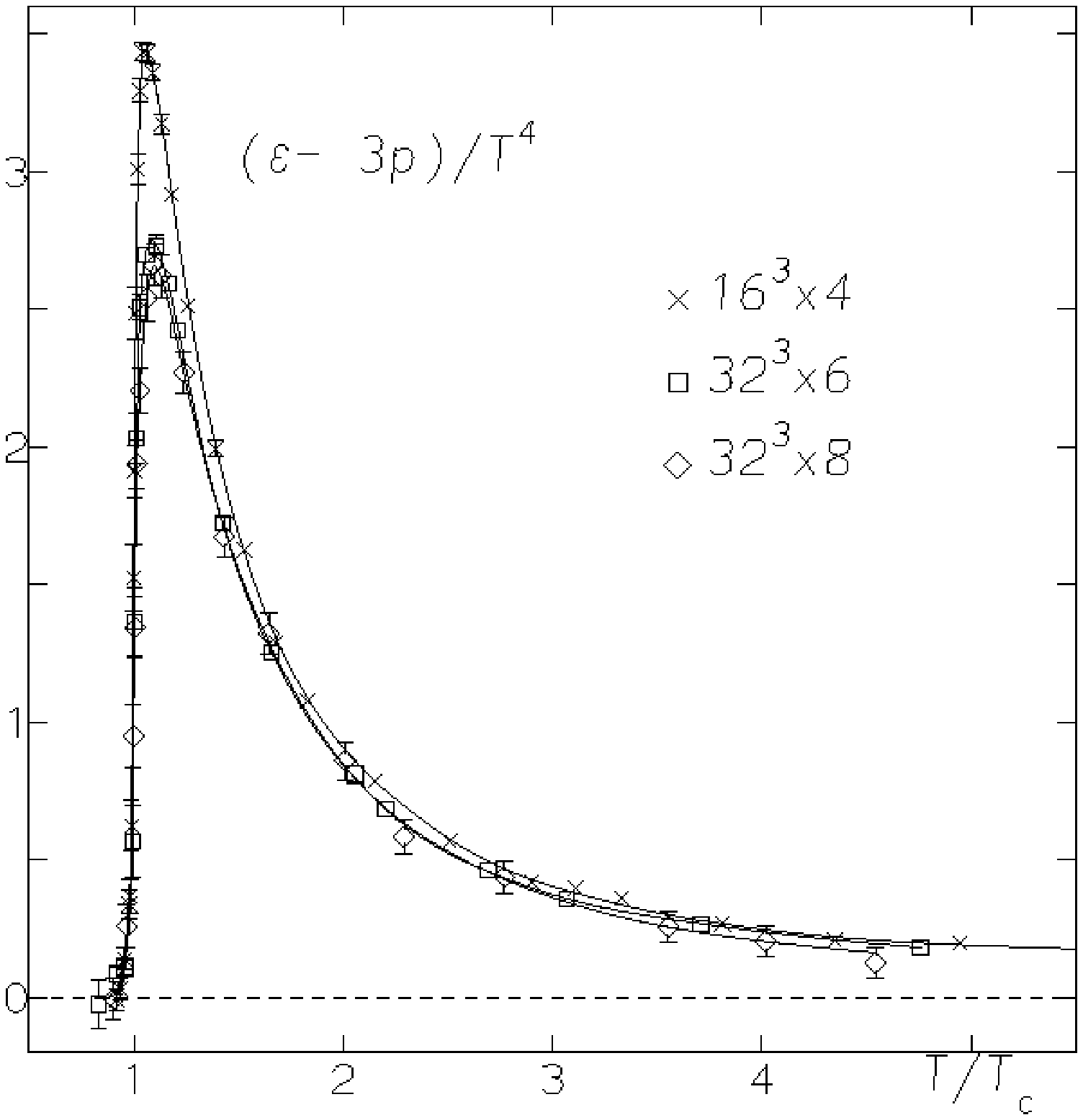}\hspace*{1.5cm}
\includegraphics[width=0.5\textwidth]{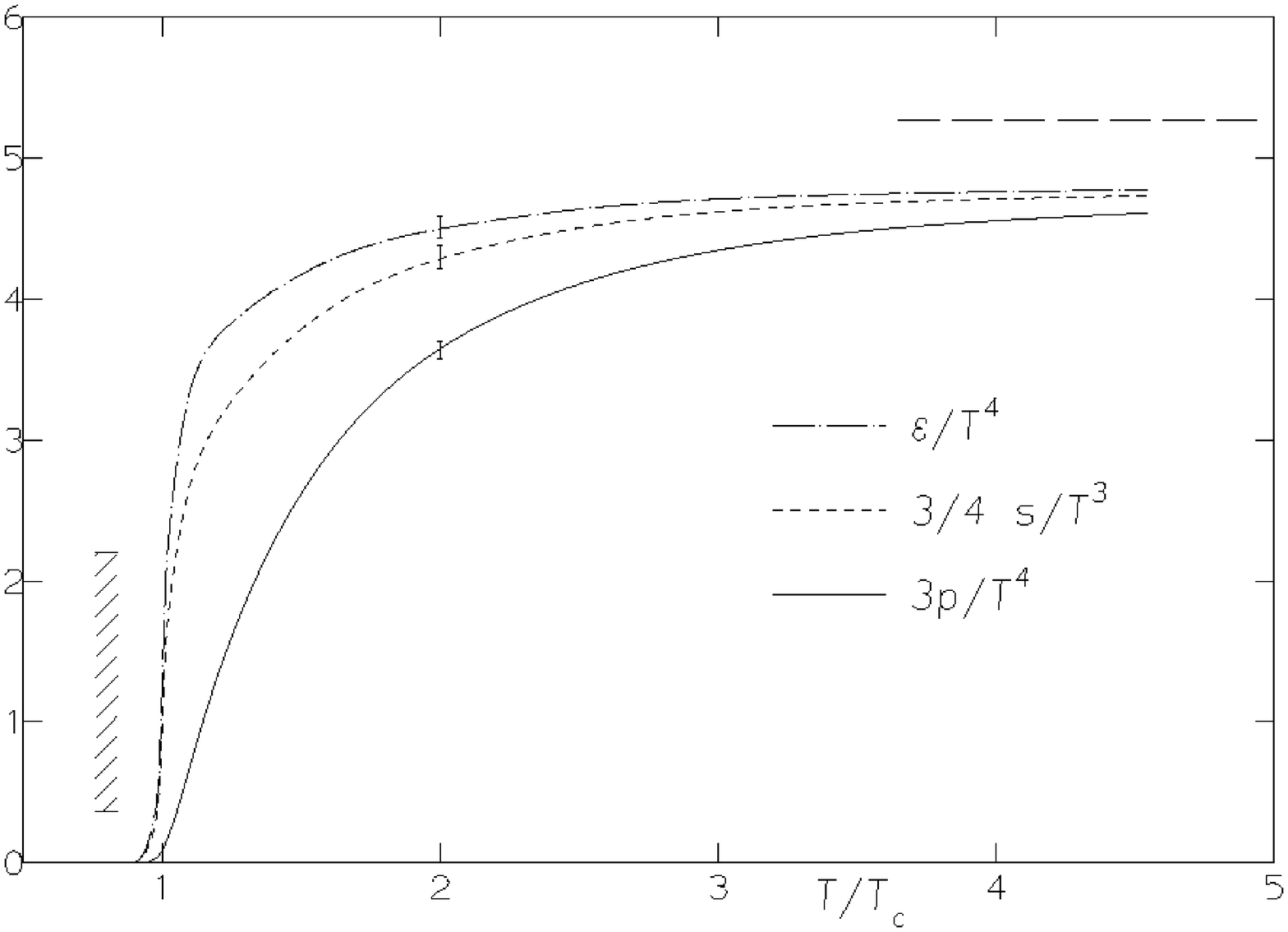}
\caption[]{Left: The trace anomaly $(\epsilon -3p)/T^4$ for pure gauge theory
calculated on lattices with temporal extent $N_\tau =4,6,8$. Right: Continuum extrapolated results for
pressure, energy and entropy density.
From \cite{boyd}.}
\label{fig:ym}
\end{figure}

The numerical results show significant cut-off effects for $N_\tau=4$.
A continuum extrapolation is thus performed for the pressure using the $N_\tau=6,8$ data based on the ansatz
\be
\biggl({p \over T^4}\biggr)_a = \biggl({p \over T^4}\biggr)_0
+ {c(T) \over N_\tau^2}\;,
\label{cfit}
\ee
with a fit parameter $c(T)$. The resulting continuum curve, as well as those for energy and entropy densities, are shown in \fig\ref{fig:ym} (right).

\begin{figure}[t]
\begin{center}
\includegraphics[height=6.0cm]{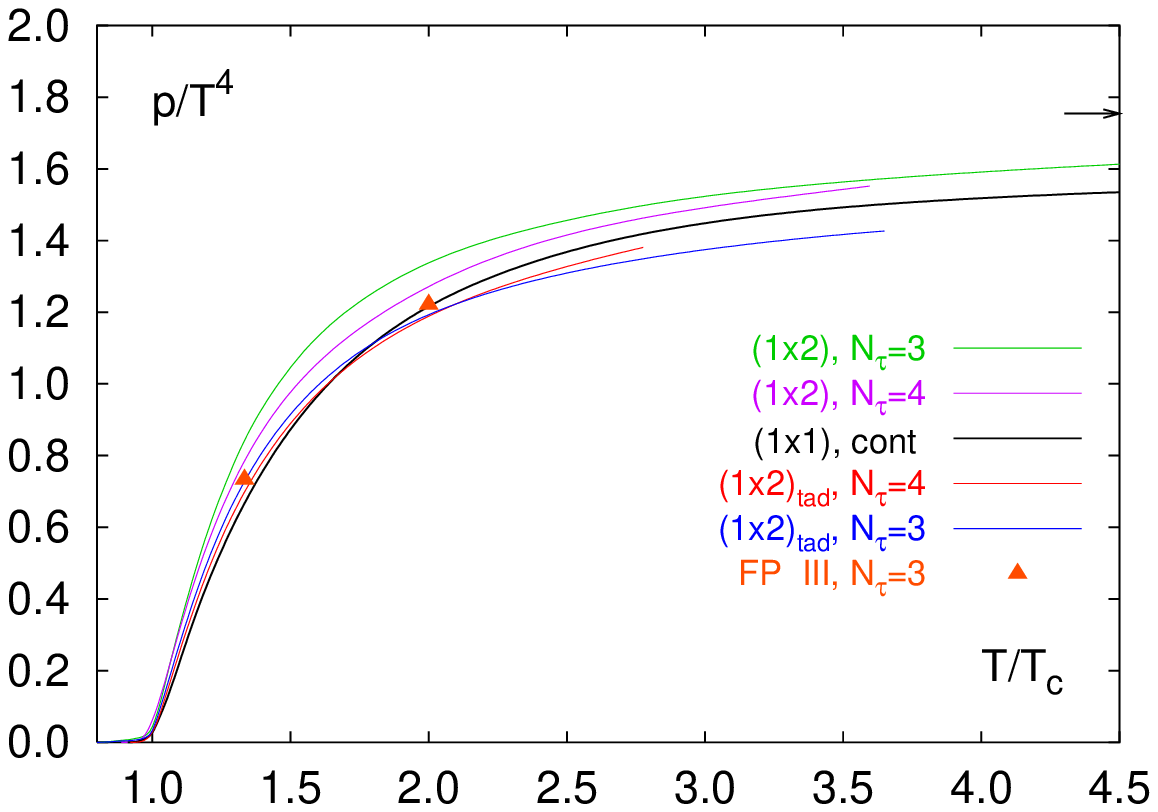}\hspace*{1cm}
\includegraphics[height=6cm]{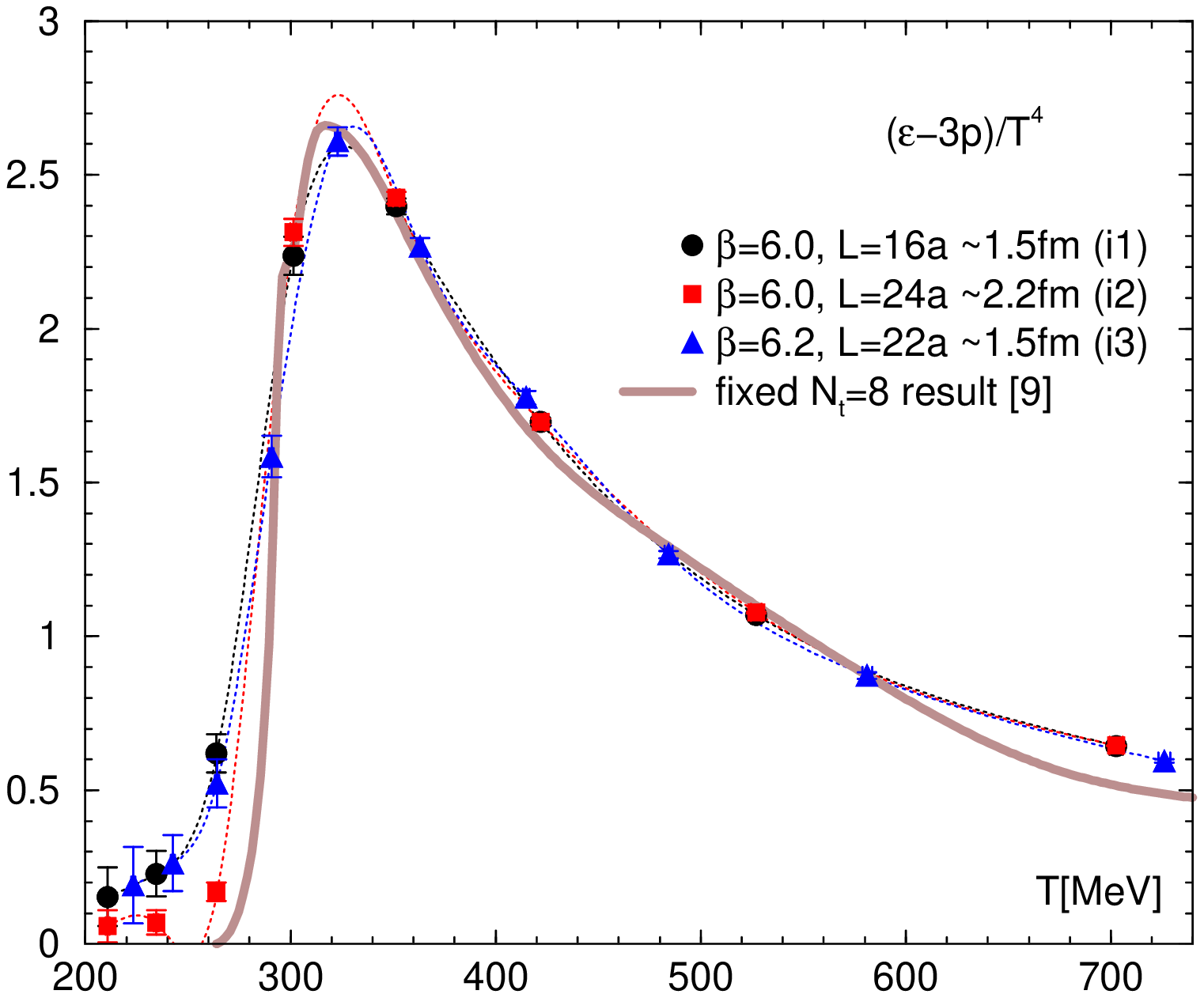}
\end{center}
\vspace*{-0.5cm}
\caption[]{Left: Comparison of pure gauge calculations with
tree-level improved, tadpole improved \cite{ym_imp} and a fixed point \cite{ym_fp} action on $N_\tau=3$ and 4 lattices.
The arrows indicate the ideal gas result in the continuum limit.
The solid line is the continuum extrapolation using the unimproved Wilson action \cite{boyd}. 
From \cite{ym_imp}.
Right: Trace anomaly from the fixed scale approach \cite{whot_fix}, compared to the $N_\tau=8$
data from \cite{boyd}. The dotted lines are spline interpolations, horizontal errors from insecurities
in the scale are smaller than the symbols. From \cite{whot_fix}.}
\label{fig:ym_imp}
\end{figure}
These results have proven stable against checks with improved actions, \fig\ref{fig:ym_imp} (left).
The data show that both the tree-level improved and tadpole improved Wilson actions  \cite{ym_imp}
as well as a fixed-point action \cite{ym_fp} 
approach continuum results for the pressure already on rather coarse
lattices with $N_\tau=3,4$. In \fig\ref{fig:ym_imp} (right) the trace anomaly is compared between 
the $N_\tau=8$ data
from \cite{boyd} and those produced with the fixed scale approach in \cite{whot_fix}.
In this work, the standard Wilson action was used with fixed lattice gauge couplings $\beta=6.0,6.1$.
Using $r_0$ to set the scale, this corresponds to the lattice spacings $a=0.093,0.068$, 
respectively. Temperature was then tuned by varying $N_\tau$ leading to the data points in the figure.
For the zero temperature subtraction, $N_\tau=16,22$ lattices were used for the $i2,i3$ runs in 
the figure, respectively. For the beta-function, the data from \cite{boyd} were used.

\fig\ref{fig:ym_imp} (right) shows generally good agreement between the two methods.
At low temperatures, one observes a finite volume effect, but no difference between the lattice spacings.
On the other hand, at large temperatures, the fixed-scale results overshoot the fixed $N_\tau$ results.
This is likely because for the former the cut-off effects are stronger at high temperatures, as discussed
in Sec.~\ref{sec:fixed}. A sensitive measure for cut-off effects is provided by the peak height of the
trace anomaly. It indicates that the continuum limit has indeed been reached in Yang-Mills theory,
whereas for full QCD the peak height is not yet settled between different actions, as we shall see later. 

\subsection{$SU(3)$ at  high temperatures}\label{sec:hight}

As discussed in Sec.~\ref{sec:syst}, simulations of thermal systems require $1\ll N_\tau \ll N_s$
in order to have manageable cut-off and finite size effects. This has limited simulations of the QCD
equation of state to typically $T\lsim 5T_c$. On the other hand, one would like to make contact to 
weak coupling perturbation theory, which unfortunately converges rather poorly 
until much higher temperatures are reached. Improvements in the convergence behaviour
can be achieved by employing screened perturbation theory, consisting of partial resummations 
to infinite orders rather than strict perturbation theory, cf.~\cite{blaizot}. Recent calculations using the
HTL-scheme have managed reasonable contact with lattice results at $\sim 4T_c$ for pure gauge 
theory \cite{htl1} and dynamical QCD \cite{htl2}. Nevertheless, it still is desirable to extend lattice simulations to higher temperatures where perturbation theory is fully controlled.

One possibility to achieve this is by use of the momentum distribution method discussed in 
Sec.~\ref{sec:mom}. Its feasibility for pure
gauge theory has been demonstrated in \cite{gm1}, with 
numerical results obtained from the standard Wilson action as shown in \fig\ref{fig:hight1} (left). 
Simulations have been performed for three different 
temperatures and the scale was set using $r_0$ and its related beta-function from \cite{ns}.
Within the uncertainties good agreement for the lower temperatures with earlier results from
the integral method \cite{boyd} is observed.
The continuum value for the entropy 
has again leading $\Oas$ corrections.

Another attempt to develop methods in this direction has been reported 
in \cite{bmw_hight1,bmw_hight2}, based on a modification of the integral method.
Starting point is the observation that the vacuum divergence, which is subtracted to renormalise
the free energy density or pressure, is temperature independent. 
The subtraction is therefore also
achieved at two non-vanishing temperatures. In \cite{bmw_hight1} it is 
suggested to consider the series with half-temperature steps,
\ba
\frac{p(T)_\text{ren}}{T^4}&=&\frac{p(T)-p(0)}{T^4}=\frac{p(T)-p(T/2)+p(T/2)-p(T/4)+p(T/4)\ldots -p(0)}{T^4}\nn\\
&=&\frac{p(T)-p(T/2)}{T^4}+\frac{1}{16}\frac{p(T/2)-p(T/4)}{(T/2)^4}+\frac{1}{256}\frac{p(T/4)-p(T/8)}{(T/4)^4}
+\ldots
\label{eq:steps}
\ea  
The idea is to evaluate the different terms using only two lattices with $N_\tau$ and $2N_\tau$. The
temperature for the successive terms is lowered by tuning the lattice gauge coupling. In practice, only a few terms are needed,
because of the rapid suppression from the prefactor. This is demonstrated in \fig\ref{fig:hight1} (right), which 
is based on the data from $N_\tau=6,8$ lattices \cite{boyd}, i.e.~with a suppression factor
$6/8$ instead of $1/2$. 

However, a word of caution is in order. \eq(\ref{eq:steps}) is written in continuum notation, where each term is represented by a $\beta$-integral from the lower integration limit $\beta_0$ to some
$\beta(T/x,N_\tau)$. The upper integration limits, and hence lattice spacings, are different in each term, 
leading to different cut-off effects. Thus, at finite lattice spacing the added and subtracted terms do not cancel exactly, as the continuum formula suggests.  Rather the continuum limit should be
taken for each term separately. It would be interesting to further investigate how improvement 
alleviates this problem or how small lattice spacings need to be to neglect this effect.
\begin{figure}[t]
\includegraphics[height=6.8cm]{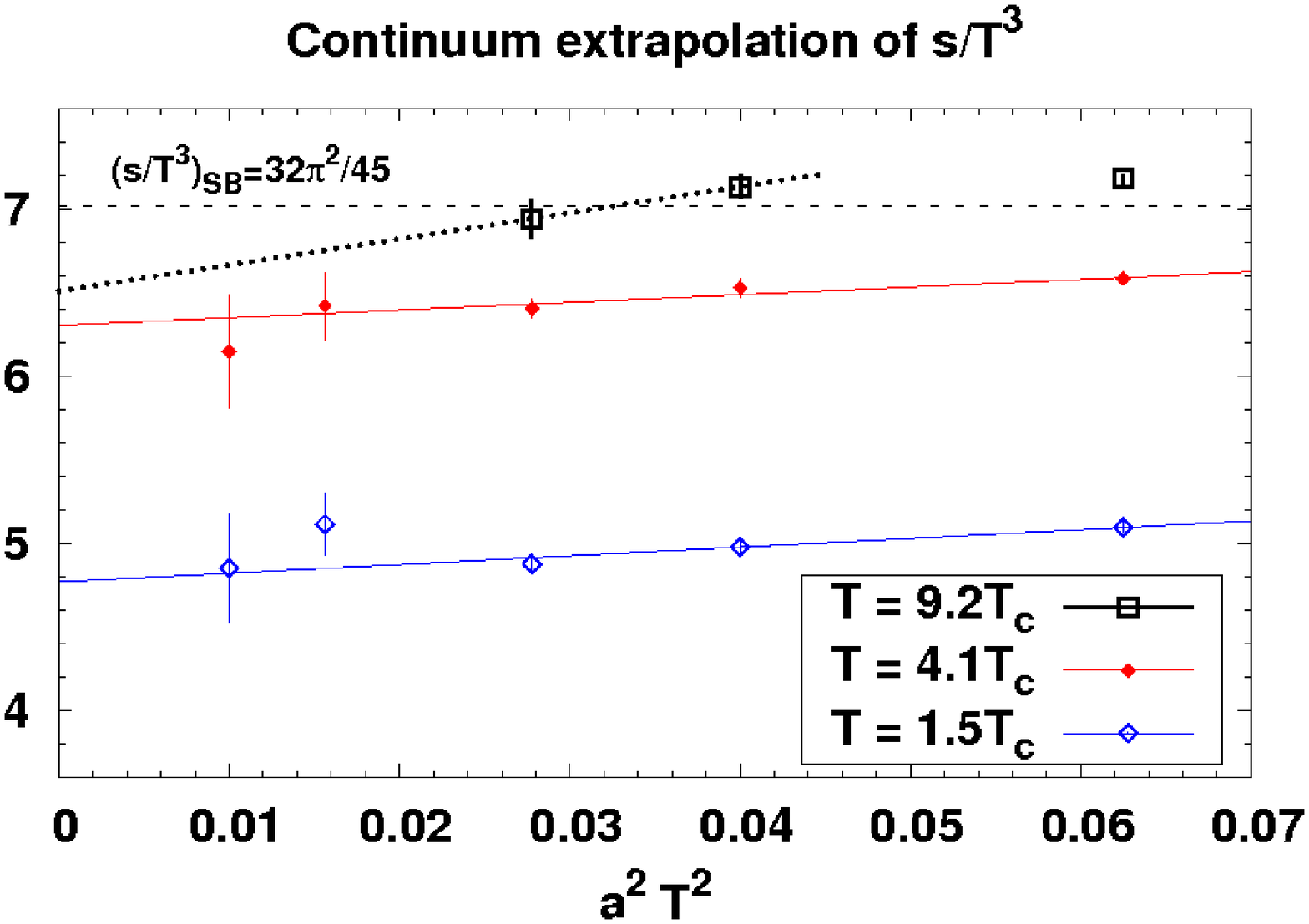}
\includegraphics[height=6.6cm]{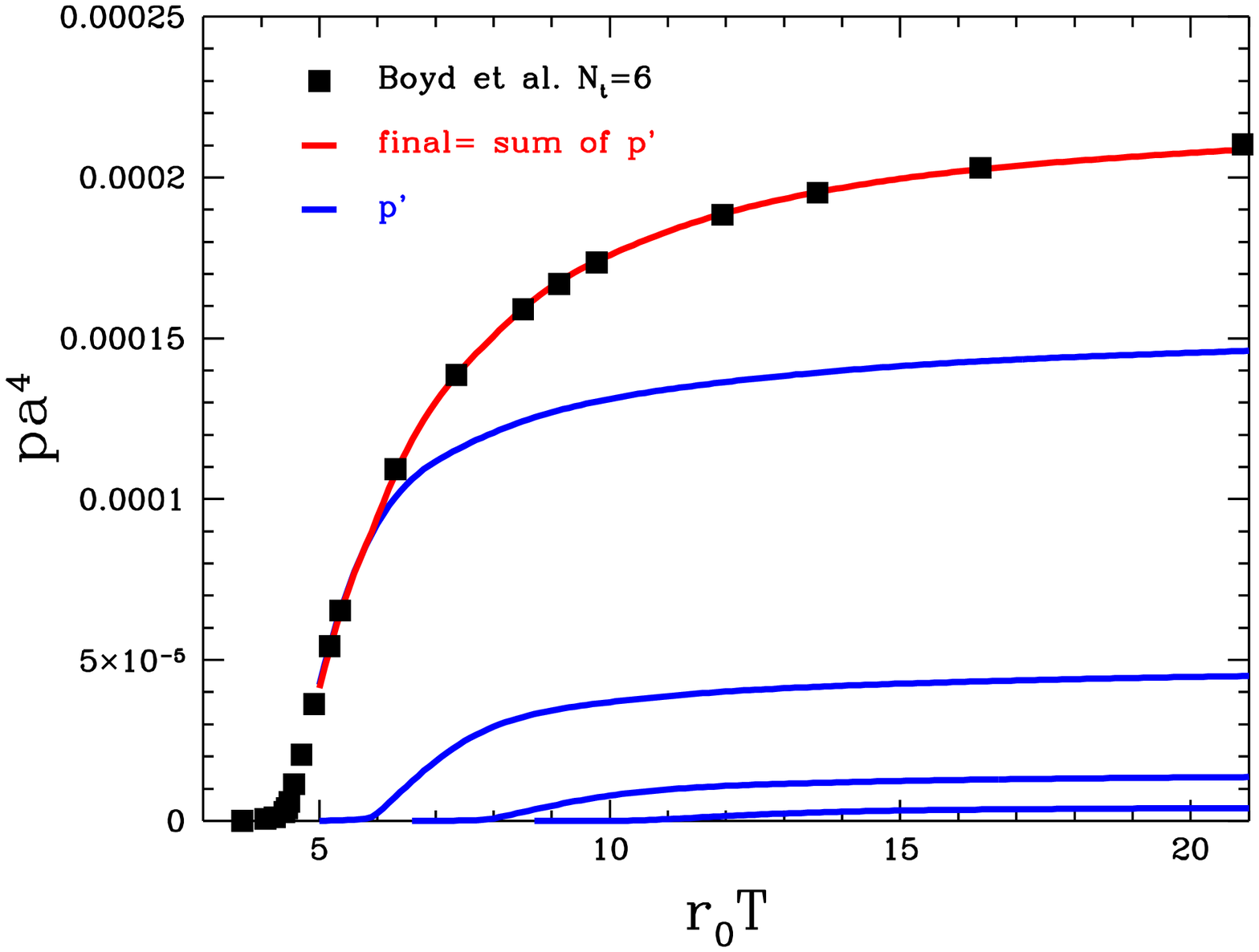}\hspace*{1cm}
\caption[]{Left: Scaling behaviour of $s/T^3$, \eq(\ref{eq:sT3}). The Stefan-Boltzmann value reached 
in the high-$T$ limit is also displayed.
From \cite{gm1}. Right: Pressure in lattice units calculated on $32^3\times 6$ and $32^4$ \cite{boyd}
compared to the sum of individual terms $p'$ of \eq(\ref{eq:steps}) using $32^3\times 6$ and $32^3\times 8$ lattices. From \cite{bmw_hight1}.}
\label{fig:hight1}
\end{figure}

This proposal was applied to a calculation of the equation of state for pure gauge theory
at high temperatures using a tree-level Symanzik improved action \cite{bmw_hight2}. 
The savings on large $N_\tau$ 
lattices were then invested in a large range of aspect ratios $N_s/N_\tau=4-24$ and dedicated finite size analyses. \fig\ref{fig:hight2} (left) shows the trace anomaly calculated on lattices with $N_\tau=5,6,7,8$,
in comparison with the known low temperature lattice data from \cite{boyd} as well as with standard 
perturbation theory through order $g^5$ \cite{klrs} and HTL-resummed perturbation theory \cite{htl1}.
\begin{figure}[t]
\includegraphics[height=7cm]{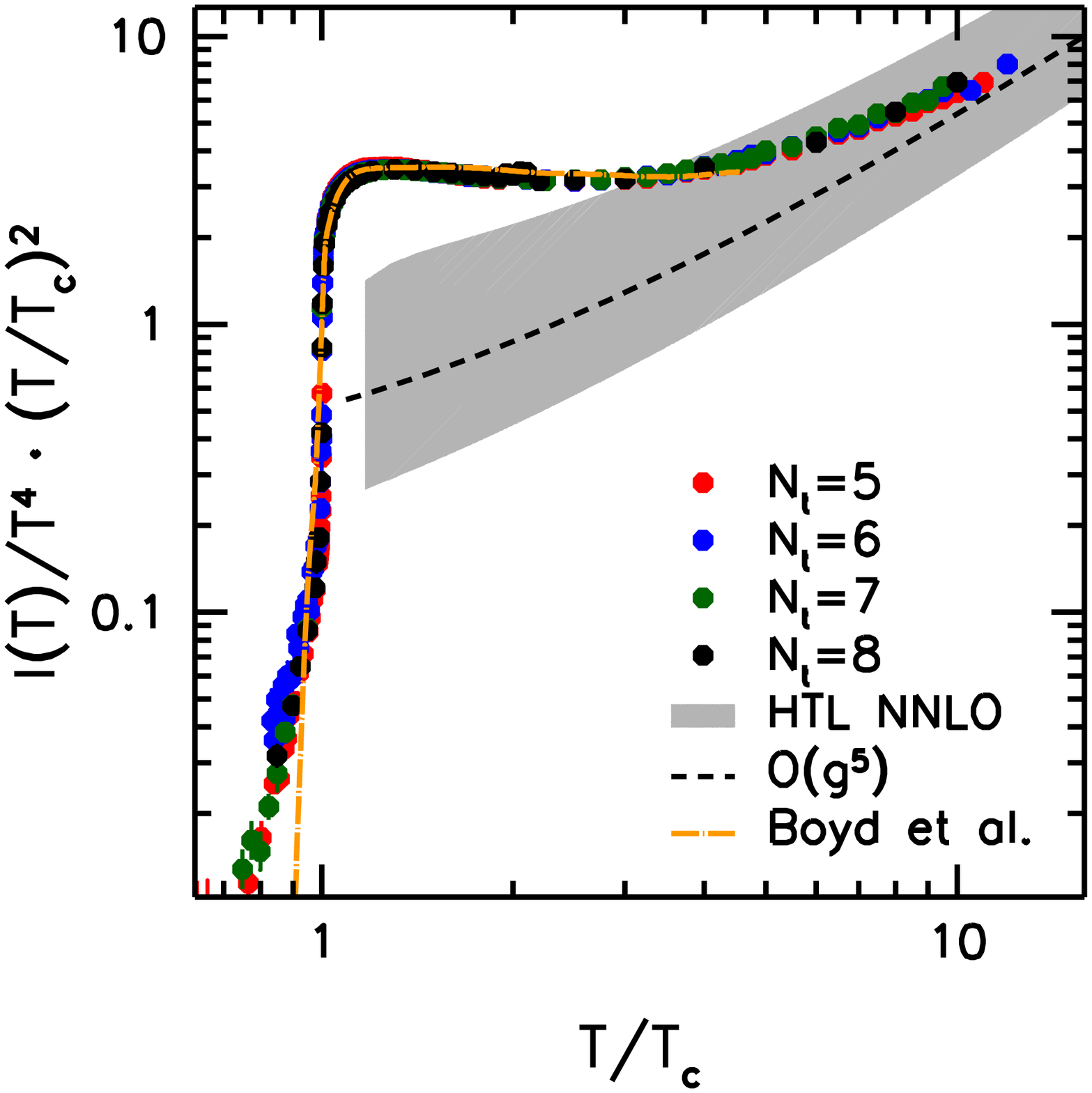}\hspace*{1cm}
\includegraphics[height=6.3cm]{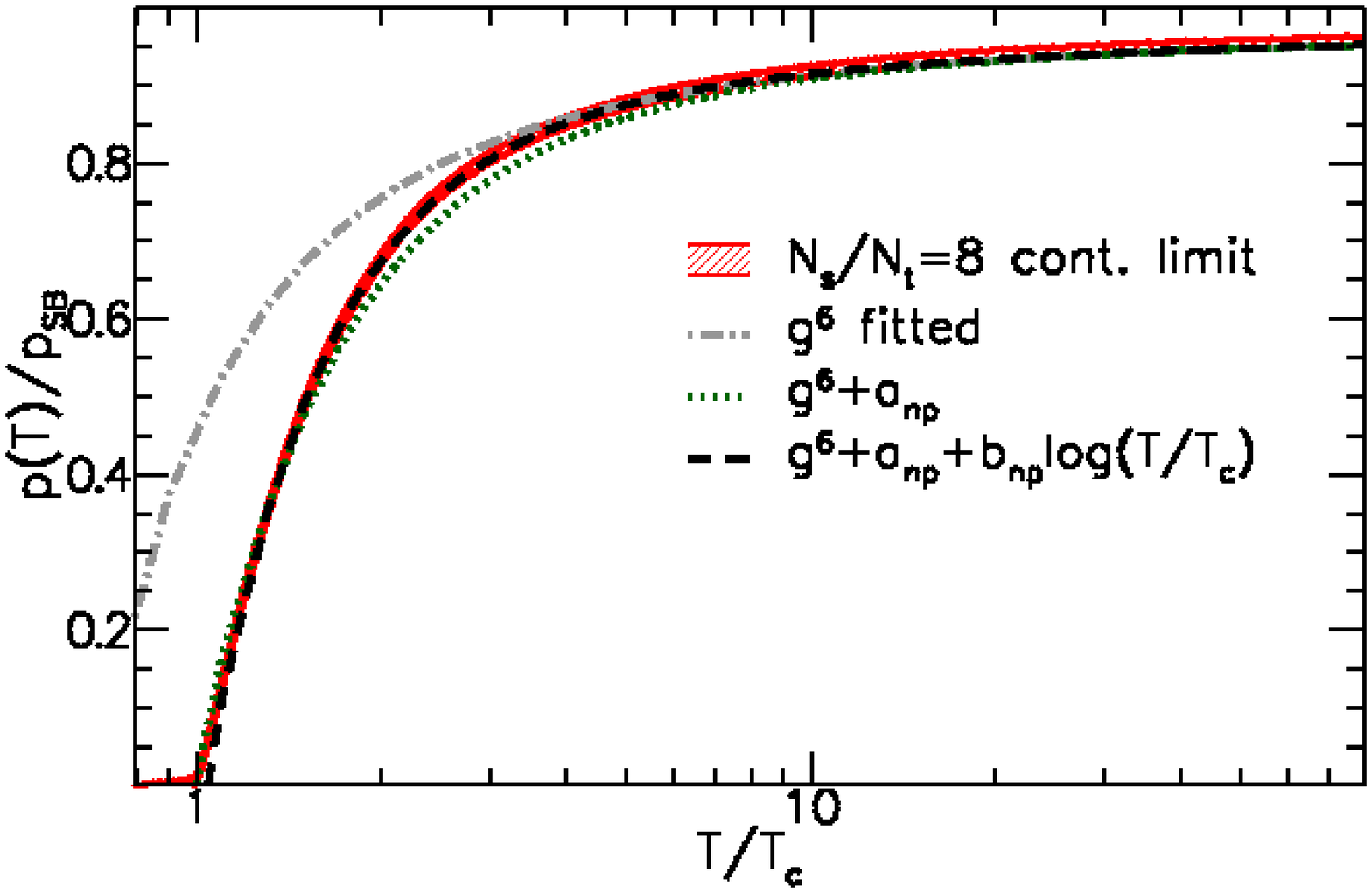}
\caption[]{Left: Trace anomaly for different $N_\tau$ compared with previous
lattice results \cite{boyd}, perturbation theory \cite{klrs} and HTL-resummation \cite{htl1}.
Right: Continuum extrapolated pressure normalised to the ideal gas value and comparison
with perturbative expressions.
From \cite{bmw_hight2}.}
\label{fig:hight2}
\end{figure}
Note that the dimensionless trace anomaly $I(T)/T^4$ has been rescaled by $(T/T_c)^2$, in order
to show deviations from perturbative behaviour. Since the effective expansion parameter for the latter
is $g^2T/m_E$, where $m_E\sim gT$ is the electric screening mass scale, at most logarithmic corrections
in temperature can modify the $T^{-4}$ behaviour of the dimensionless trace anomaly in perturbation 
theory. Hence the linear, nearly constant section up to $\sim 4T_c$ is indicative of non-perturbative
effects in the pressure, which then weakens at higher temperatures.

A study of the volume dependence and hence the requirements on the aspect ratio was also performed
on $N_\tau=5$ lattices.
From $N_s/N_\tau=6$ upwards the data do not show any finite size effects. This is an important observation, since the soft modes of the magnetic gauge fields
live on the scale $\sim g^2(T)T$ and for high temperatures correspond to a large correlation length
$1/(g^2(T)T)$. However, the corresponding screening masses are those of the lightest glueball. However,
the coefficient is large,
$m_{0^{++}}\approx 2.5g^2(T)T$ \cite{3dscr}, 
$\xi\sim 0.4 /g^2(T)T$ and $T$ grows faster than $g(T)$ shrinks. It is thus plausible that the finite size requirements are fulfilled on modest lattice volumes as observed in the simulations.

A continuum limit was then taken for the aspect ratio $N_s/N_\tau=8$ and the corresponding 
pressure is shown in \fig\ref{fig:hight2} (right). The perturbative regime can thus be reached at about 
$T\gsim 10T_c$ and the data
can be used to fit the $g^6$-coefficient in the perturbative expression for the pressure \cite{klrs}. 
The dashed curves in the figure represent a modelling of the non-perturbative temperature
regime $T_c<T\lsim 4T_c$, by adding to the perturbative formula up to $g^6$ the contributions,
\ba
p&=&p_\text{pert}(g^6)+a_\text{np}\;,\qquad \hspace*{2.5cm} a_\text{np}=0.879(2)(40)\nn\\
p&=&p_\text{pert}(g^6)+ a_\text{np}+ b_\text{np}\ln(T/T_c)\qquad\; a_\text{np}=1.371(1)(50),\;
b_\text{np}=-0.618(2)(4)\;.
\ea
The parametrisation including the log gives a good description of the data all the way down to the
critical temperature. Together with the perturbative expression from \cite{klrs}, we then have a 
parametrisation of the equation of state for all of the deconfined phase at our disposal. Note that even
at temperatures $T\sim 10T_c$ the pressure is still $\sim 10\%$ below the ideal gas limit. This is of course
consistent with the logarithmic running of the coupling constant. It illustrates however, that there
are still significant interactions of the soft modes in the plasma and
one must be very careful with intuition based on the Stefan-Boltzmann limit. The latter becomes accurate 
only at exceedingly high temperatures.

\subsection{$SU(N)$ with $N>3$}

$SU(N)$ pure gauge theory is also of theoretical interest  in the limit $N\rightarrow \infty$.
Firstly, one would like to understand how far $SU(3)$ is from this limit, for which considerable calculational
simplifications apply and analytic treatments become feasible \cite{hooft}. Secondly, the large $N$ limit
is a crucial ingredient for the modelling of QCD-like theories
based on  the AdS/CFT conjecture \cite{erd}. In order to see to which extent $SU(3)$ can be described
by large $N$ theories,
a non-perturbative calculation of the equation of state by lattice methods is useful also for $N>3$.

Calculations
for the critical temperature $T_c$ have been extended to $N=10$ \cite{dattag}, the equation of state
exists for $N=3,4,5,6,8$ \cite{panero1,panero2}.
The simulations for the latter were run on $20^3\times 5$ lattices for $N=3$ and $16^3\times 5$ lattices
for $N>3$. The calculation employs the integral method and the scale for $SU(3)$ was set using
$r_0$ \cite{ns}. For larger $N$, it was based on interpolated string tension values from \cite{teper}. 
\begin{figure}[t!]
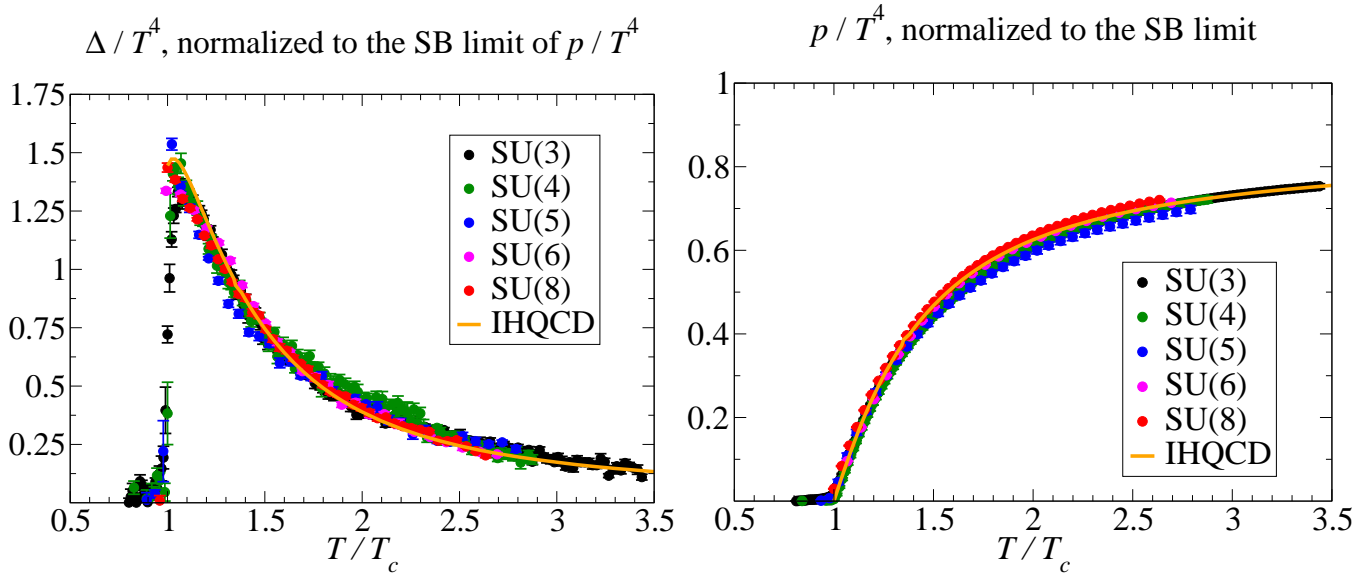

\begin{center}
\includegraphics*[width=0.48\textwidth]{plots/rescaled_trace}
\includegraphics*[width=0.48\textwidth]{plots/pressure}
\vspace{-0.4cm}
\caption{\label{fig:largen} Trace anomaly and pressure, obtained for various numbers of colours, $N$.
From \cite{panero2}.}
\end{center}
\end{figure}
The numerical results are displayed in \fig\ref{fig:largen}. A rather remarkable scaling is observed
for all thermodynamic potentials. Also shown are results from a prediction from a holographic
QCD model \cite{string} which agrees with the data remarkably well. (An earlier  
model compared to $SU(3)$ is in \cite{andreev}).

\section{Numerical results with staggered fermions}

As described in Sec.~\ref{sec:stag}, staggered fermions are computationally the cheapest because
they correspond to one-component fields in  the action \eq(\ref{stag}) rather than four-spinors. 
Their leading discretisation errors in the equation of state are $\Oas$ in the standard formulation and 
suppressed when using improved actions. For those reasons simulations using staggered
fermions are most advanced, in the sense of being closest to the physical masses and using the finest
lattices. Extended series of simulations have provided evidence that, at the long last, systematic errors
are getting under control and which actions have the best scaling behaviour in a given 
temperature range.

\subsection{The pseudo-critical temperature}

Let us recall that, for QCD with physical masses and a large neighbouring region, there
is no chiral or deconfinement phase transition, merely an analytic crossover. Hence
there is no uniquely defined critical temperature,  only a pseudo-critical ``$T_c$'' depending on the observable used to define it. 
 
 The determination of the pseudo-critical temperature for QCD at the physical point has been
the subject of some apparently contradictory results. Based on simulations with the 
asqtad action on $N_\tau=4,6,8$, $T_c= 169(12)(4)$ MeV was quoted \cite{milc_05}.
This was followed by the prediction $T_c=192(7)(4)$ MeV based on the p4 action on 
$N_\tau=4,6$ \cite{cheng_06}. In both cases simulations were performed at several light quark masses
larger than physical and then extrapolated simultaneously to the physical point and the continuum limit.
In \cite{cheng_06} the extrapolation formula is
\be
r_0T_c(m_l,m_s^\text{phys},N_\tau)=r_0T_c(m_l^\text{phys},m_s^\text{phys},\infty)+
A (r_0 m_\text{ps})^d+\frac{B}{N_\tau^2}\;.
\ee 
Here $O(4)$ scaling in the chiral limit is assumed, the data being consistent with the associated critical exponent $d$. Moreover, the values for $T_c$ based on susceptibilities of the Polyakov loop or chiral
observables were found to be within $185-195$ MeV, the quoted value representing an average.

By contrast, in a series of simulations using a stout link improved staggered action over a wide range of $N_\tau=4,6,8,10,12,16$ \cite{bmw_tc1,bmw_tc2,bmw_tc3}, consistently lower
values were quoted. The simulations were performed at the physical point and hence no extrapolation 
in the quark mass was necessary. For the continuum extrapolation the four largest $N_\tau$ 
corresponding to the finest lattices were used, providing moreover 
different values when extracted from the chiral condensate $\Delta_{l,s}$, \eq(\ref{eq:deltals}), or the Polyakov loop, 
\eq(\ref{eq:poly}), \cite{bmw_tc3,bmw_tc2}:
\be
T_c=\left\{\begin{array}{cc}  157(3)(3) {\rm MeV}& \Delta_{l,s}\\ 170(4)(3) {\rm MeV} & L
\end{array}\right.\;.
\ee
The continuum extrapolation is based on four lattice spacings, which allows to monitor that 
cut-off effects are small and truly scale as $\sim1/N_\tau^2$.
Note that the observable-dependent difference in $T_c$ 
is completely consistent with the fact that there is
no true phase transition at all, neither a chiral nor a deconfining one, merely a transition region
in which the chiral and confinement properties change gradually and smoothly.

In the meantime the differences between these actions are understood and nearly resolved, especially
since a new set of simulations with asqtad and HISQ/tree actions on $N_\tau=8,12$  and
an additional lighter quark mass \cite{hot_hisq} is available. A combined chiral and continuum extrapolation, based on the observable $\Delta_{l,s}$ gives
$T_c=154(9)$ MeV, in complete agreement with the stout results above. 
The previous differences are largely due to underestimated cut-off effects due to 
taste splitting in the asqtad and p4
actions.  Moreover, the corresponding data obtained for $N_\tau=4,6$ are not in the $N_\tau^{-2}$
scaling region and could not yet be extrapolated with the formula used. 
The same effects are visible in the equation of state, and we shall discuss them together
when we compare results between different actions in detail in Sec.~\ref{sec:comp}.

\subsection{Flavour dependence and interpretation of the equation of state}

The results of a computation of the pressure with the p4 action \cite{Karsch:2000ps}
are shown in \fig\ref{fig:eos1_hot} (left). The data have been obtained for
$N_f=2,3$ with (bare) mass $m_q/T = 0.4$ as well as
for $N_f=2+1$ with a heavier mass $m^s_q/T = 1$ on a very coarse lattice, $N_\tau=4$.
For comparison, 
continuum extrapolated pure gauge results are also included. Without entering the details of
the calculation, the plot tells us some interesting qualitative features.
As in the pure gauge case we see a rapid rise of the pressure in a narrow transition region.
The pseudo-critical temperature as well as the
magnitude of $p/T^4$ rise with the number of degrees of
freedom liberated at the transition. 
This last conclusion is firm, since the pressure also rises for fixed temperature
when light quarks are added to the theory, consistent with the expectations based on 
the Stefan-Boltzmann limit. We may thus conclude that the rapid rise in the pressure signals
deconfinement, in the sense that there are more and lighter degrees of freedom in the system.
On the other hand, these are not yet free quarks and gluons, as the pressure falls significantly
short of its Stefan-Boltzmann limit.

\subsection{Equation of state for almost physical quark masses: hotQCD}\label{sec:almost}

The hotQCD collaboration employs and continues previous work of the MILC and RBC-Riken collaborations.  They have 
been working with asqtad and p4 actions over several years. 
Simulations exist for $N_f=2+1$ using three different lattice spacings, with $N_\tau=4,6$ published in \cite{milc_07} (asqtad)
and \cite{cheng_08} (p4), while combined results for both actions and $N_\tau=8$ have been presented
in \cite{hotp4asq}.

The actions used for those simulations are those specified in Sec.~\ref{sec:fat}.
The strange quark mass $m_s$ is fixed close to its
physical value while $m_s/m_{ud}=10$. This corresponds to a pion mass of $m_\pi\approx 220$ MeV.
The lines of constant physics are defined by keeping the meson masses in Table \ref{tab:LCP} fixed,
where the scale is set by the Sommer parameter $r_0$. The latter is related to the 
$\Upsilon (2S-1S)$ splitting 
in calculations with the asqtad action \cite{ups,gray}. This gives $r_0=0.469(7)$~fm.  For $N_\tau=8$,
an aspect ratio $N_s/N_\tau=4$ was used throughout. 
\begin{figure}[t]
\includegraphics[width=0.5\textwidth]{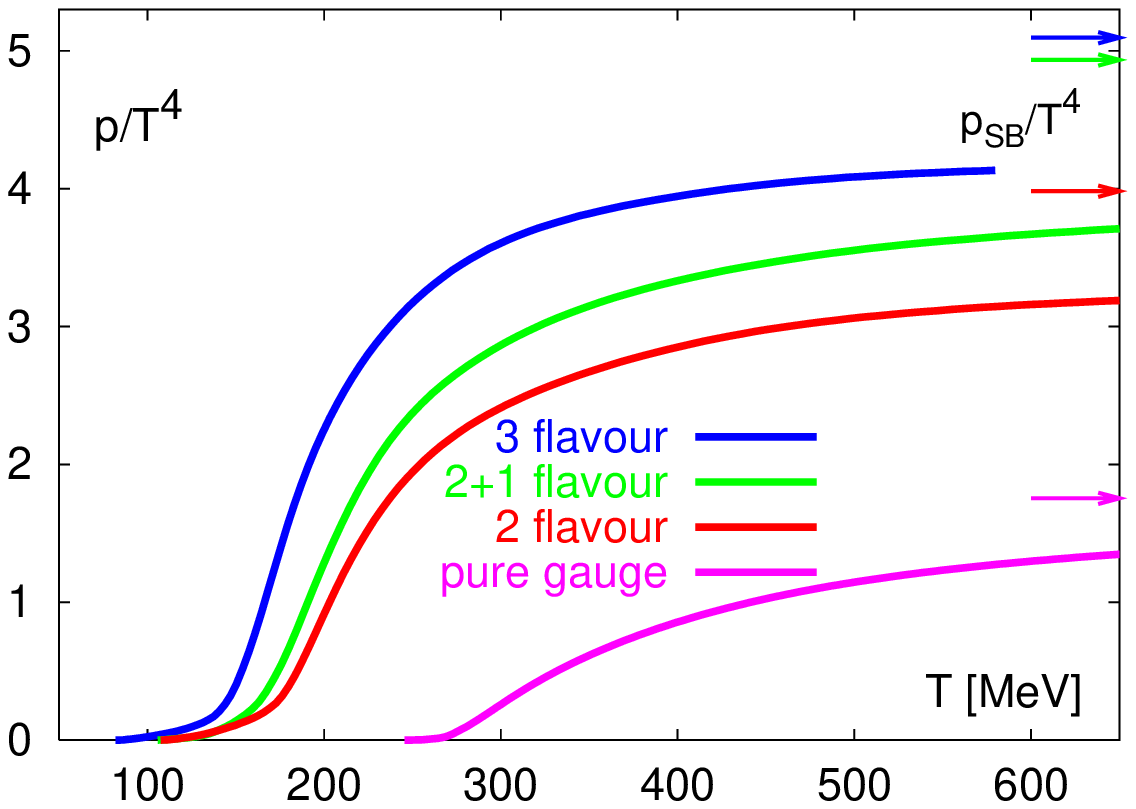}
\includegraphics[width=0.5\textwidth]{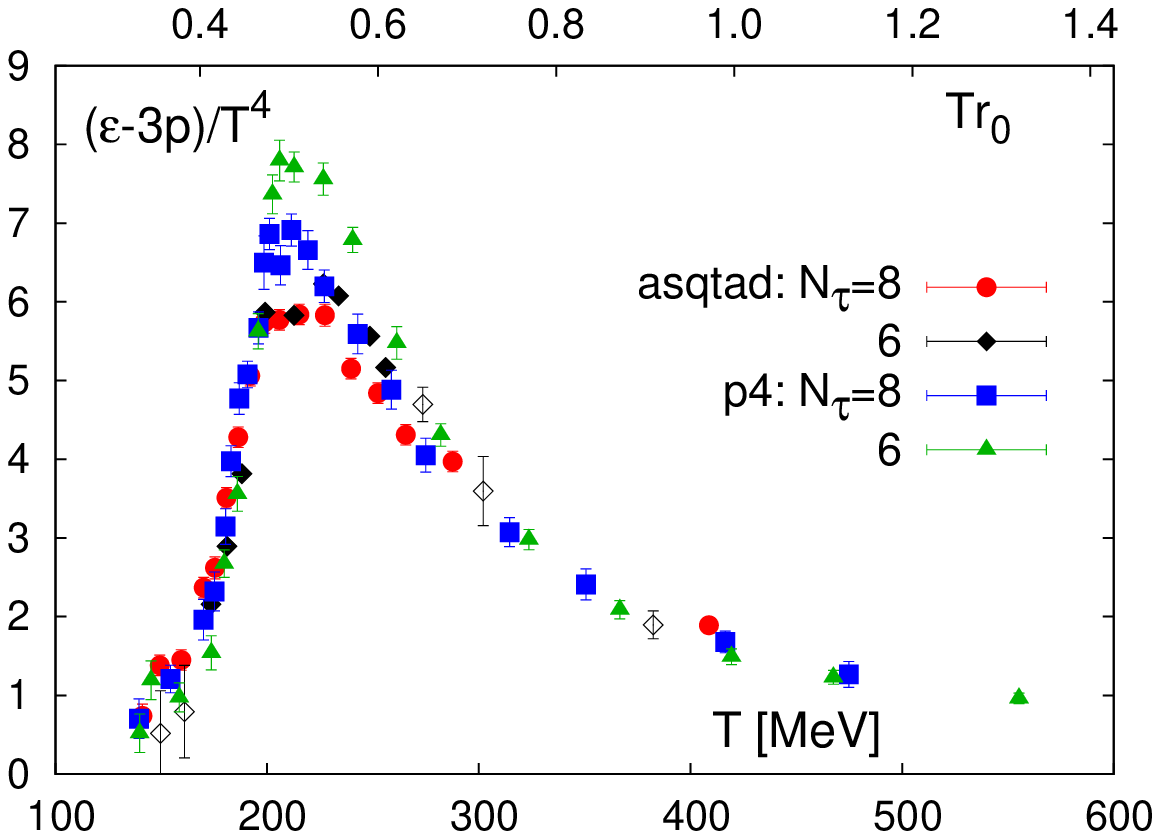}
\caption[]{Left: Flavour dependence of the pressure
for $N_\tau=4$ lattices compared to the continuum pure gauge result. From 
\cite{Karsch:2000ps}.
Right: The trace anomaly for $m_l/m_s=0.1$
calculated on lattices with temporal extent $N_\tau =6,8$.
From \cite{hotp4asq}.
}
\label{fig:eos1_hot}
\end{figure}
\begin{table}[bp]
\begin{center}
\vspace{0.3cm}
\begin{tabular}{|c|c|c|}
\hline
~ & p4-action & asqtad action \\
\hline
$m_{\bar{s}s} r_0$ &  1.59(5) &  1.83(6) \\
$m_{\bar{s}s}/m_K$ &  1.33(1) & 1.33(2) \\
$m_{\pi}/m_K$ &  0.435(2) &  0.437(3) \\
\hline
$m_\pi r_0$& 0.371(3)& \\
$m_{\bar{s}s} r_0$& 1.578(7)& \\
$m_K r_0$ & 1.158(5)& \\
\hline
\end{tabular}
\end{center}
\caption[]{The strange pseudo-scalar mass
$m_{\bar{s}s} \equiv \sqrt{2m_K^2 - m_\pi^2}$ in units of $r_0$ and mass ratios 
that characterise lines of constant physics for fixed 
ratios of light and strange quark masses, $m_l/m_s = 0.1$, above, and $m_l/m_s=0.05$, below. 
Errors represent the variation over the lines of constant 
physics.
}
\label{tab:LCP}
\end{table}

The beta-function for the lattice gauge coupling was fixed by fitting the $\beta$-dependence of $r_0$
to the ansatz
\be
\frac{r_0}{a} = 
\displaystyle{\frac{1+ e_r \hat{a}^2 (\beta) +f_r \hat{a}^4 (\beta)}{a_r 
R_2(\beta) 
\left(1+ b_r \hat{a}^2 (\beta) +c_r \hat{a}^4 (\beta) +d_r \hat{a}^6 (\beta)
\right)}}
\;,
\label{fit}
\ee
where $R_2(\beta)$ is the two-loop beta-function for 3-flavour QCD and $\hat{a}(\beta) =  R_2(\beta)/R_2(3.4)$. 
Similarly the beta-function for the quark mass was obtained from a parametrisation of the bare quark
mass 
\be
 m_l r_0/a = m^{RGI} r_0 \left(\frac{12 b_0}{\beta}\right)^{4/9} P(\beta)\;,
\ee
with a sixth order rational fit function $P(\beta)$ and the renormalisation group invariant $m^{RGI}$.

\begin{figure}[p]
\includegraphics[width=0.5\textwidth]{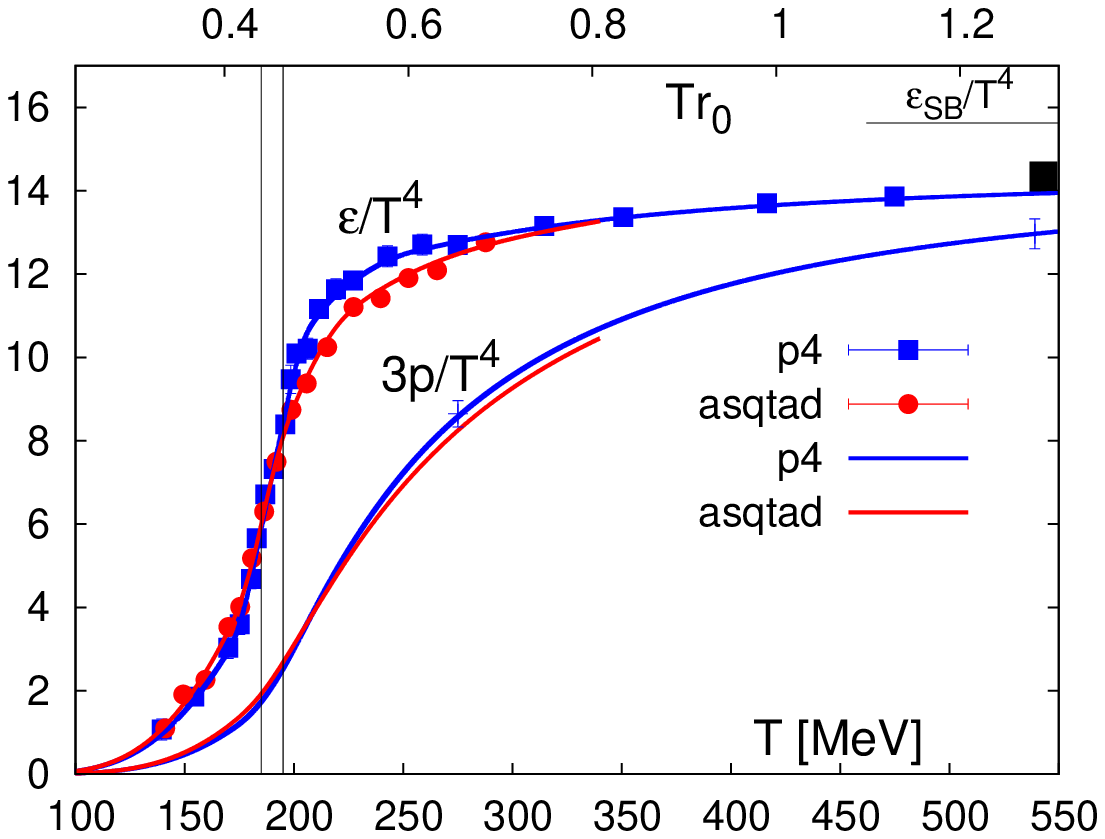}
\includegraphics[width=0.5\textwidth]{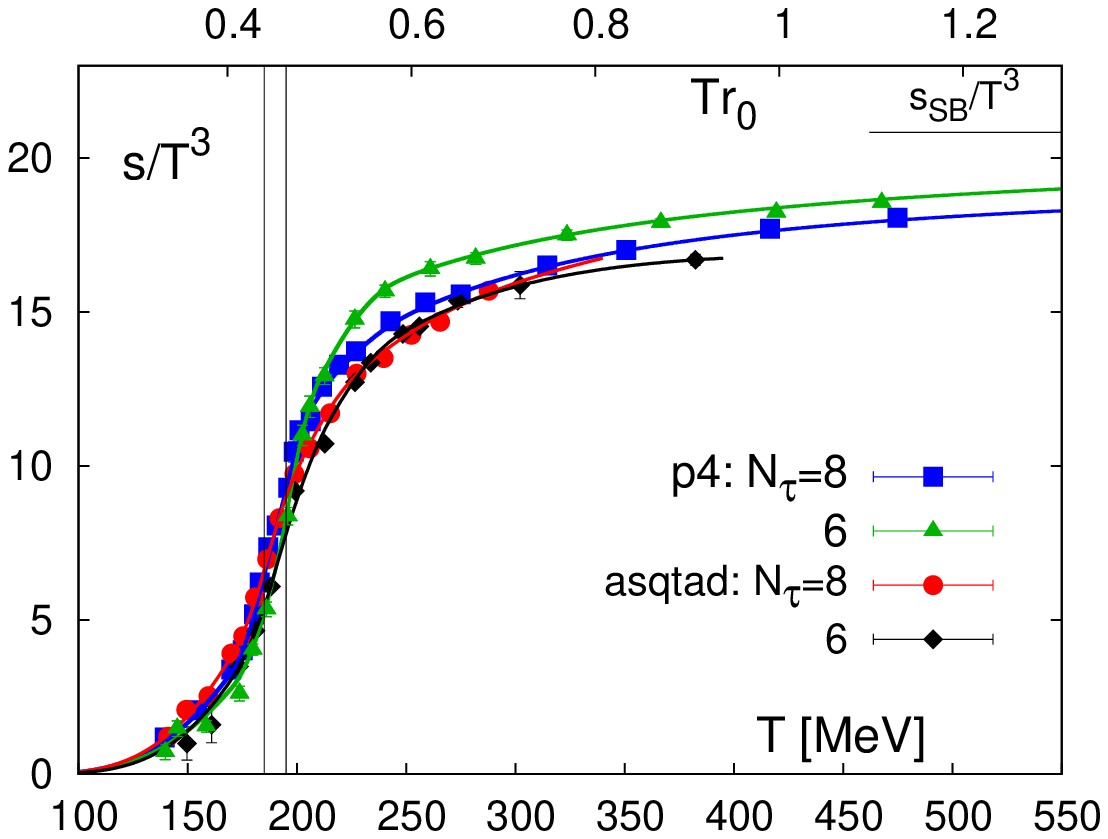}
\caption[]{\label{fig:eos3_hot} 
Left: Asqtad and p4 results for $m_l/m_s=0.1$ on $N_\tau=8$. Crosses with error bars indicate the systematic error from different integration schemes.
The black bar at high temperature indicates the systematic shift of data
when matching to a hadron resonance gas at $T=100$~MeV.
The vertical band indicates the transition region $185\; {\rm MeV} < T < 195\; {\rm MeV}$.
Right: Entropy density. 
From \cite{hotp4asq}.}
\end{figure}
\begin{figure}[p]
\includegraphics[width=0.5\textwidth]{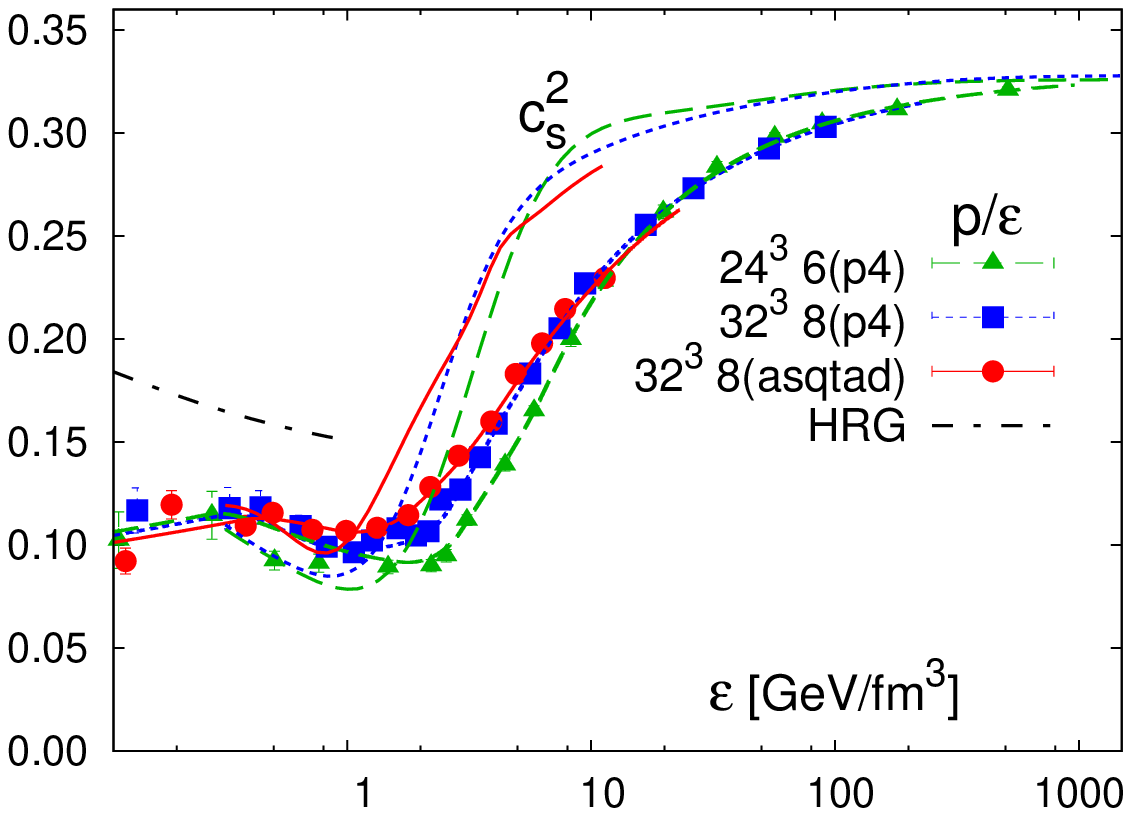}\hspace*{1.0cm}
\includegraphics[width=0.34\textwidth]{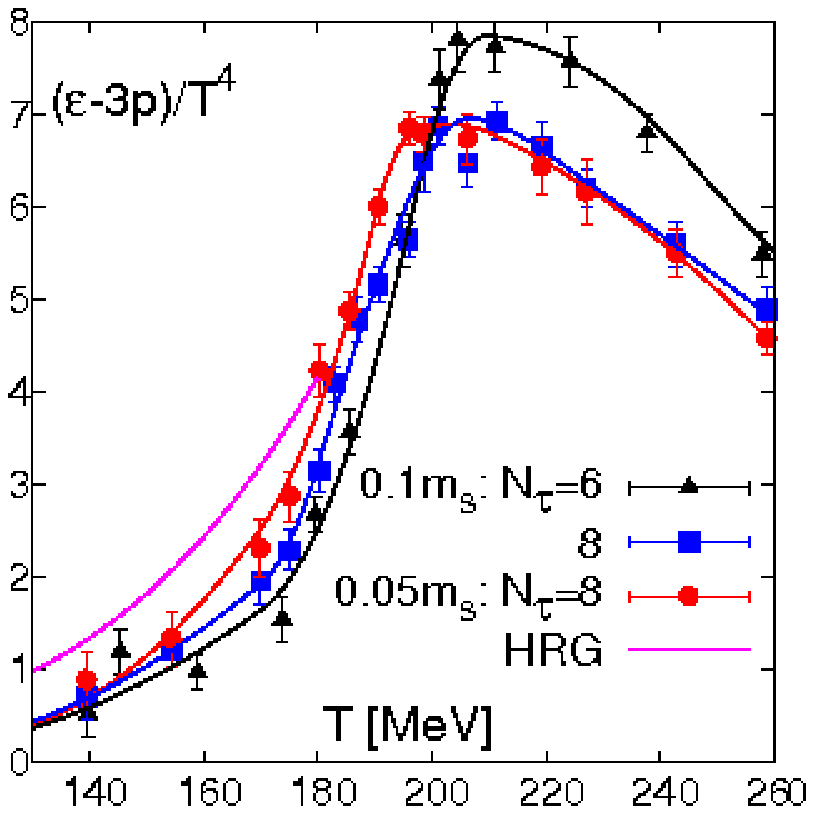}
\caption[]{\label{fig:eos2_hot}  
Left: Pressure divided by energy density ($p/\epsilon$) 
and velocity of sound ($c_s^2$)  
Lines without data points give $c_s^2$ calculated from the interpolating curves for $\epsilon/T^4$ and 
$p/T^4$. From \cite{hotp4asq}. Right: 
Comparison of $m_l=0.1m_s, 0.05m_s$ 
as well as with the HRG model which
includes all the resonances up to $2.5$GeV. Also shown are interpolations
of the lattice data. 
From \cite{cheng_09}. 
}
\end{figure}

The trace anomaly calculated for both types of actions is shown in \fig\ref{fig:eos1_hot} (right). Good agreement between different lattice spacings and actions is observed for low and high temperatures,
but differences remain in the peak region, indicating that there are still significant discretisation effects.
In order to obtain a smooth curve, the data are interpolated numerically.
The remaining
thermodynamic quantities are then obtained by numerical integration, 
\eq(\ref{freediv}), and using
the thermodynamic relations \eq(\ref{thermo}). 
This gives the results for the pressure, energy and the entropy,  \fig\ref{fig:eos3_hot}, and the speed of sound, \fig\ref{fig:eos2_hot} (left). The integration requires
fixing the lower integration constant by normalising the pressure at some low temperature to a prescribed
value. The figures correspond to setting $p=0$ at $T_0=0$ and integrating the fits to the trace anomaly.
Normalising the pressure to the hadron resonance gas prediction at $T_0=100$ MeV instead leads to
a constant shift marked by the black bar in the upper right of the figures.
The shape of the velocity of sound curve appears to depend on the details of the numerical interpolation
of the data for the trace anomaly, as is evinced by the difference between the data points and the smooth
curves. This difference should be interpreted as a measure of the 
systematic error associated with the processing of the original data for the trace anomaly.

These results were supplemented subsequently with an additional set of simulations using the p4 
action with lighter quark masses, $m_l/m_s=0.05$, corresponding to $m_\pi\approx156$ MeV, using
$32^3\times 8$ finite temperature and $32^4$ zero temperature lattices \cite{cheng_09}.
The line of constant physics in this case was defined by the second set of quantities in Table \ref{tab:LCP}.
The resulting trace anomaly  is shown in \fig\ref{fig:eos2_hot} (right) and compared to the previous
calculations for heavier light quark masses. In order to see the mass effect, the $N_\tau=8$ data
should be compared. Differences are visible for $T\lsim 200$ MeV, where the lighter quark values
sit systematically higher. This is consistent with the decrease of the 
pseudo-critical temperature for the crossover (within 
a given definition)  with decreasing quark mass. On the other hand, for temperatures above the
transition the quark mass dependence largely disappears. This is consistent with the fact that
hadrons are deconfined and the screening masses corresponding to the same quantum number 
channels are proportional to $\sim 2\pi T$, compared to which the bare quark masses
are negligible. 
 
Note that, for $T\lsim 180$ MeV, data from both masses are significantly lower than the 
hadron resonance gas prediction, pointing at the presence of discretisation errors in the light 
hadron sector. This is corroborated by a comparison with hadron resonance gas predictions using
the distorted lattice spectrum including cut-off effects such as taste splitting \cite{pasi10}, as outlined in 
Sec.~\ref{sec:taste}, and is shown in \fig\ref{fig:hrg_comp} (right) in Sec.~\ref{sec:comp}.
When the hadron resonance gas model includes the taste splitting and cut-off effects in the mass
values, excellent agreement is found between its predictions
and the lattice data for $T\lsim T_c$. One is thus led to conclude that the observed deviations from the
continuum hadron resonance gas predictions is to a smaller extent caused by larger than physical quark
masses, and to a large extent by discretisation effects in the low lying hadron spectrum. In the final
limits of physical masses and vanishing lattice spacing one would thus expect agreement with the 
continuum hadron resonance gas predictions.

\subsection{The equation of state for physical quark masses: BMW}\label{sec:bmw}

\begin{figure}[t]
\includegraphics[width=0.55\textwidth]{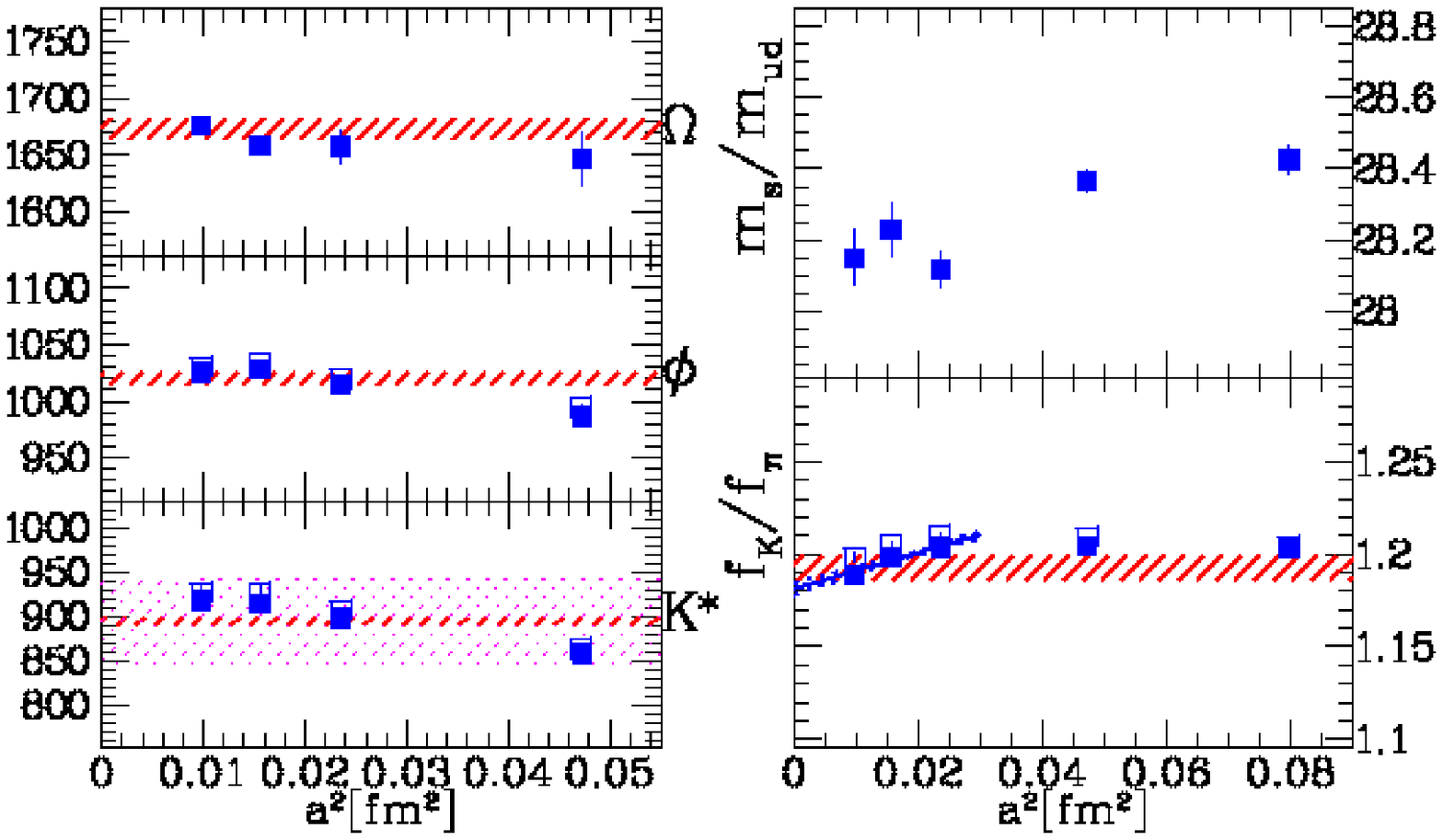}\hspace{1.5cm}
\includegraphics[width=0.35\textwidth]{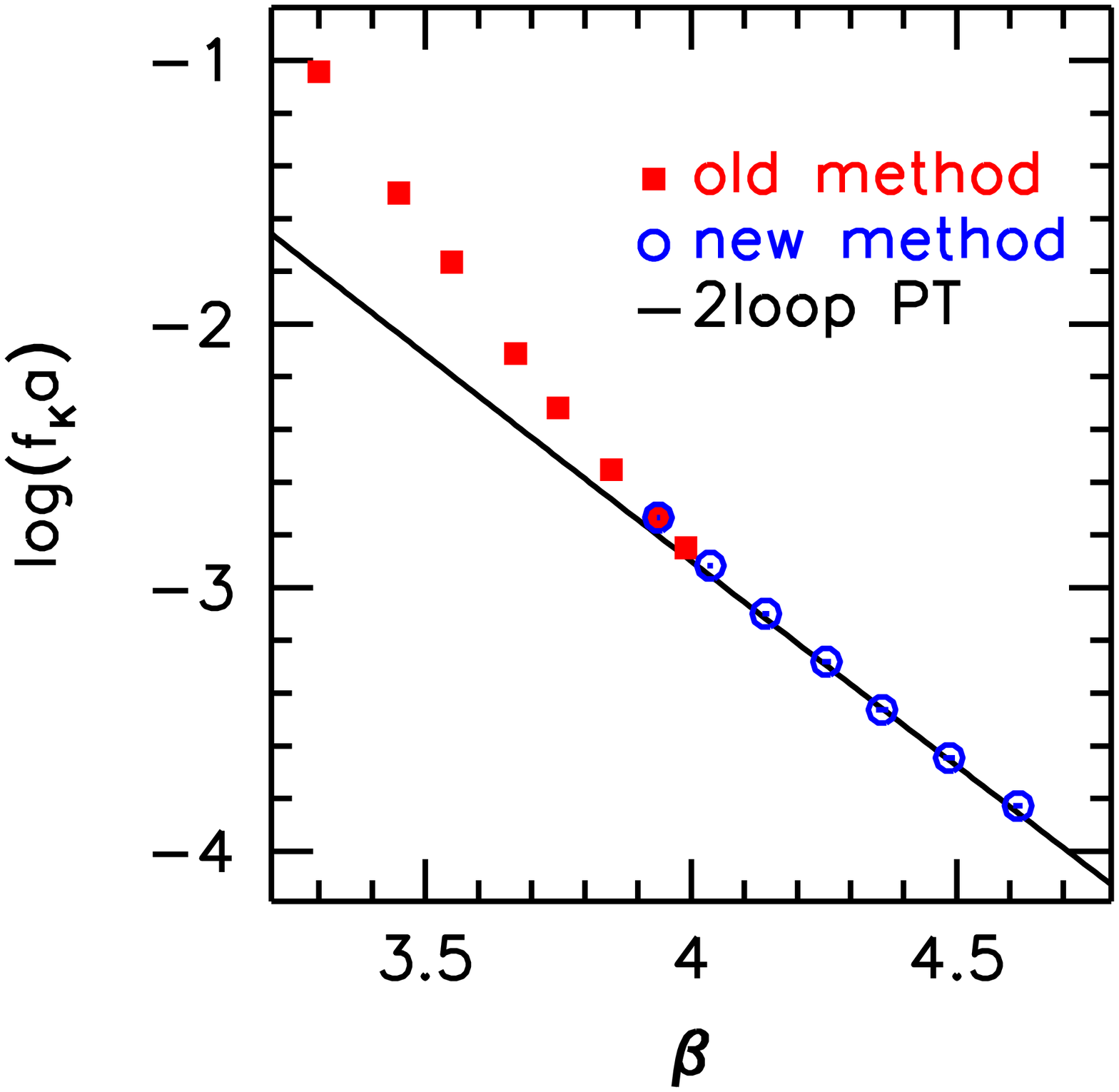}
\caption[]{Left: Meson masses and ratios of quark masses and decay constants on the finest lattices
denoted by squares in the left plot. Horizontal bands give experimental results. From \cite{bmw_tc2}.
Right: Lattice spacing as a function of $\beta$ for $N_f=2+1$ simulation. 
The curve corresponds to 2-loop perturbative running. From \cite{bmw_eos2}.
}
\label{fig:lcp_bmw}
\end{figure}
At the time of writing, only the Budapest-Marseille-Wuppertal collaboration has provided 
a series of calculations for physical quark masses.
They are also working on the finest lattices used in this context and provide a 
full continuum extrapolation for a growing number of temperature values. 
Moreover, comparison between different lattice spacings
suggests that cut-off effects are indeed small. 
Results for the equation of state on $N_\tau=4,6$ can be found in \cite{bmw_eos1}, $N_\tau=8,10,12$
were added in \cite{bmw_eos2, bmw_charm}. 

The actions used in this series of simulations are a tree-level Symanzik improved gauge action, 
Sec.~\ref{sec:impr_gauge}, and the one-link staggered action, \eq(\ref{stag}), where the link
is stout-smeared, Sec.~\ref{sec:fat}. The link smearing was applied twice, with parameter $\rho=0.15$.
The bare lattice parameters are fixed using as input the physical values $m_\pi=135$ MeV,
$f_K=155.5$ MeV, $m_K=495$ MeV \cite{pdg} to define the line of constant physics.
This translates to a fixed bare quark 
mass ratio $m_s/m_{ud}=28.15(1)$ along the line $m_s(\beta)$.  
The required beta-functions were computed in two different ways. For $3.45 < \beta < 3.85$ the hadronic
observables were calculated on zero temperature lattices $24^3\times 32$ up to 
$48^3\times 64$, setting the scale using $f_K$. The masses of the $\Omega, \phi, K^*$ mesons
were then monitored to stay at their physical values, \fig\ref{fig:lcp_bmw} (left). For the larger couplings
$3.9 < \beta < 4.6$ corresponding to the smallest lattice spacings, spectrum calculations would have
been too expensive. Instead Creutz ratios were used to define an effective pure gauge coupling 
\cite{bilgici}, which was then modified by including perturbative one-loop running of the quark masses.
For a few $\beta\geq 3.99$, control simulations were run to ascertain that the deviation from the true
line of constant physics in this way remained smaller than 2\%.    
The lattice gauge couplings employed reach the weak coupling 
region, where the corresponding beta-function is described by perturbative two-loop running, 
\fig\ref{fig:lcp_bmw} (right). 

A check for finite volume effects was run for $N_\tau=6$ with 
aspect ratios $N_s/N_\tau=3,6$, corresponding to box sizes of $\sim 3.5,7$ fm, respectively. 
In addition, aspect ratios $N_s/N_\tau=3,4$ were compared on $N_\tau=8$ \cite{bmw_charm}. 
No differences beyond the statistical errors were found,
so all simulations on finer lattices were
performed with $N_s/N_\tau=3$.

The main results characterising the equation of state are shown in \fig\ref{fig:eos1_bmw}, where tree-level
improvement on the level of the observable has been applied in order to reduce cut-off effects.
For the trace anomaly, there are data with four different lattice spacings. The $N_\tau=8,10,12$ data
essentially lie on top of each other, indicating that cut-off effects are very small in this regime of 
lattice spacings. Only the $N_\tau=6$ data show some deviation on the right flank of the peak. The inset
shows a comparison with the continuum hadron resonance gas model using physical masses, which describes the data well up to 
$T\sim 140$ MeV. The panel showing the pressure consists of three lattice spacings corresponding
to $N_\tau=8,10,12$. The two coarser lattice spacings cover a temperature range 
over an order of magnitude, $T\sim 100-1000$ MeV. Since the lattice spacings shrink with $T$, the 
results are most reliable at high temperatures. Interestingly,
$N_\tau=6,8$ show no difference at higher temperatures and
one may consider these lattices fine enough, even though stout fermion actions are not optimised
for the ideal gas. This is consistent with the
observation that for the highest temperatures, the pressure is still significantly short of the Stefan-Boltzmann limit. This means interactions still play an important role for 
those temperatures, even though the rise of the pressure signals 
deconfinement. Of course, this is to be expected based
on the slow logarithmic running of the coupling $\alpha_s(T)$.

On the other hand, at the lowest temperature $T\approx 100$ MeV, the lattice spacing is $a\sim 0.2$ fm
on $N_\tau=10$, which is too coarse to be in the scaling region. This is also indicated by the fact that
for this lowest temperature the pressure from the lattice is only about two thirds of what the hadron
resonance gas predicts. While the latter of course does not constitute a complete result either, it
should be better the lower the temperature. The authors quote the discrepancy to the hadron resonance gas as a measure of the remaining systematic error for low temperatures. Simulations at an additional $N_\tau=12$ followed by a continuum extrapolation should help to provide a final answer, see below. Finally, \fig\ref{fig:eos1_bmw}  also shows the entropy and energy densities, while the speed of sound is depicted in \fig\ref{fig:eos2_bmw} (left).
\begin{figure}[t]
\includegraphics[width=0.47\textwidth]{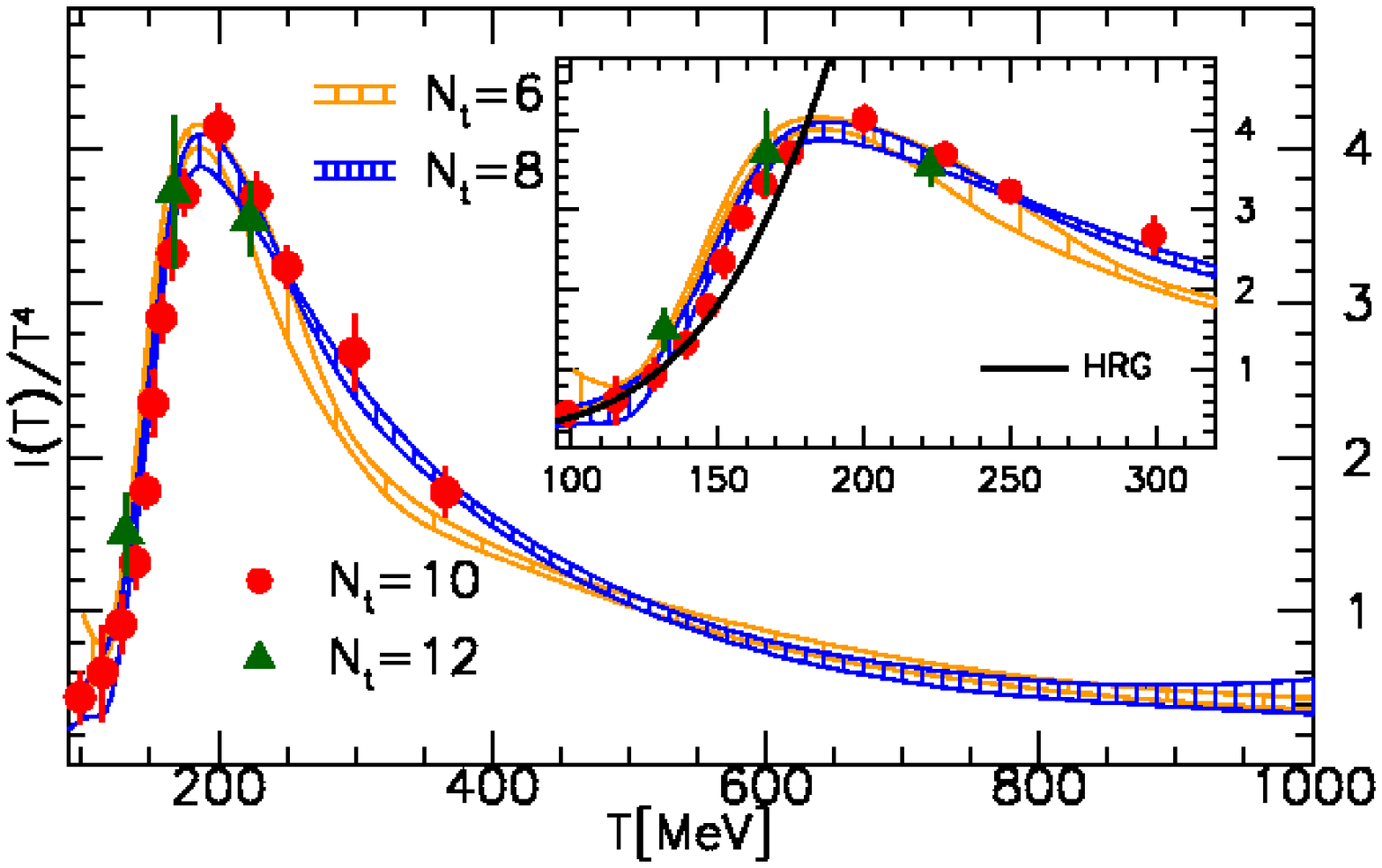}\hspace*{0.5cm}
\includegraphics[width=0.47\textwidth]{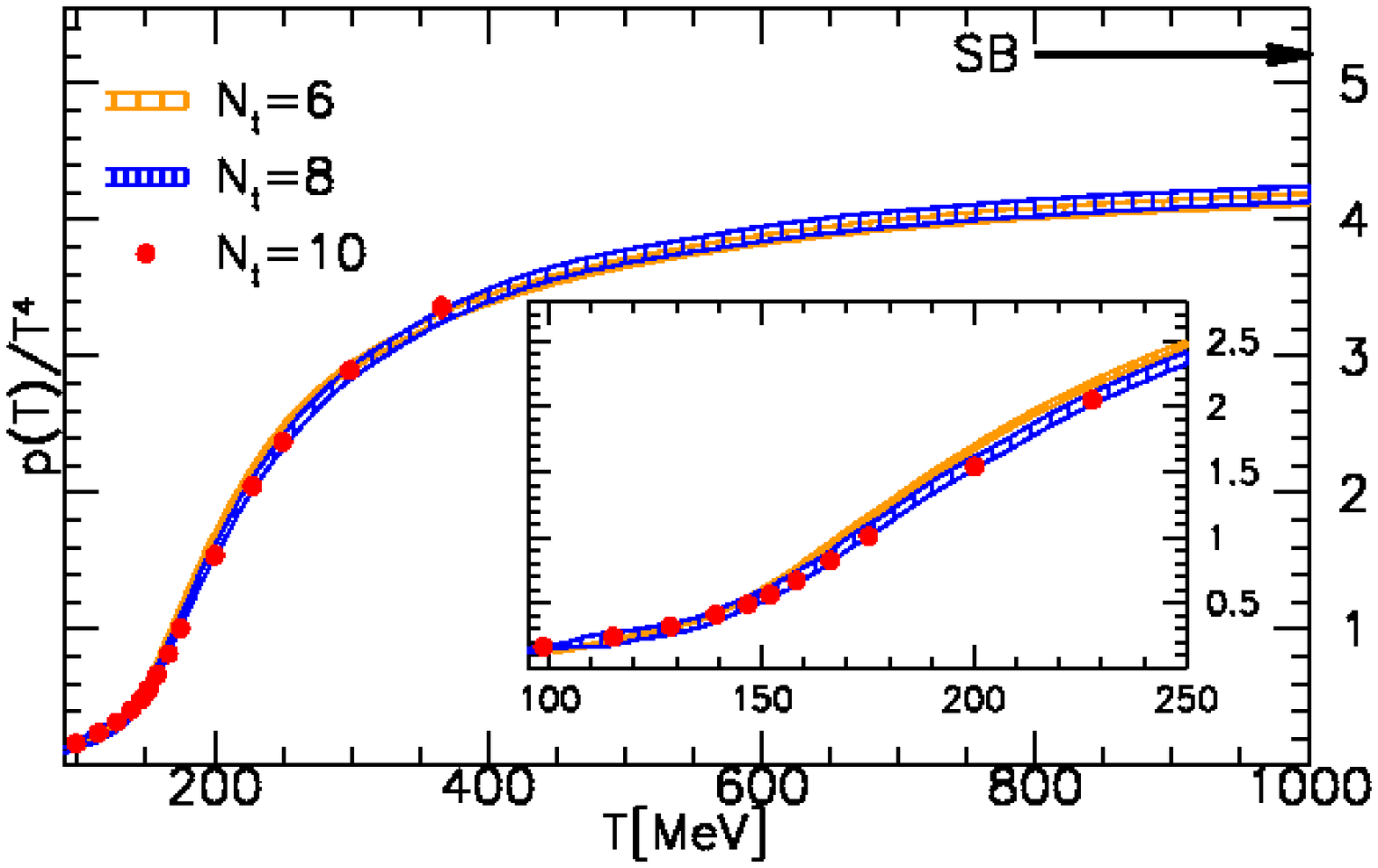}
\includegraphics[width=0.47\textwidth]{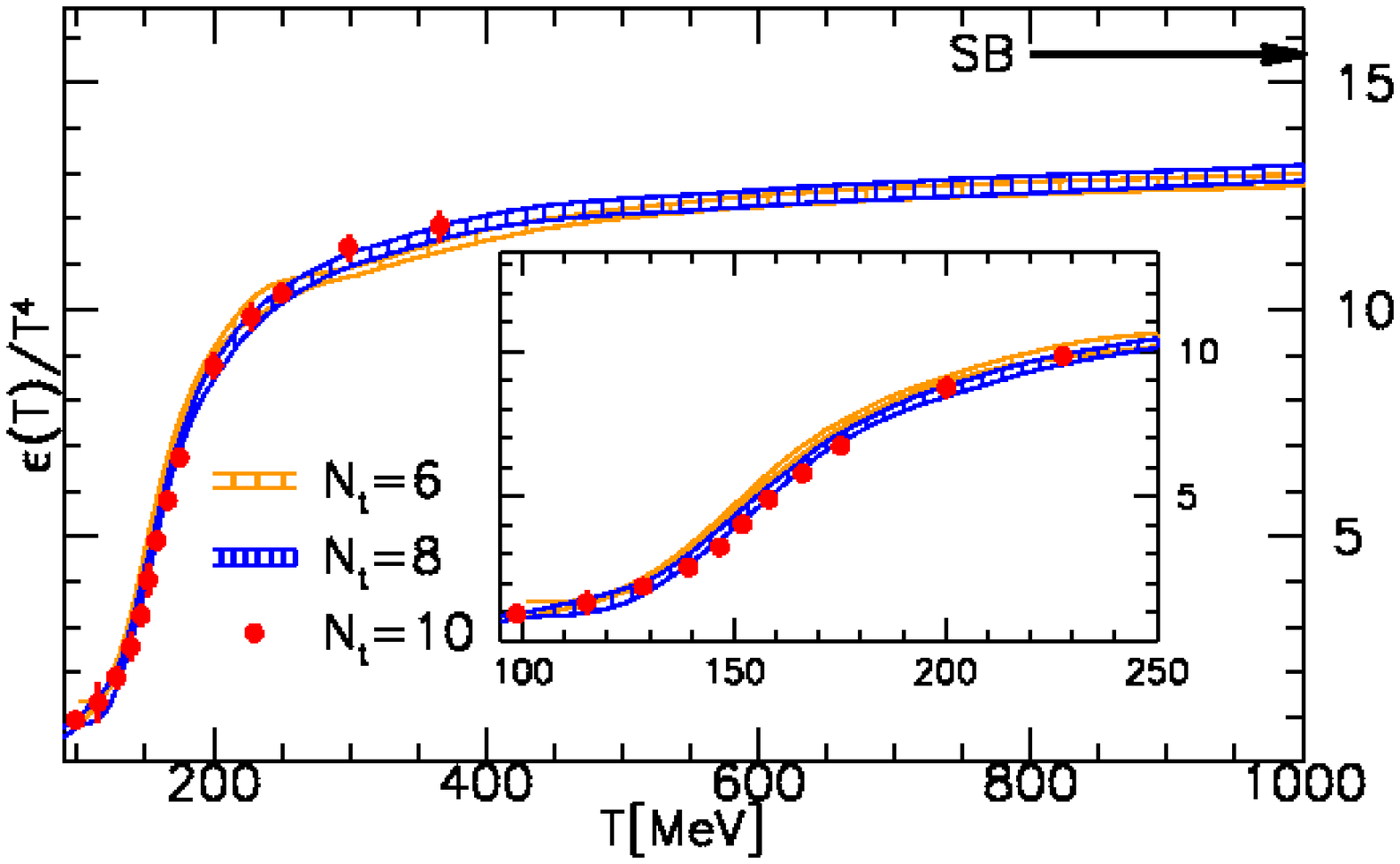}\hspace*{1.2cm}
\includegraphics[width=0.47\textwidth]{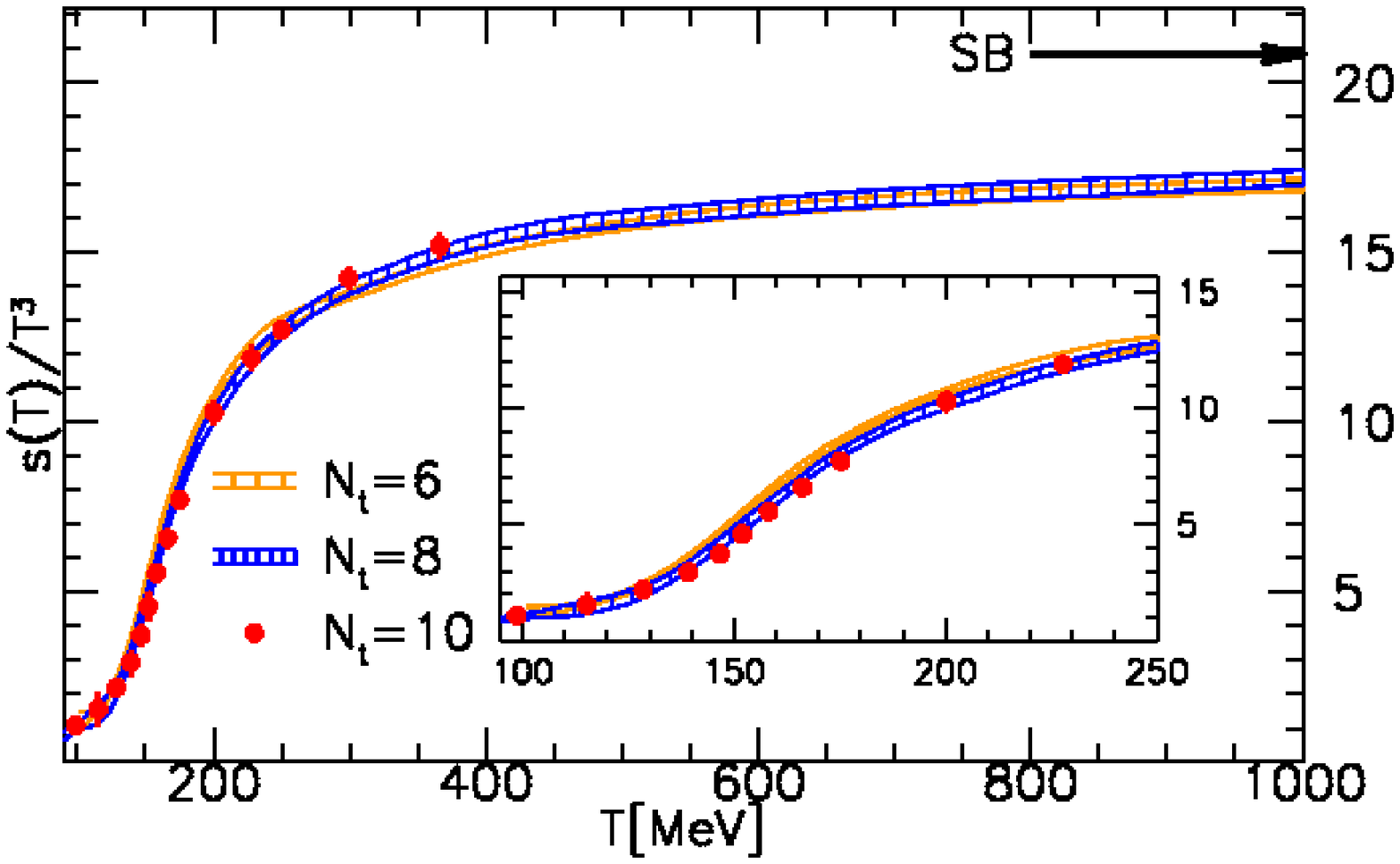}
\caption[]{Left: Equation of state as a function of temperature on $N_\tau=6,8,10,12$.
Upper Left: Trace anomaly. Upper Right: Pressure. Lower Left: Energy density. 
Lower Right: Entropy density.
From \cite{bmw_eos2}.}
\label{fig:eos1_bmw}
\end{figure}
\begin{figure}[th]
\includegraphics[width=0.5\textwidth]{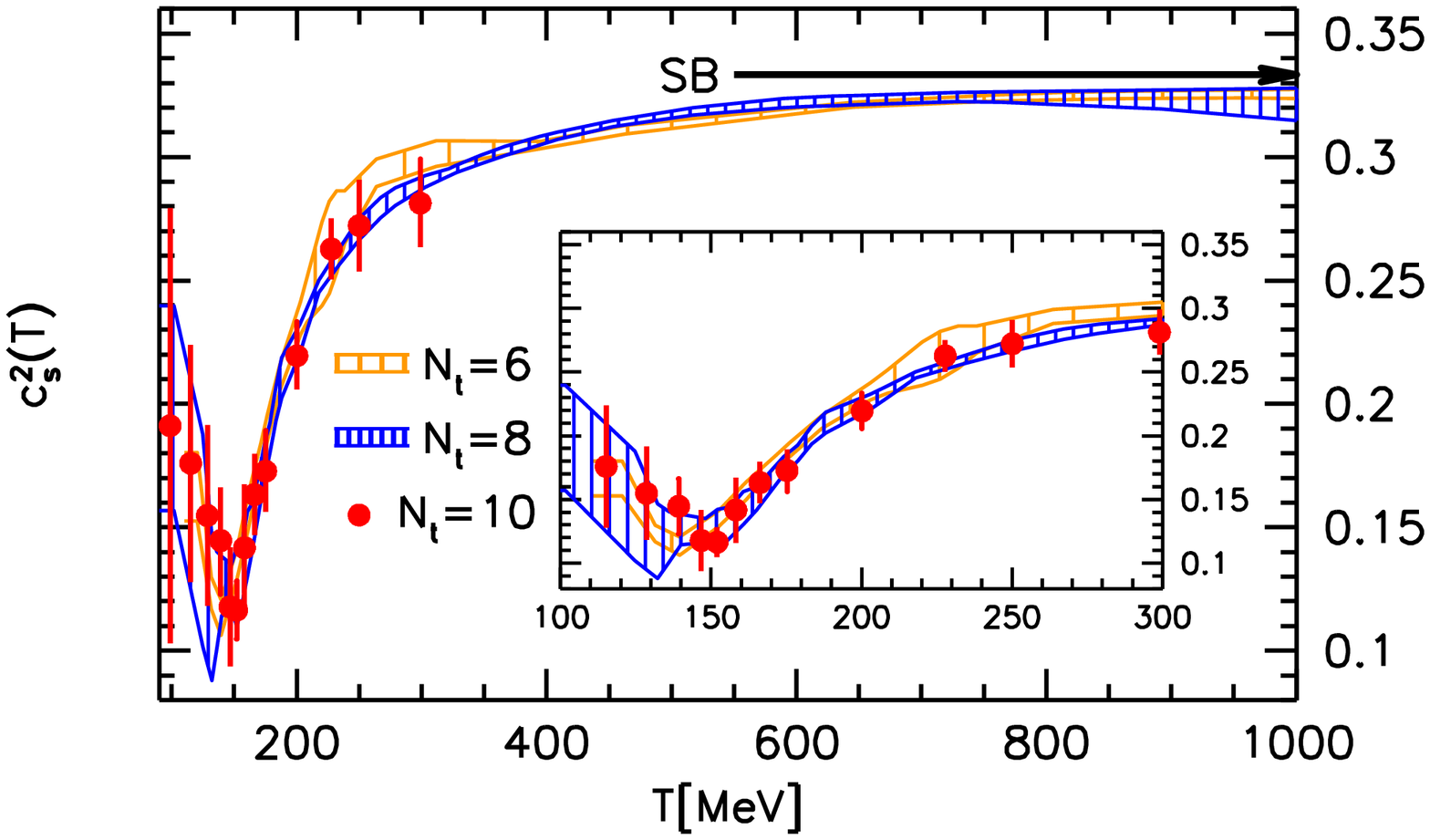}
\includegraphics[width=0.5\textwidth]{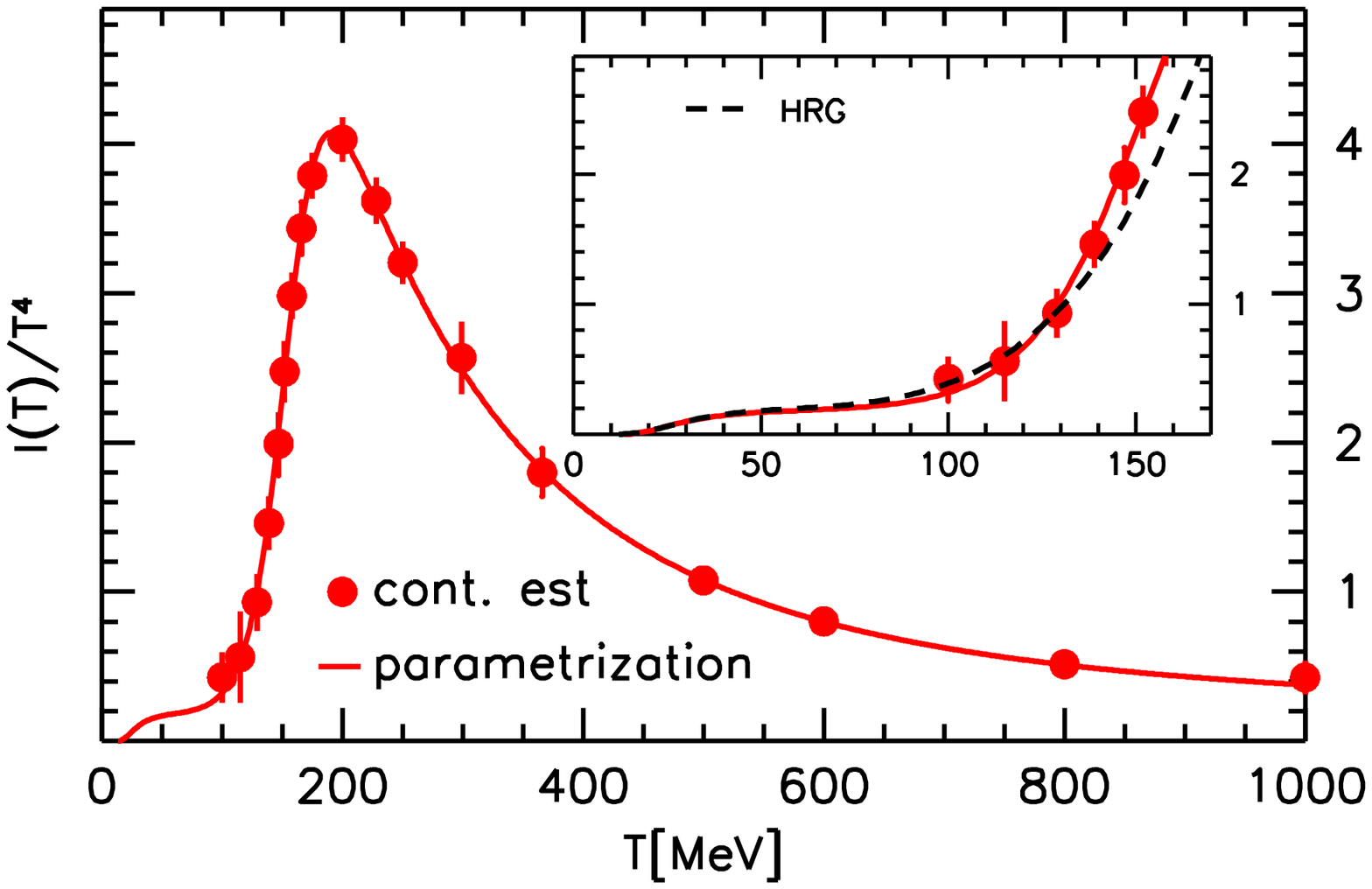}
\vspace*{-4cm}
\caption[]{Left: Speed of sound on $N_\tau=6,8,10$. Right: Trace anomaly based 
on the average of $N_\tau=6,8$ data as ``continuum estimate''. Also shown is the parametrisation 
\eq(\ref{eq:par}). 
From \cite{bmw_eos2}.}
\label{fig:eos2_bmw}
\end{figure}
\begin{table}[th]
\begin{center}
\begin{tabular}{|c||c|c|c|c|c|c|c|c|}
\hline
$N_f$ & $h_0$ & $h_1$ & $h_2$ & $f_0$ & $f_1$ & $f_2$ & $g_1$ & $g_2$ \\
\hline
$2+1$ & \multirow{2}{*}{0.1396} & \multirow{2}{*}{-0.1800} & \multirow{2}{*}{0.0350} & 2.76 & 6.79 & -5.29 & -0.47 & 1.04\\
\cline{1-1}\cline{5-9}
$2+1+1$ & & & & 5.59 & 7.34 & -5.60 & 1.42 & 0.50\\
\hline
\end{tabular}
\end{center}
\caption{\label{tab:par}
Parameters of the function in Equation (\ref{eq:par}) describing the trace anomaly in the $N_f=2+1$
and in the $N_f=2+1+1$ flavour cases.
}
\end{table}

Note that the results presented in the plots do not yet constitute a continuum extrapolation over the entire 
temperature range, which are however provided for $T=132,167,223$ MeV.
Nevertheless, on the grounds that no systematic shift with lattice spacing is visible, the authors provide an estimate for the continuum result by averaging the two finest lattices $N_\tau=8,10$, with the low temperature discrepancy from the hadron resonance gas
result quoted as a systematic error over the whole range. The resulting trace anomaly is shown in 
\fig\ref{fig:eos2_bmw} (right), together with a fit function which provides a convenient closed-form parametrisation over the entire temperature range,
\be
\label{eq:par}
\frac{I(T)}{T^4}= 
\exp(-h_1/t - h_2/t^2)\cdot \left( h_0 + \frac{f_0\cdot [\tanh(f_1\cdot t+f_2)+1]}{1+g_1\cdot t+g_2\cdot t^2} \right)\;.
\ee
Here, $t=T/(200 \text{MeV})$ and the parameter values describing the simulation data are given in Table \ref{tab:par}.

Since the publication of \cite{bmw_eos2}, further data were collected on $N_\tau=12$ lattices, working
towards a true continuum extrapolation, with preliminary results presented in \cite{bmw_charm}.
The data points for all $N_\tau$ were interpolated by standard cubic splines and then extrapolated
to the continuum using a $\sim 1/N_\tau^2$ fit ansatz. The results are shown 
in \fig\ref{fig:bmw_charm} (left).
Note that tree-level lattice corrections were subtracted from the plotted data, but
not from the data entering the continuum extrapolation. Excellent agreement with the hadron resonance
gas prediction using physical masses is obtained.
\begin{figure}[t]
\centerline{
\includegraphics[width=0.5\textwidth]{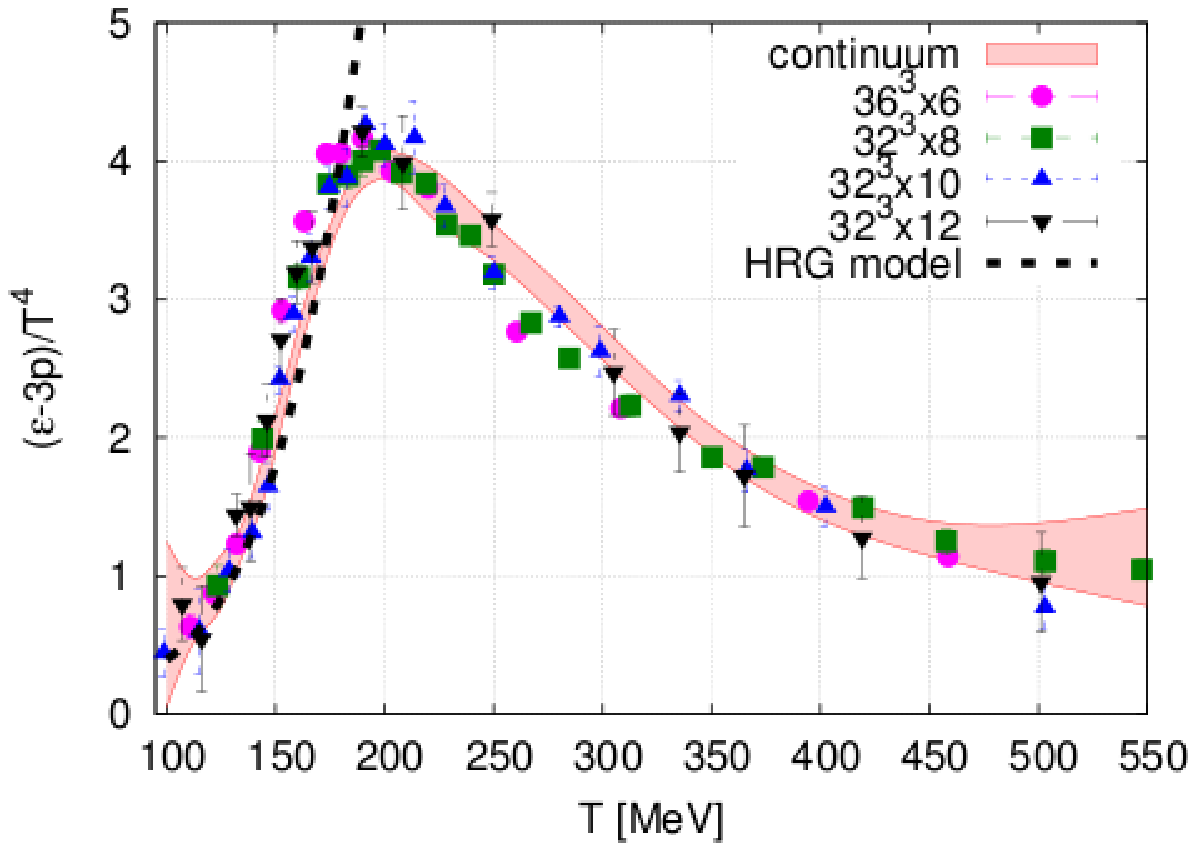}
\includegraphics[width=0.55\textwidth]{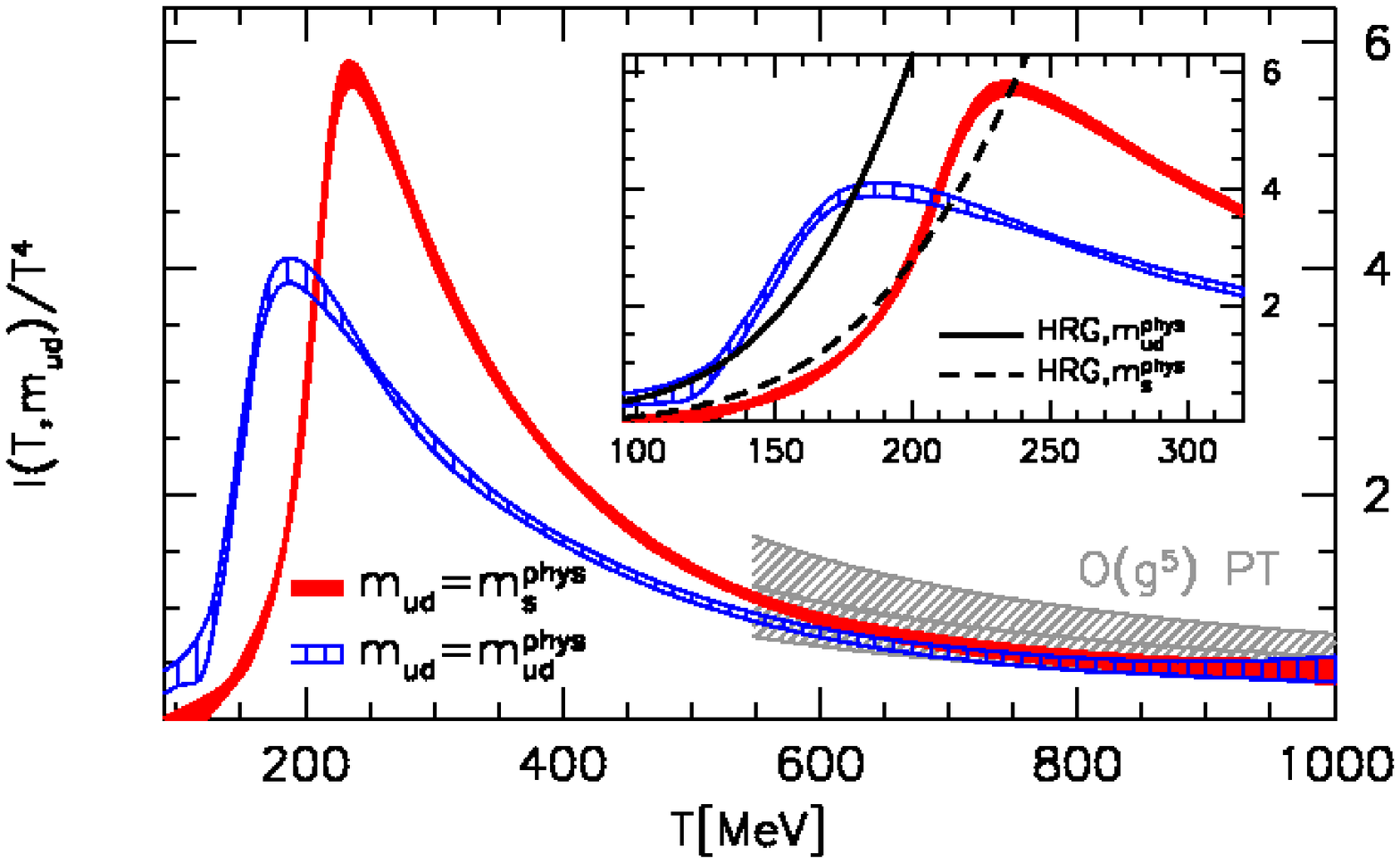}
}
\caption[]{Left: Attempt for a continuum extrapolation based on four lattice spacings. The plotted data points are tree-level improved, but not the data entering the continuum extrapolation. 
From \cite{bmw_charm}. Right: Mass dependence of the trace anomaly. From \cite{bmw_eos2}.}
\label{fig:bmw_charm}
\end{figure}

Finally, for purposes of comparing with simulations at different parameter values also the effect of quark 
mass has been studied on an $N_\tau=8$ lattice. This is shown in \fig\ref{fig:bmw_charm} (right),
where the results for the physical parameter values for all three quarks are compared with the
mass-degenerate $N_f=3$ case, where $m_{u,d}=m_s^\text{phys}$. This corresponds to 
a rather heavy $m_\pi\sim 720$ MeV. Qualitatively, similar changes as in \fig\ref{fig:eos2_hot} are 
observed, only much more pronounced as the mass difference is larger. The peak shifts to lower temperatures and decreases in height with shrinking light quark masses. Note that for large temperatures
the quark mass dependence disappears. This is consistent with the fact that confinement is lost while
the leading order screening masses in the corresponding quantum number channels are $\sim 2\pi T$,
against which the bare quark masses are negligible. This is also why it makes sense to compare
with massless perturbation theory in the plot.

\subsection{Towards inclusion of the charm quark}\label{sec:charm}

Several years ago, based on next-to-leading order 
perturbative investigations it has been suggested that the equation of state gets modified by
thermalised charm quarks at temperatures starting at $T\gsim 2T_c$ \cite{ls}, 
as depicted in \fig\ref{fig:oldcharm} (left). If thermalisation happens sufficiently fast, this would affect
heavy ion collisions at LHC, where such temperatures will be reached, and certainly the equation of state
of the Standard Model in the early universe \cite{hp}. The behaviour predicted in perturbation theory 
was confirmed by initial lattice studies where the charm was treated in the quenched approximation \cite{lev,bmw_eos2}, \fig\ref{fig:oldcharm} (right). One would imagine this to be a good 
approximation, since quantum fluctuations of the charm (quark loops) are suppressed by its mass. 
On the other hand, the same argument would suggest a small effect on the physics, so the question
merits a closer look.
\begin{figure}[t]
\includegraphics[width=0.39\textwidth]{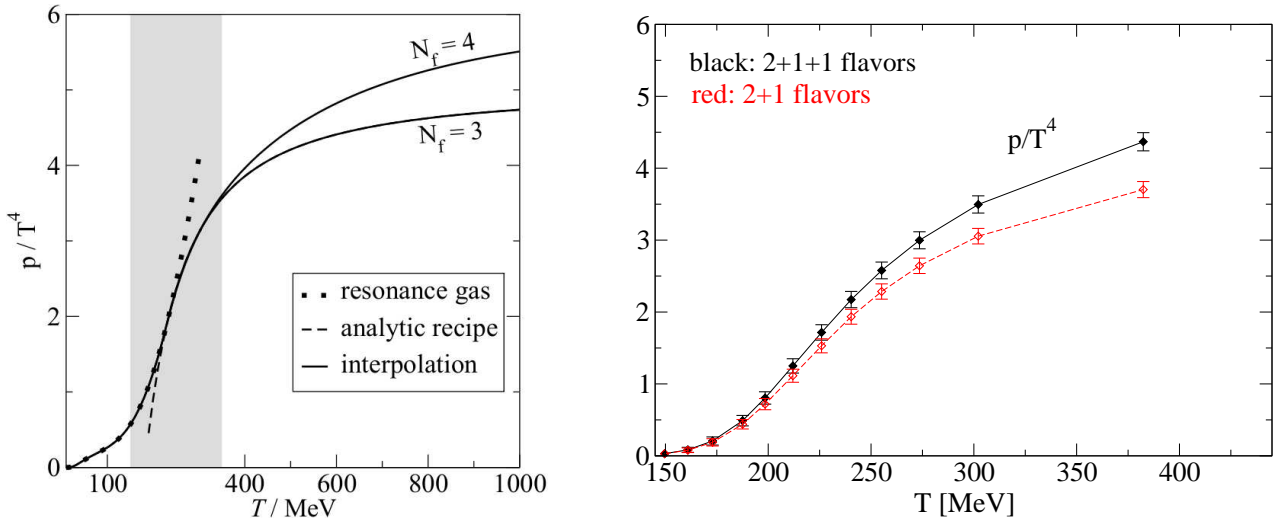}\hspace*{1cm}
\includegraphics[width=0.46\textwidth]{plots/p_charm}
\caption[]{Left: Perturbative calculation of the pressure matched to the resonance gas for three and four flavours of quarks. From \cite{ls}. Right: Lattice results for $N_f=2+1+1$ with charm in the quenched approximation  
compared to $N_f=2+1$ (dynamical) \cite{lev}.
From \cite{lev1}.
}
\label{fig:oldcharm}
\end{figure}
\begin{figure}[h!!]
\includegraphics[width=0.48\textwidth]{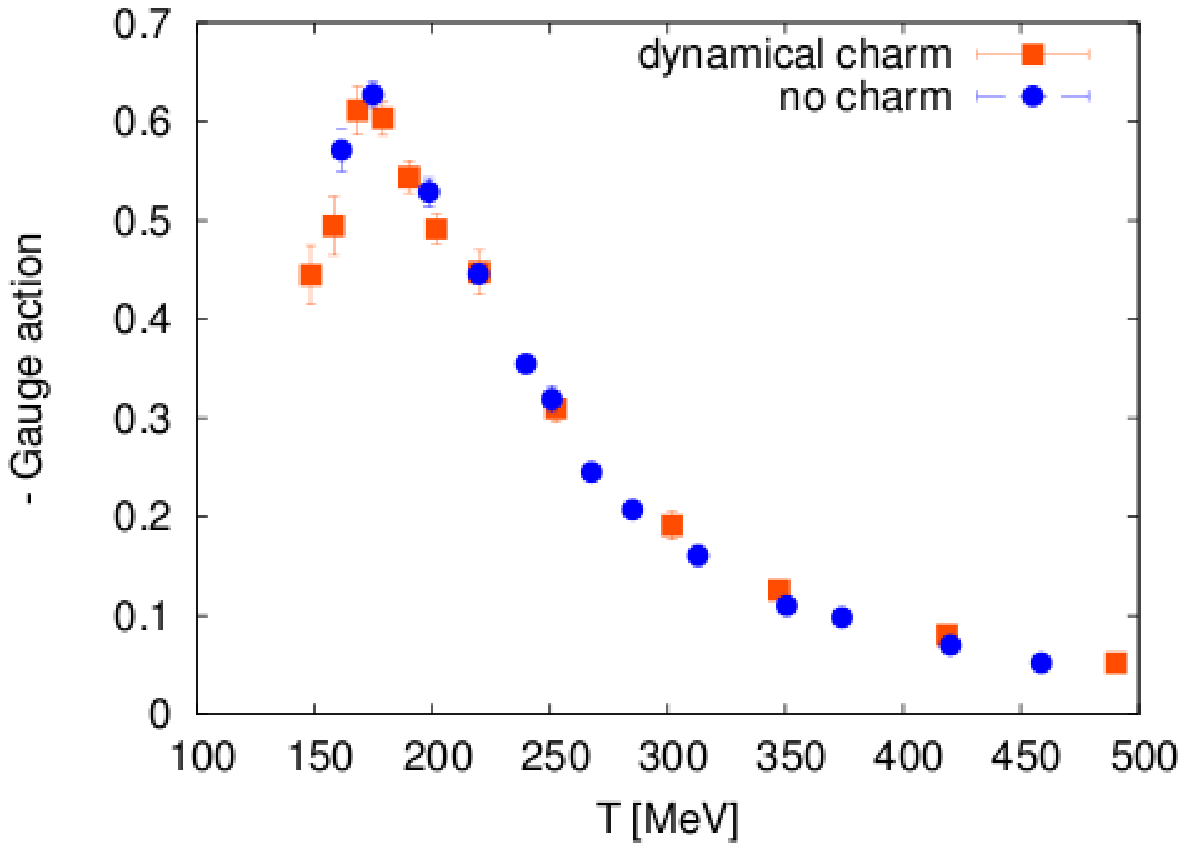}
\includegraphics[width=0.5\textwidth]{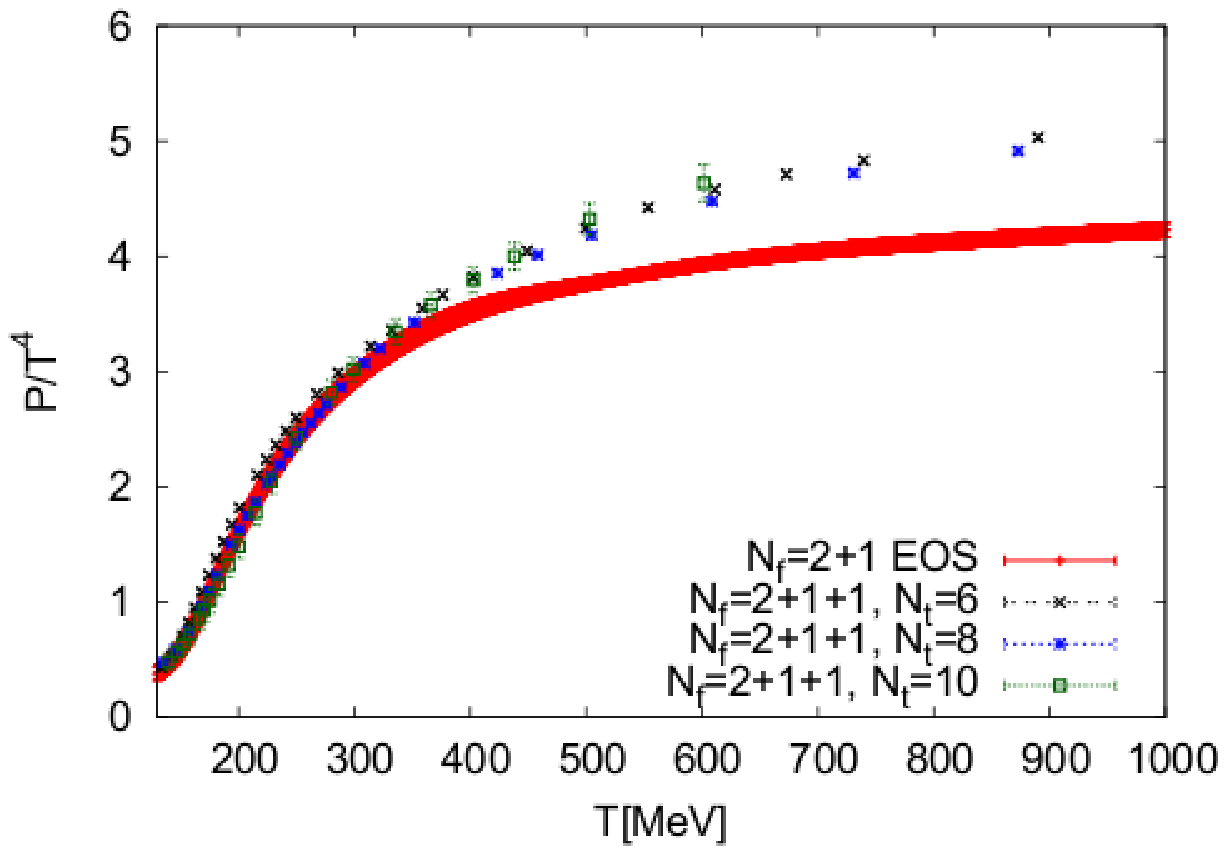}
\caption[]{Left: Gauge action used in the integral technique from the $N_f=2+1$ data set, as in a partially quenched calculation, 
compared to the preliminary results for $N_f=2+1+1$.
Right: $N_f=2+1+1$. The data points are tree-level corrected and compared to the $N_f=2+1$ continuum estimate.  
From \cite{bmw_charm}.
}
\label{fig:newcharm}
\end{figure}

A detailed investigation, comparing quenched and dynamical simulations, was recently started in \cite{bmw_charm}.
According to the formula used for the integral method, \eq(\ref{freediv}), 
the charm quark may
modify the expectation values of the gauge action and the chiral condensate. Its influence on the 
beta-function is known to be small. 
\fig\ref{fig:newcharm} (left) compares the contribution of the gauge action with and without a dynamical
charm quark, with practically no difference visible, and the same finding for the quark contribution 
to the pressure. The effect of the charm quark on the pressure then enters likely through the integration, i.e.~solely through the modification of the line of constant physics.
Nevertheless, in \cite{bmw_charm} the full dynamical calculation was performed, with its result in 
\fig\ref{fig:newcharm} (right).

The result shows clear differences to the quenched calculation discussed above. The main observation
is that the effect of the charm quark is delayed to $T\sim 300$ MeV now, similar to what is seen in 
the perturbative estimate. This is a very interesting observation, calling for a closer investigation. 
In particular, one would now like to see an explicit evaluation of the effect of the charm on the lines
of constant physics. Another interesting question concerns cut-off effects. It is well known that 
at zero temperature heavy quarks are beset by large discretisation errors when their 
mass gets too close to the cut-off. By contrast, the data in \fig\ref{fig:newcharm} (right) appear
to show no significant cut-off effects even at low temperatures, where the lattices are coarsest.
This is somewhat surprising, since a $N_\tau=6$ lattice at low temperatures corresponds to 
$a\gsim 0.2$ fm. At high temperature the lattice spacing is finer and the situation should be 
better. It will be interesting to see future investigations of these questions.
In any case, the conclusion emerges that for temperatures $T\gsim 400$ MeV indeed a thermalised
charm quark significantly affects the equation of state of QCD and the Standard Model.

\section{Numerical results with Wilson fermions}\label{sec:reswil}

The Wilson discretisation of the fermionic action gets rid of the doublers dynamically, by providing them
with masses $\sim a^{-1}$ which lead to decoupling as the continuum is approached. There is no reduction
of spin degrees of freedom on the level of the partition function, as for staggered fermions, and hence
the simulations require both more memory and time. To a large part this explains why
there are less Wilson results at finite temperature. 
For a long time, a study of the equation of state existed only for two flavours on relatively coarse lattices
\cite{wilson}. Recently some efforts were made to study the quark hadron transition on finer lattices up to 
$N_\tau=12$ and lighter quark masses, down to $m_\pi\sim 400$ MeV \cite{born,twist,twist1}, but these
are again for $N_f=2$ and so far without results 
for the equation of state.
The first thermodynamics studies with $N_f=2+1$ flavours of Wilson fermions appeared only very recently \cite{whot,bmw_w1,bmw_w2}. 

In \cite{whot} the trace anomaly was calculated based on the equation,
\begin{eqnarray}
\frac{\epsilon-3p}{T^4}&=& 
\frac{N_t^3}{N_s^3}
\left(
{a\frac{d\beta}{da}}
\left\langle
\frac{\partial S}{\partial\beta}
\right\rangle_{\rm \! sub}
+
{a\frac{d\kappa_{ud}}{da}}
\left\langle
\frac{\partial S}{\partial \kappa_{ud}}
\right\rangle_{\rm \! sub}
+{a\frac{d\kappa_s}{da}}
\left\langle
\frac{\partial S}{\partial \kappa_s}
\right\rangle_{\rm \! sub}
\right) 
\label{eq:tranom}
\end{eqnarray}
The authors work with the fixed-scale approach, cf.~Sec.~\ref{sec:fixed}, using an RG-improved 
gauge action, Sec.~\ref{sec:impr_gauge}, and non-perturbatively clover-improved Wilson fermions,
Sec.~\ref{sec:clover}. 
For the finite temperature configurations, lattice $32^3\times N_\tau$ 
with $N_\tau=4,6,8,\ldots 16$ are used.
For the zero temperature subtraction, they use data from the CP-PACS+JLQCD
collaboration \cite{cppacs}. These correspond to a physical strange quark mass, defined by 
$m_{\eta_{\bar{s}s}}/m_\Phi\approx 0.74$ and light quark masses defined by $m_\pi/m_\rho=0.63$, 
with pions of $m_\pi\sim 620$ MeV. 
On the $28^3\times 56$ vacuum lattices, the lattice spacing was determined
as $r_0/a=7.06(3)$ \cite{whot_scale}, or $a\approx 0.07$ fm.

A main ingredient for  the study of the equation of state is a good knowledge of the beta-functions along
the line of constant physics. In a direct evaluation one would measure the values of 
$am_\rho, m_\pi/m_\rho, m_{\eta_{\bar{s}s}}/m_\phi$ and fit those as functions of the three lattice couplings
$\beta,\kappa_{ud},\kappa_s$. 
However, the quark mass beta functions, $a d\kappa_i/da$, are much smaller than
the gauge part and the errors in the matrix inversion are often too large for an accurate determination.
The authors of \cite{whot} therefore directly fit the lattice couplings $\beta,\kappa_{ud},\kappa_s$ 
as a function of the meson masses using a third order polynomial ansatz in 
$am_\rho,(m_\pi/m_\rho), (m_{\eta_{\bar{s}s}}/m_\phi)$.
As an example, the result for the lattice gauge coupling is shown in \fig\ref{fig:beta_whot} (left), with similar
plots for the other lattice couplings \cite{whot}. These fits can then be numerically converted to the required beta functions. 
\begin{figure}[t]
\includegraphics[width=0.45\textwidth]{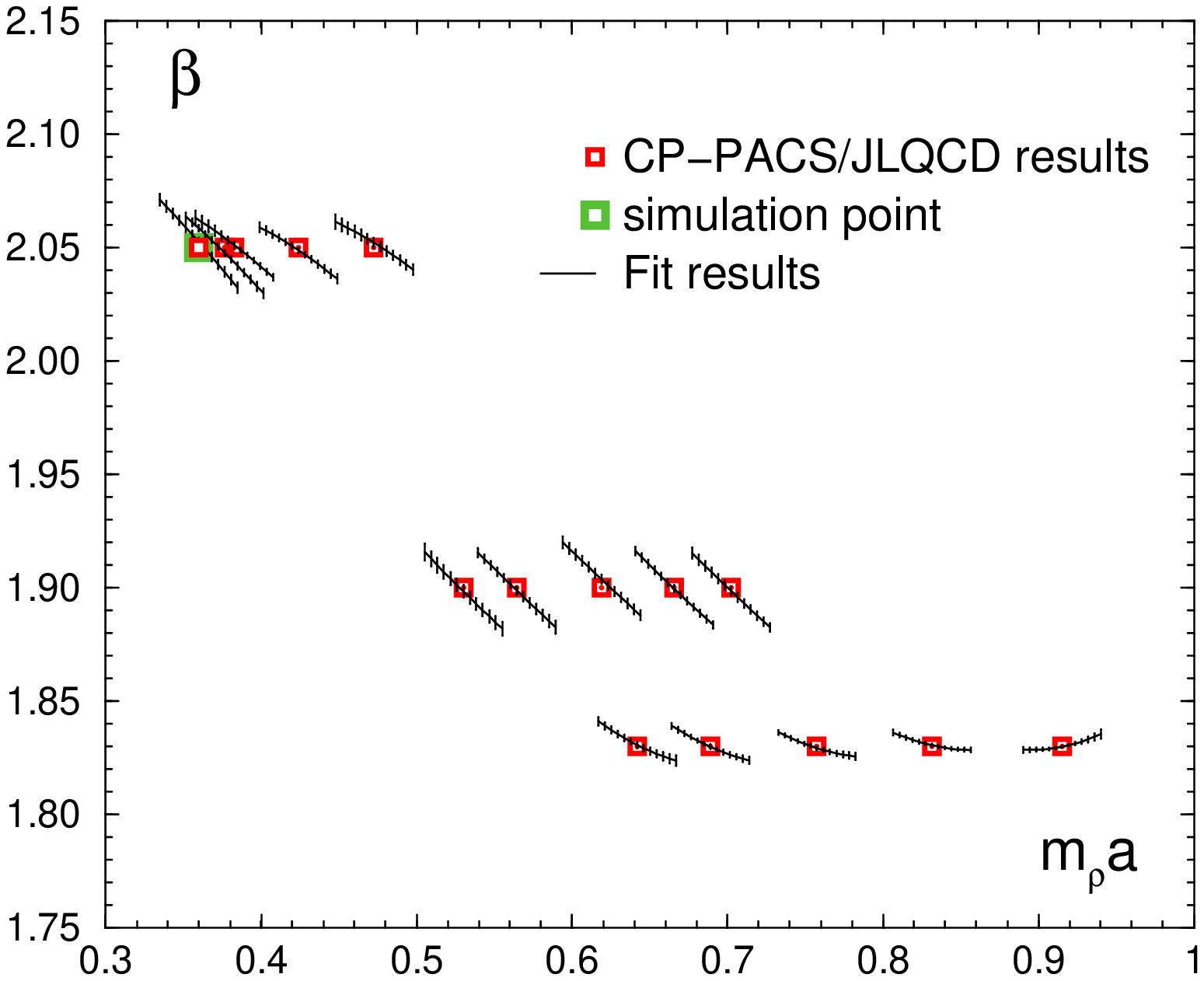}\hspace*{1cm}
\includegraphics[width=0.45\textwidth]{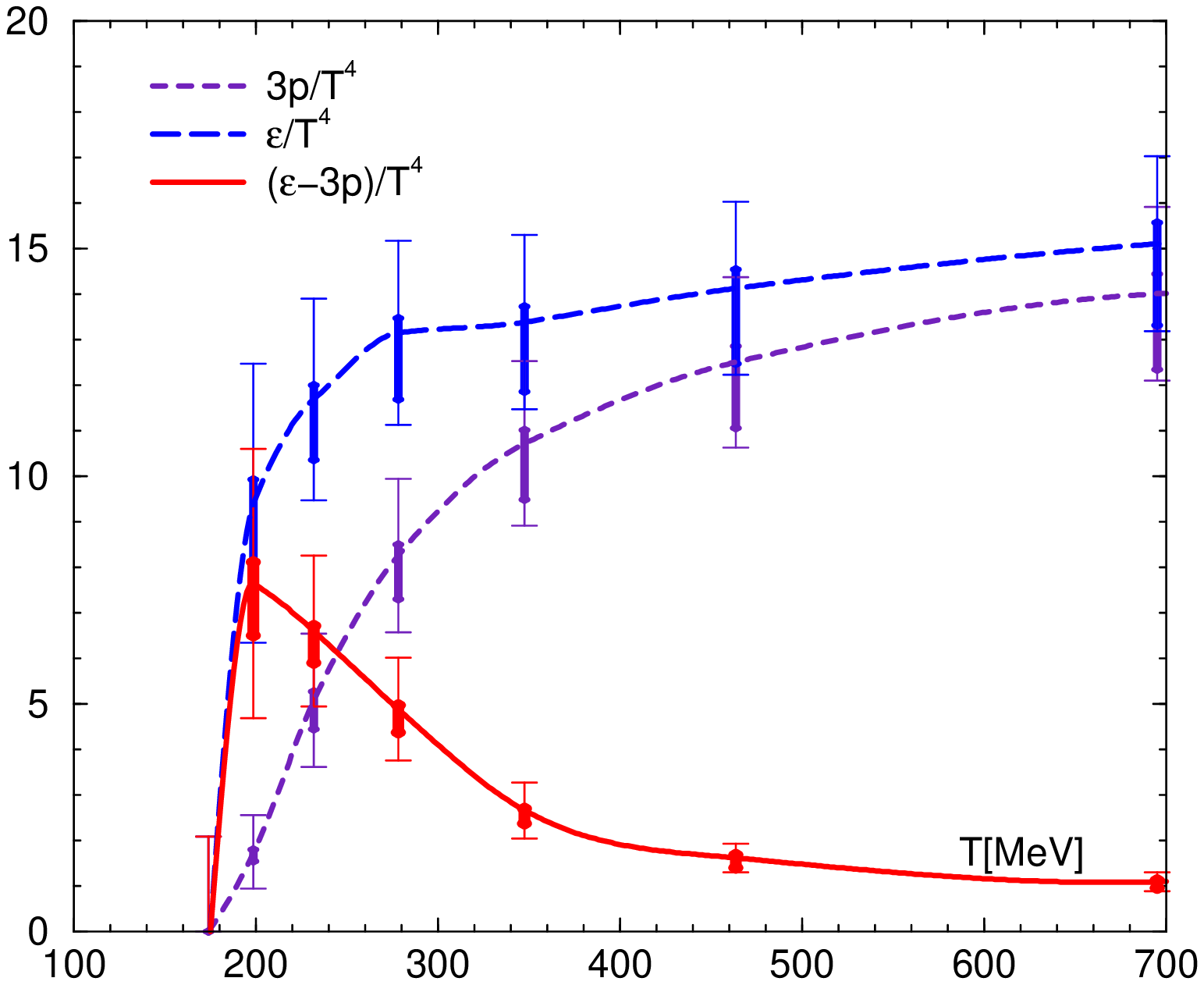}
\caption[]{Left: Global fit for $\beta$ 
    as a function of $m_\rho a$. Square symbols show coupling 
    parameters in the CP-PACS/JLQCD study. Solid lines show global
    fit results at each simulation point with corresponding 
    $m_\rho/m_\pi$ and $m_{\eta_{ss}}/m_\phi$. 
Right: Trace anomaly, energy density
    and pressure for QCD with clover improved Wilson fermions, $m_\pi \sim 620$ MeV. 
    Thin and thick vertical bars represent statistic and systematic errors, respectively.
    The curves are drawn by the Akima spline interpolation.    
From \cite{whot}.}
\label{fig:beta_whot}
\end{figure}

The resulting trace anomaly is depicted in \fig\ref{fig:beta_whot} (right). The thick error
bars give an estimate of the systematic error due to variations of setting the scale by different quantities.
Akima splines are used for interpolation and the numerical integration of the trace anomaly to yield
the pressure. Starting point for the temperature integration is $N_\tau=16$, for which $T$ is low 
enough for the trace anomaly to be statistically compatible with zero. The authors also find that the variation associated with using different interpolation methods for the integration is smaller than the error bars. 
This is an interesting observation since, in the fixed scale method, simulation points for
different temperatures can only be varied in discrete steps. The current spacing of temperature values
could be halved by inserting odd values for $N_\tau$. 
In addition, the temperature resolution can of course be improved
by simulating on finer lattices in the future.

There are also preliminary results for the equation of state based on the twisted mass formulation of 
$N_f=2$ Wilson fermions. Using also tree-level Symanzik improvement, the relevant formula for the trace anomaly is
\ba
    \frac{\epsilon - 3 p}{T^4} 
    &=& \left ( a \frac{d \beta}{d a} \right ) N_\tau^4 \Bigg (\frac{ c_0}{3} \vevsub{\Real \Trace U_{\mu\nu}} + 
       \frac{c_1}{6} \vevsub{\Real \Trace U_{\mu\nu}^{1\times 2} }\nn\\
    &&+ \frac{\partial \kappa_c}{\partial \beta} \vevsub{\bar \psi H[U] \psi} 
     - \left ( 2 a \mu \frac{\partial{\kappa_c}}{\partial{\beta}} + 2 \kappa_c
     \frac{\partial (a \mu)}{\partial \beta} \right ) \vevsub{\bar \psi i \gamma_5 \tau^3 \psi} \Bigg ) \;.
\ea
Note that $\kappa$ is tuned to $\kappa_c(\beta)$ to ensure maximal twist and the quark mass
is tuned with the twist parameter $\mu$.

Preliminary numerical results are based on simulations reported in \cite{twist1} as well as so far unpublished data, running on $32^3\times N_\tau$ lattices with $N_\tau=6,8,10,12$, where
the standard integral method is used. The pion mass is $m_\pi\sim 400$ MeV. The required
beta function is obtained from 
interpolations of the chirally extrapolated $r_0$ parameter
provided by the $T=0$ simulations of the ETMC collaboration \cite{etmc}. This is shown in 
\fig\ref{fig:twist} (left). Similarly, the vacuum expectation values needed for renormalisation
are obtained from interpolations of ETMC data.
The resulting trace anomaly is shown in \fig\ref{fig:twist} (right).
While it is too early to draw conclusions at this stage, scaling appears to set in at $N_\tau\gsim 8$
which would permit a continuum limit once enough statistics has been collected.
\begin{figure}[t]
\includegraphics[width=0.43\textwidth]{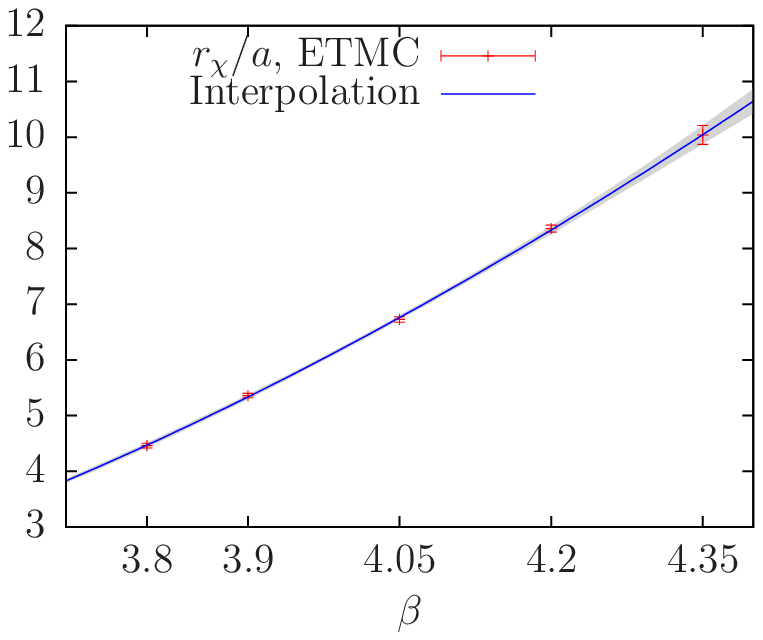}\hspace*{1cm}
\includegraphics[width=0.45\textwidth]{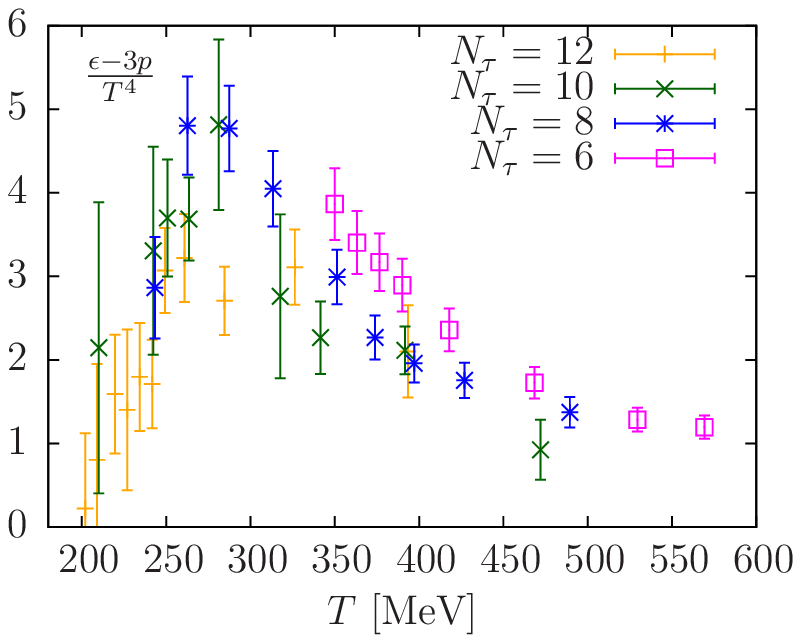}
\caption[]{Left: Interpolation of the chirally extrapolated scale parameter $r_0$ from the ETMC
collaboration \cite{etmc}.
Right: Trace anomaly for $N_f=2$ twisted mass fermions with $m_\pi\sim 400$ MeV.}
\label{fig:twist}
\end{figure}

\section{Comparing results from different discretisations}\label{sec:comp}

Once a controlled continuum extrapolation has been performed, all lattice actions must, for fixed
physical parameters and observables, give the same answers, lest a particular lattice formulation is fundamentally flawed. 
Working with different discretisation schemes therefore is a necessary and powerful check of the validity of the results. It is however natural to also compare results even before the continuum limit has been taken.  Different actions have
different cut-off effects. Notable discrepancies are therefore an indicator that the quantity under
investigation has not yet reached its continuum value in at least one of the studied schemes.

\subsection{Staggered vs. staggered fermions}\label{sec:compstag}

\begin{figure}[t]
\includegraphics[width=0.5\textwidth]{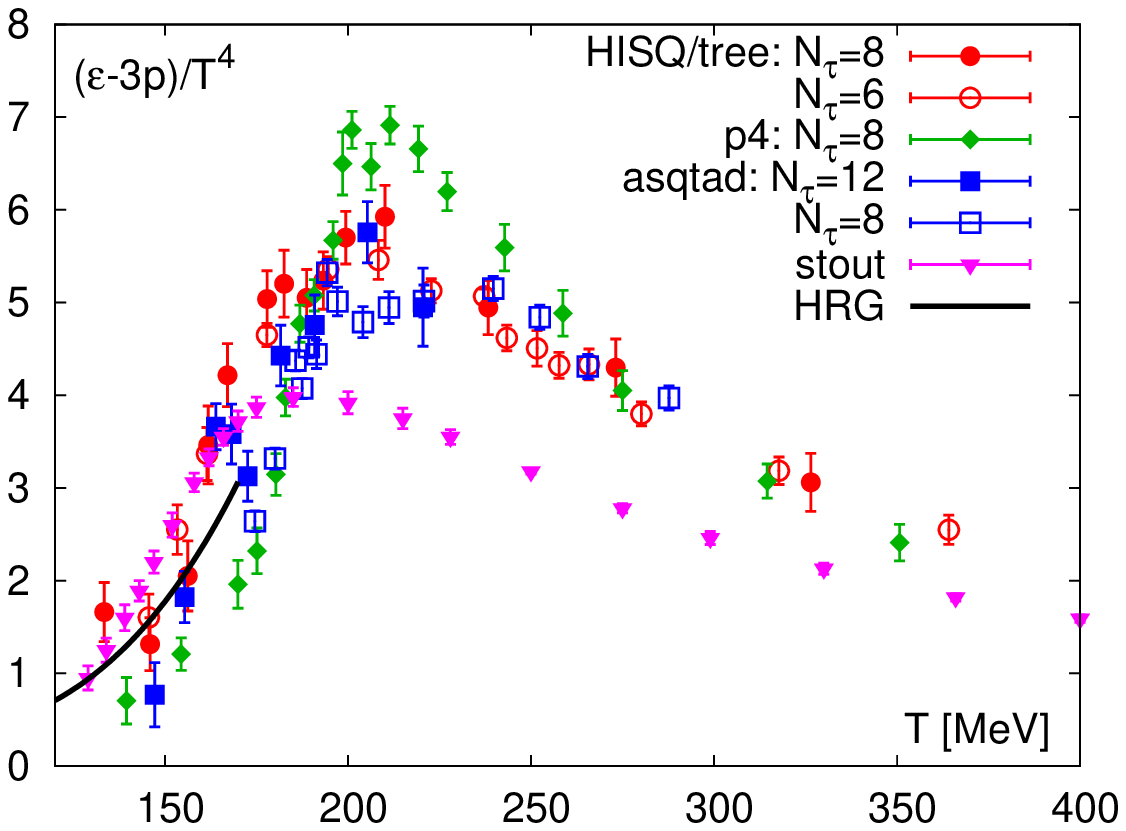}
\includegraphics[width=0.5\textwidth]{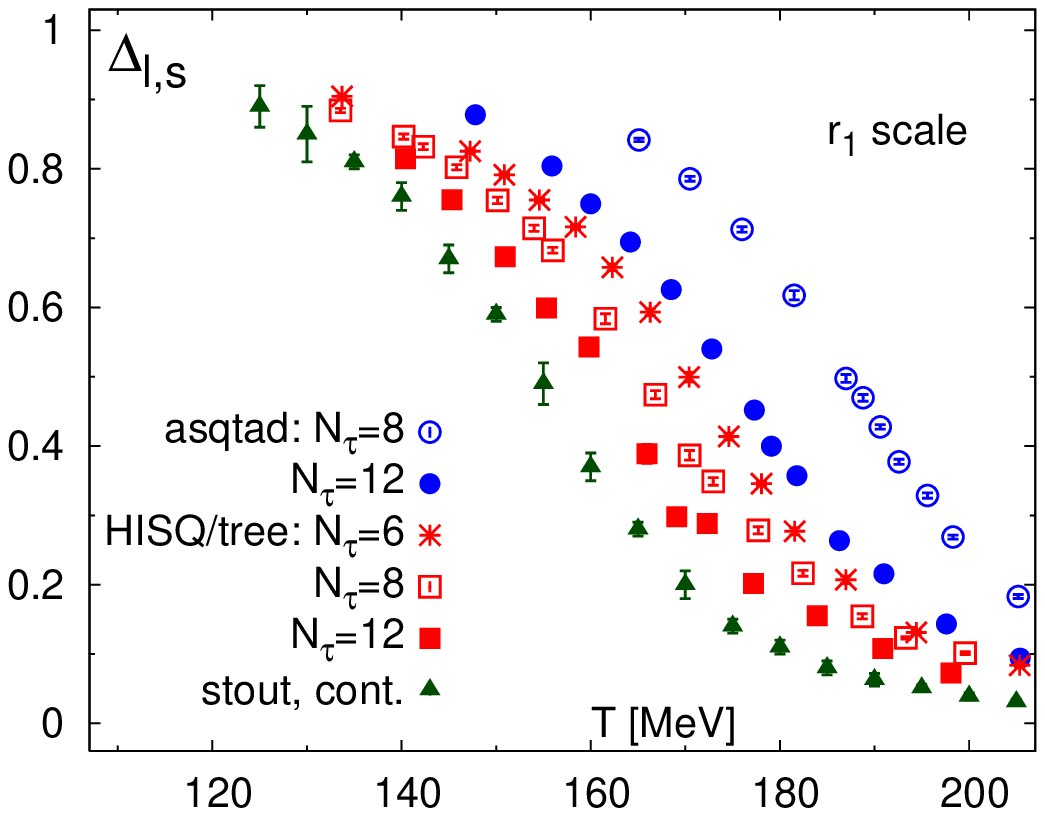}
\caption[]{Left: Trace anomaly computed by different actions on $N_\tau=8$. Data for p4fat and asqtad are from \cite{hotp4asq,cheng_09}, HISQ/tree from \cite{taste}, 
those for stout correspond to the continuum estimate based on $N_\tau=6,8,10$ in \cite{bmw_eos2}.
From \cite{taste}. Right: The subtracted chiral condensate for the asqtad and HISQ/tree
  actions \cite{hot_hisq} with $m_l=m_s/20$ compared with the continuum extrapolated
  stout action results~\cite{bmw_eos2}.  The temperature
  $T$ is converted into physical units using $r_1$. 
From \cite{hot_hisq}.}
\label{fig:eos_comp}
\end{figure}
Comparisons of the trace anomaly between different staggered fermion actions were made in \cite{bmw_eos2,taste}. Note that in pure gauge theory, $N_\tau=6,8$ with an unimproved action
proved to be fine enough for a continuum extrapolation in the $\sim 1/N_\tau^2$ scaling region, passing  later tests by improved actions, as discussed in Sec.~\ref{sec:ym}. This might have been the reason for
an overly optimistic approach to dynamical simulations. A comparison between p4, asqtad, stout and 
HISQ/tree actions from \cite{taste} 
is shown  for $N_\tau=8$ in \fig\ref{fig:eos_comp} (left). Striking differences are apparent.
Note that the pions have physical masses for the stout data, but are slightly heavier for the other 
discretisations, $m_\pi\approx 160$ MeV. However, the quark mass dependence quickly diminishes
for $T>T_c$, since there are no more hadrons and the screening masses in the same quantum number
channels are dominated by $\sim 2\pi T$, compared to which the vacuum masses of the pions are small. Hence, the quark masses should only affect the data below $T_c$.  
By contrast and 
consistent with the calculations of $T_c$ alluded to earlier, features like the rising flank and the peak position of the hotQCD results appear shifted towards 
higher temperatures. Much larger than this shift, however, is the difference
in the peak height, which at the maximum corresponds to almost $50 \%$. If it is not the difference in quark
mass, the difference must be accounted for by discretisation effects.
It is interesting to note the evolution of the different discretisations with $N_\tau$ and compare with 
the corresponding evolution of the stout data shown in \fig\ref{fig:bmw_charm}.
However, at least for the $N_\tau=6,8$ data shown, neither the HISQ/tree nor the asqtad data decrease the peak in the trace anomaly noticeably. It will be 
necessary and highly interesting to see $N_\tau=12$ data with the HISQ/tree action to clarify this 
discrepancy.

Similarly, differences are observed when looking at the chiral condensate 
and the strange quark number susceptibility. 
This is shown in \fig\ref{fig:eos_comp} (right) and \fig\ref{fig:eos_comp2} (left), respectively.
In these observables a very clear trend can be seen. The data of both, the asqtad and the 
HISQ/tree action, move significantly towards the stout results with increasing $N_\tau$, i.e.~on finer lattices.
A smaller discrepancy is observed in the renormalised Polyakov loop, \fig\ref{fig:eos_comp2} (right). Note
the gradual decrease in cut-off effects in these three observables. 
The light quark condensate $\Delta_{l,s}$ is  particularly sensitive to the chiral properties
of the action. These are dominated by the pion sector and the associated cut-off effects by taste splitting. Accordingly, the actions with the smallest taste splittings, i.e.~the
stout and HISQ/tree actions are expected to do best for this observable.
By contrast, the Polyakov loop is a pure gauge observable and couples only indirectly to the sea quark effects. It is therefore less sensitive to taste splitting. For all observables, the scaling seems to be
improved when the scale is set with $f_K$ rather than $r_1$ \cite{bmw_tc2,hot_hisq}.
\begin{figure}[t]
\includegraphics[width=0.5\textwidth]{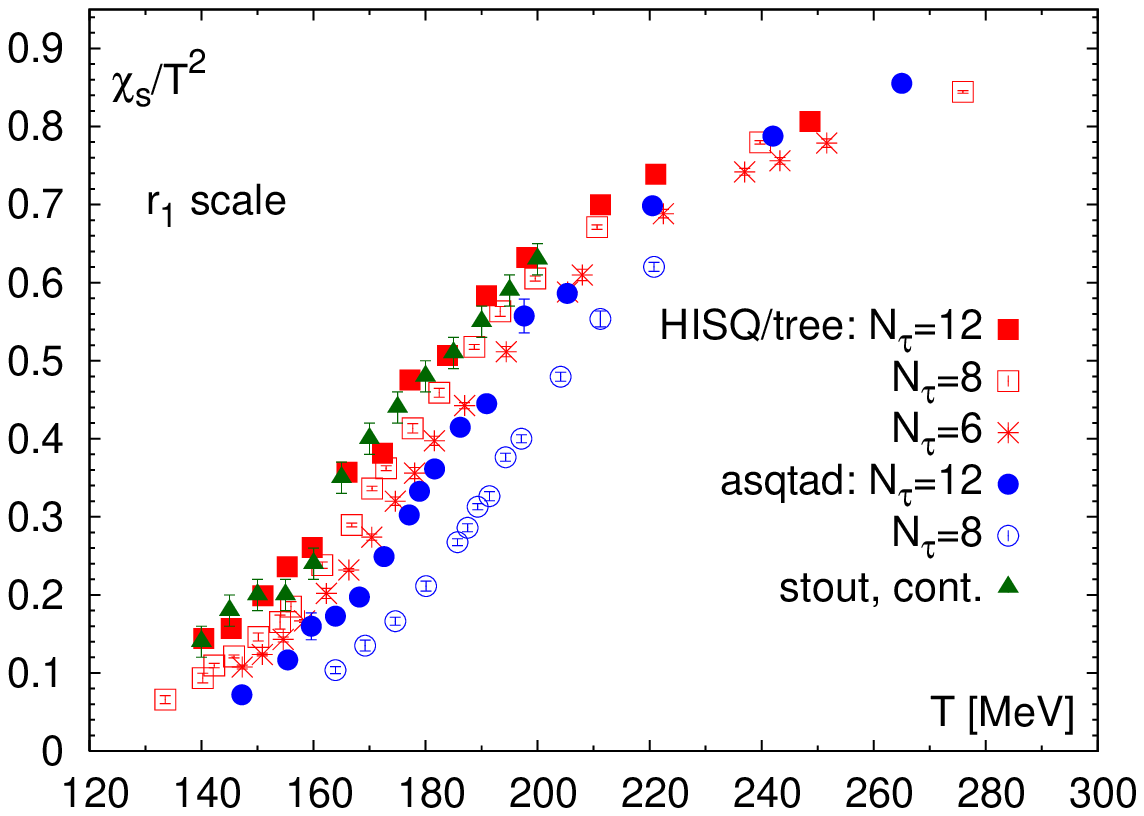}
\includegraphics[width=0.4\textwidth]{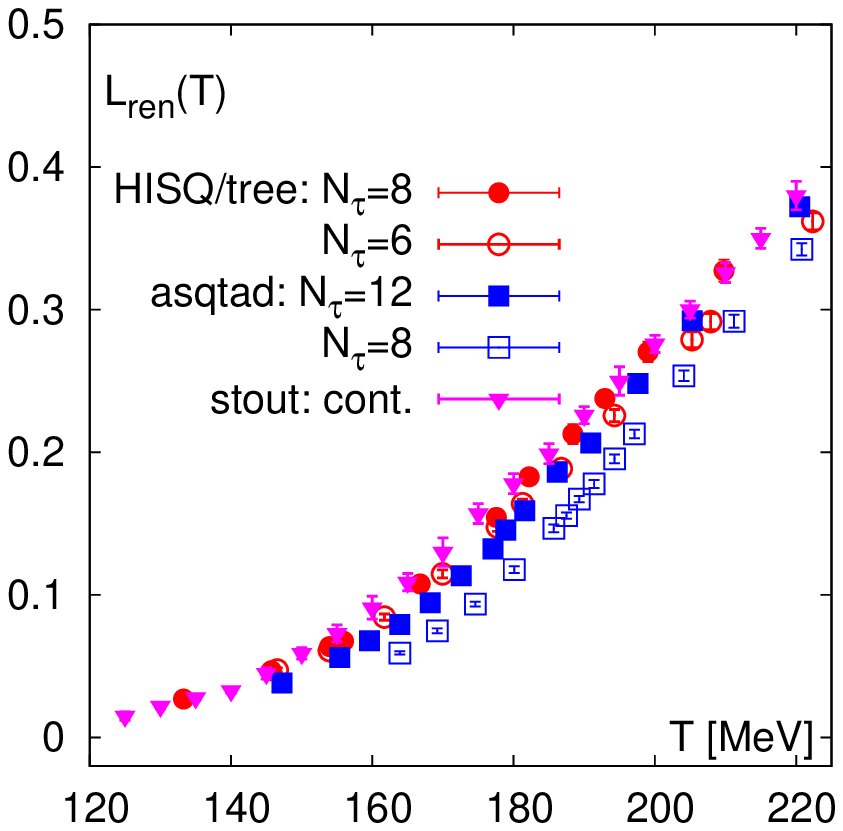}
\caption[]{Left: Strange quark number susceptibility for $m_\pi\approx 160$ MeV
with the asqtad and HISQ/tree actions compared to 
the continuum extrapolated stout results at physical masses \cite{bmw_eos2}. From \cite{hot_hisq}.
Right: Renormalised Polyakov loop for the asqtad and
  HISQ/tree actions \cite{taste} compared with the continuum extrapolated stout
  result \cite{bmw_eos2}.
From \cite{taste}.}
\label{fig:eos_comp2}
\end{figure}
\begin{figure}[t]
\includegraphics[width=0.5\textwidth]{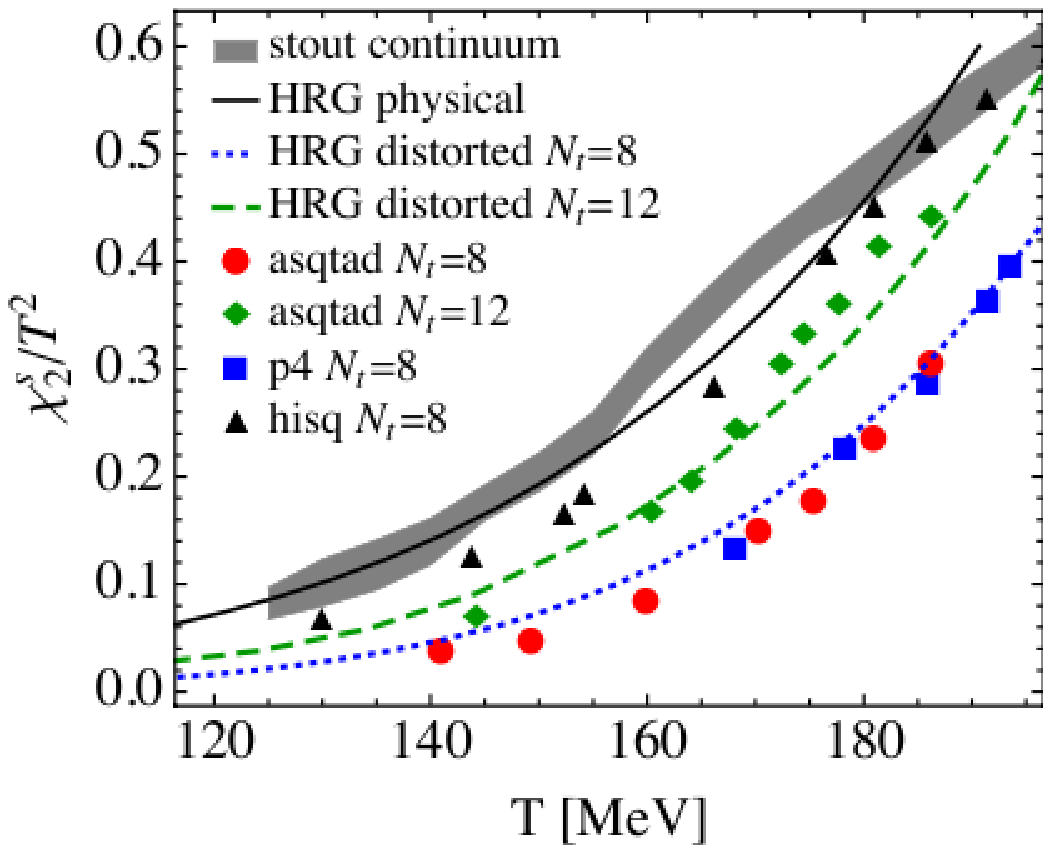}
\includegraphics[width=0.5\textwidth]{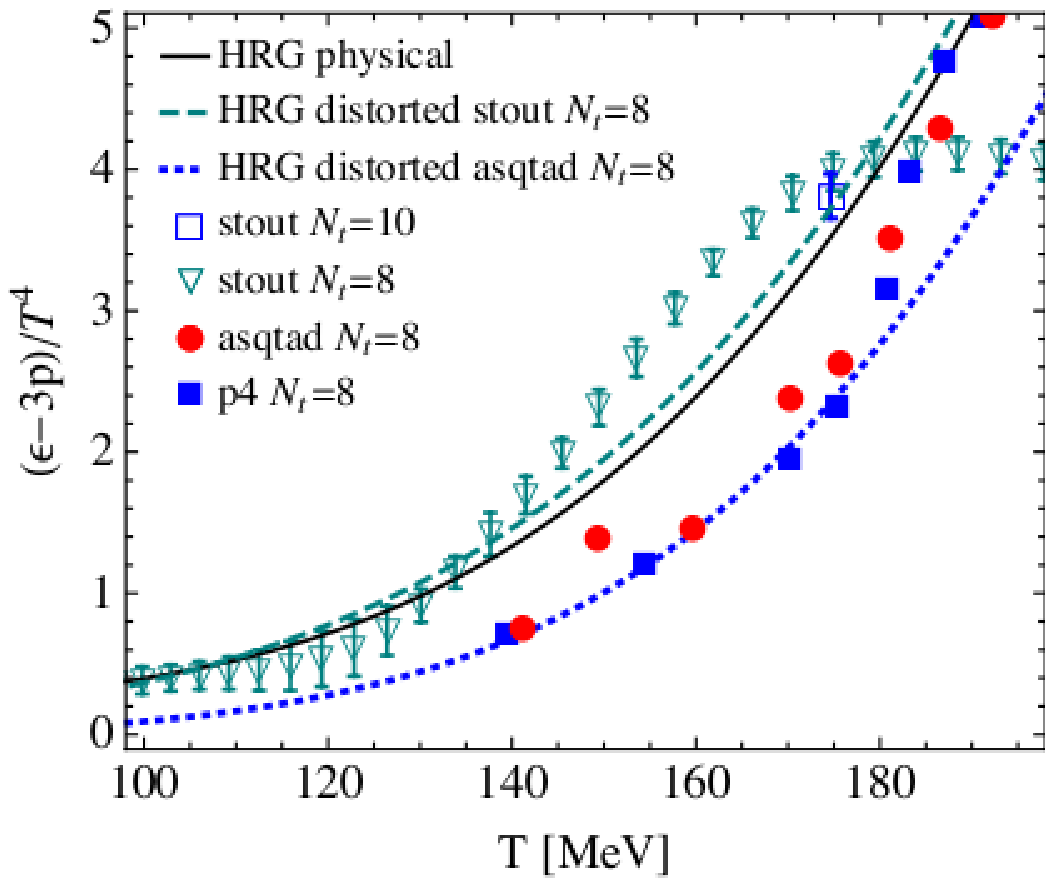}
\caption[]{Left: Strange quark number susceptibility from continuum extrapolated 
stout data \cite{bmw_tc3}  at physical masses, compared to HISQ/tree and asqtad \cite{hotp4asq,taste}.
Also shown are hadron resonance gas predictions using physical and cut-off-distorted
spectra. 
Right: Renormalised Polyakov loop for the asqtad and
  HISQ/tree actions \cite{hot_hisq} compared with the continuum extrapolated stout
  result \cite{bmw_eos2}.
From \cite{bmw_tc3}.}
\label{fig:hrg_comp}
\end{figure}

Finally, it is instructive to compare the numerical results below $T_c$ with the hadron resonance gas
prediction. In Sec.~\ref{sec:hrgsc} we have seen that the hadron resonance gas is more than just
a model description, but becomes increasingly accurate the lower the temperature, and hence the
lattice gauge coupling at fixed $N_\tau$. We have also seen how to adapt the 
hadron resonance gas to finite lattice spacings, by including the taste splittings as well as 
the hadron spectrum shifted by finite cut-off effects, Sec.~\ref{sec:taste}. Following \cite{pasi10},
the BMW collaboration performed a comparative analysis of hadron resonance gas descriptions with 
and without inclusion of cut-off effects \cite{bmw_tc3}. This is shown in \fig\ref{fig:hrg_comp}. 
Up to $T\sim T_c$, the continuum extrapolated stout data are in complete agreement with the 
hadron resonance gas using the physical continuum spectrum. On the other hand, similar agreement
is achieved for the asqtad action when appropriate cut-off effects for the hadron masses are included
in the hadron resonance gas description, corresponding to the ``distorted'' curves in \fig\ref{fig:hrg_comp}.

Let us attempt to draw some conclusions. 
If we take agreement with the hadron resonance gas for a criterion,  we obtain indeed one explanation
for all discussed observables.  The  trend for the p4 and asqtad actions to 
push features of the curves to slightly larger temperatures is then due to cut-off effects manifested 
in the hadron spectrum and mainly 
in the pion splitting. On the other hand, the cut-off effects in the dispersion relation appear to matter less in 
the transition region. This is not surprising, their improvement effect is relevant for 
the ideal or weakly interacting gas and hence at very high temperatures. The fact that the equation of state up to almost $T_c$ is described by the hadron resonance gas shows to the contrary that it
is strong coupling physics that is at work here, and hence the cut-off effects on the hadron spectrum that
matter most. 
For this reason the stout and HISQ/tree actions perform best in this range of 
temperatures. (Note that an improvement scheme that is complete to some order in the lattice spacing
works of course on all scales).
All of this is of course also consistent with the notion of a
strongly coupled quark gluon plasma just above $T_c$. 
Finally the observation that, for all observables
considered, the stout data for $N_\tau\geq 8$ appear to sit on top of each other within small errors
indicates that, for the temperature range considered, they have reached the scaling regime and
a reliable continuum extrapolation can be performed once the $N_\tau=12$ data are complete.
This constitutes the last step in the determination of the equation of state within that discretisation scheme.
It is now most desirable to confirm this with an independent discretisation, and the HISQ/tree action is the most promising candidate.

\subsection{Staggered vs. Wilson fermions}

Wilson and staggered quarks have different systematic errors and hence make for a valuable
one-to-one comparison of results in physical units. In the continuum limit, both must give the same
results of course, if there are no fundamental problems with either discretisation. Conversely,
deviations at finite lattice spacings give valuable insight into systematic errors and limitations, since
in this case we are not only comparing cut-off effects of the same kind with different size, but really
two conceptually different ways of dealing with the doubling problem. 
The difficulty is to find simulations performed
at the same physical masses so as to be directly comparable. So far the equation of state itself
is not yet available for this purpose, but the behaviour of the chiral condensate and the Polyakov 
loop through the transition have been studied by the BMW collaboration in a direct comparison
between Wilson and staggered quarks  \cite{bmw_w1, bmw_w2}.

For the Wilson simulations they use the fixed scale approach, employing
a tree-level Symanzik improved gauge action, Sec.~\ref{sec:impr_gauge}. The Wilson
fermion action has a tree-level clover improvement term, $c_{SW}=1$ Sec.~\ref{sec:clover}, and employs stout smeared links with six iterations and smearing parameter $\rho=0.11$, Sec.~\ref{sec:stout}. 
The scale was set by $m_\Omega=1672$ MeV and the line of constant physics was determined
by the mass ratios in Table \ref{tab:bmw_w}, utilising data from previous spectrum calculations
at zero temperature \cite{bmw_spec1,bmw_spec2,bmw_spec3}. This translates to 
$m_\pi\approx 545$ MeV and $m_K\approx 614$ MeV.
On the other hand, the staggered results entering in the comparison are those obtained at 
fixed $N_\tau$, following the same procedure to fix the lines of constant physics as in 
\cite{bmw_nature,bmw_tc1,bmw_tc2,bmw_tc3}, cf.~Sec.~\ref{sec:bmw}. Essentially the beta function
for the strange quark was determined while keeping the bare quark mass ratio fixed at $m_l/m_s=2/3$.
The resulting ratios $m_\pi/m_\Omega, m_K/m_\Omega$ were then found to be within 3\% of the Wilson results.
In order to renormalise the Polyakov loop, the authors of \cite{bmw_w1,bmw_w2} 
simply fix a value $L^*=1.2$ at 
a temperature $T^*=0.143m_\Omega$. The renormalised Polyakov loop is then
\be
L_R(T) = \left( \frac{L_*}{L(T_*)} \right)^{\frac{T_*}{T}} L(T)\;.
\ee
\begin{table}
\begin{center}
\begin{tabular}{|c|c|c|c|c|c|c|c|}
\hline
$\beta$ & $m_\pi/m_\Omega$ & $m_K/m_\Omega$  &  $m_s/m_{ud}$ & $a m_\Omega$    & $a\; [fm]$ \\
\hline                                                                                      
\hline                                                                                      
3.30 &   0.332(3)          &   0.373(3)      & 1.529(2)    & 1.16(1)        &  0.139(1)     \\
\hline                                                                                       
3.57 &   0.319(6)          &   0.359(4)      & 1.531(2)    & 0.777(9)       &  0.093(1)     \\
\hline                                                                                      
3.70 &   0.326(5)          &   0.369(5)      & 1.531(3)    & 0.586(8)       & 0.070(1)      \\
\hline                                                                                      
3.85 &   0.314(7)          &   0.358(6)      & 1.528(4)    & 0.480(8)       & 0.057(1)       \\
\hline
\end{tabular}
\end{center}
\caption{Spectroscopy from zero temperature Wilson simulations. 
The quark mass ratios are $m_s/m_{ud}=(2m_K^2-m_\pi^2)/m_\pi^2$. The lattice spacings are 
set by $m_\Omega = 1672$ MeV.}
\label{tab:bmw_w}
\end{table}
\begin{figure}
\includegraphics[width=0.5\textwidth]{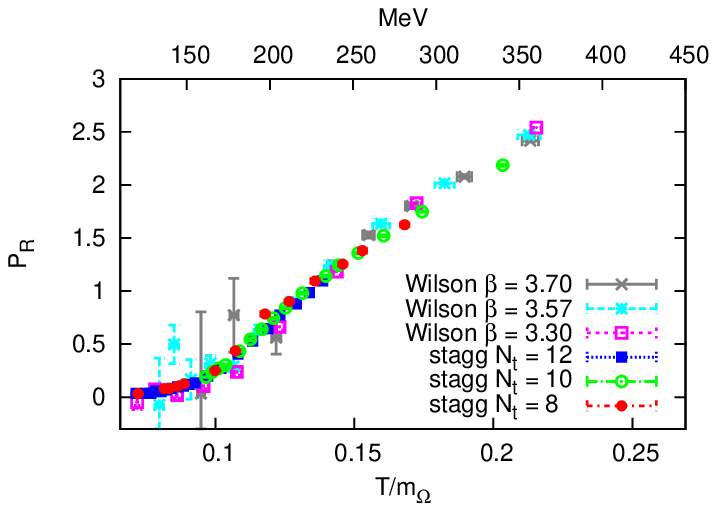}
\includegraphics[width=0.5\textwidth]{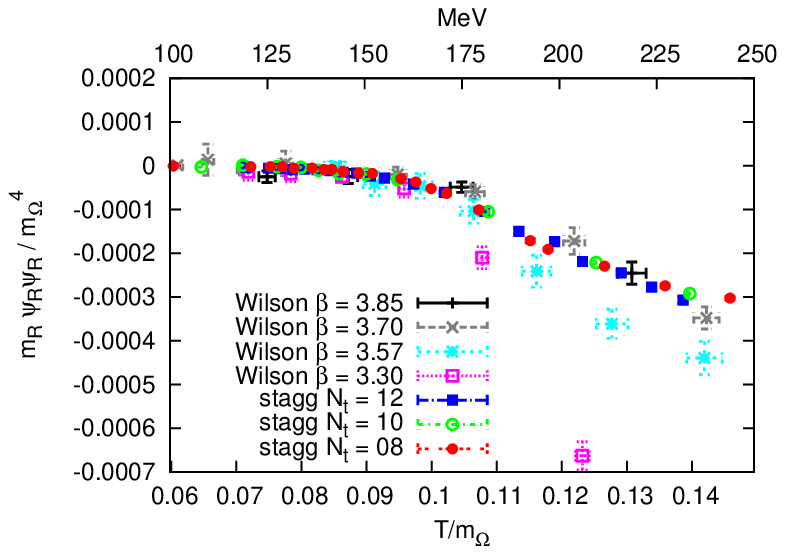}
\vspace*{-1.0cm}
\caption[]{Comparison of the renormalised Polyakov loop (left) and chiral condensate (right)
between simulations using staggered and Wilson fermions. From \cite{bmw_w1}.}
\label{fig:stagw}
\end{figure}

A comparison between results from staggered and Wilson fermions is shown in \fig\ref{fig:stagw} 
\cite{bmw_w1}.  For the Polyakov loop good agreement is observed in the entire range.
For the chiral condensate, the Wilson data corresponding to the two larger $\beta$-values  
nicely fall on the curve of the staggered data, while
the two coarser lattice spacings show large deviations that get stronger with temperature, as expected
in the fixed scale approach.
Note that the Polyakov loop appears to be less sensitive to discretisation effects than
the chiral condensate, which is presumably because the latter is sensitive to the chiral symmetry breaking of the Wilson action at finite lattice spacing.

Interestingly, this cut-off effect of the Wilson action can be largely reduced by employing the definition
\eq(\ref{eq:defmR}) for the renormalised chiral condensate. For an $\Oa$-improved action, its corrections
start only at $\Oas$, with leading order corrections cancelling between $m_R$ and 
$\langle \bar{\psi}\psi\rangle_R$. Of course, this definition requires in addition the definition and calculation 
of $Z_A$ and $m_{PCAC}$.
With the combined tree-level clover  and stout link improvement 
this appears to be almost achieved and a much improved scaling behaviour for the chiral condensate
of the Wilson data is observed \cite{bmw_w2}. This allows for a continuum extrapolation of both types
of discretisations over the whole temperature range, providing perfectly agreeing continuum limits,
\fig\ref{fig:stagw2}. 
\begin{figure}
\includegraphics[width=0.5\textwidth]{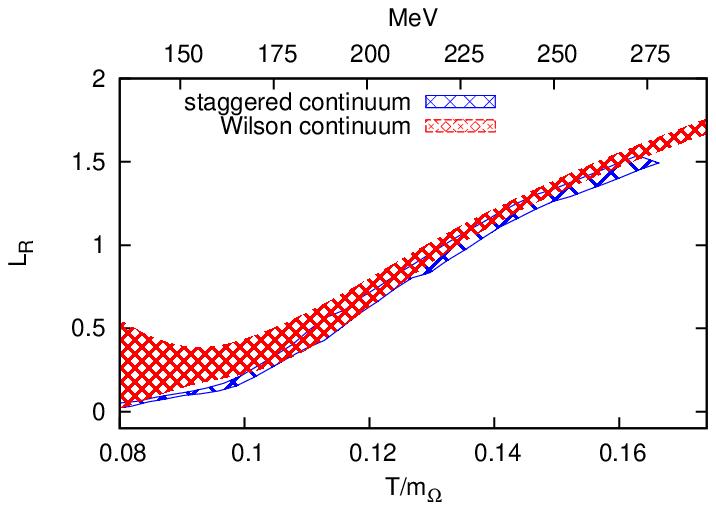}
\includegraphics[width=0.5\textwidth]{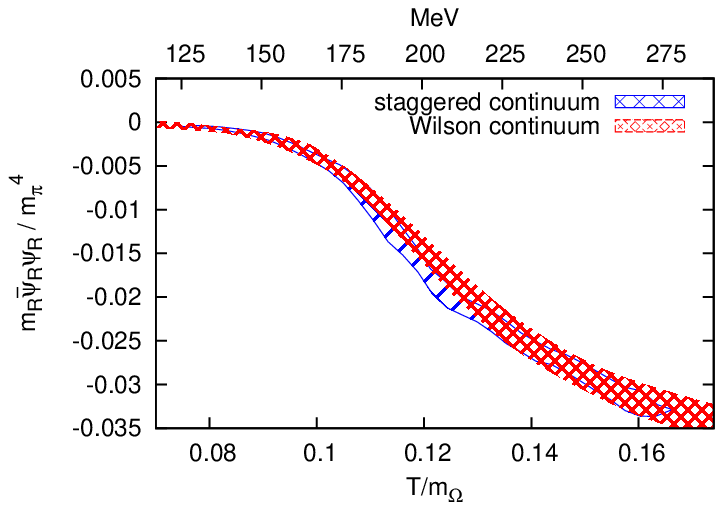}
\vspace*{-1.0cm}
\caption[]{Continuum extrapolated results for the renormalised Polyakov loop (left) and the chiral 
condensate (right). Perfect agreement is found between Wilson and staggered fermion discretisations. From \cite{bmw_w2}.}
\label{fig:stagw2}
\end{figure}

We may thus conclude that, for pions with $m_\pi\gsim 500$ MeV, thermodynamics simulations using
the staggered or Wilson discretisation schemes indeed arrive at the same continuum limit, i.e.~the different
lattice artifacts disappear from each formulation as desired. However, there is one caveat left prohibiting
us from drawing the same conclusion for all quark masses. Concerns about a potentially wrong
continuum limit of staggered fermions employing the rooting trick are to a large extent based
on the $U(1)_A$ anomaly and its relation to the fermionic zero modes \cite{stag3}. This potential
problem is not an issue for the relatively heavy quarks considered here, but becomes relevant when 
quarks are light such as in the physical case. It is thus mandatory to continue such comparisons
until Wilson simulations either reach or can be extrapolated to the physical mass values. 

\section{The equation of state at finite density}\label{sec:mu}

Heavy ion collisions proceed at a given baryon number, depending on the choice of nuclei. Within
the grand canonical potential, this corresponds to a non-vanishing chemical potential for baryon number,
which can also be tuned by the collision energy of the ions. A long term goal for future facilities like
FAIR at GSI (Darmstadt, Germany), is to also 
probe cold high density matter that is relevant for the interior of compact stars. 
Unfortunately, few first-principles predictions exist for these regimes, because
of the so-called sign problem of lattice QCD. For a complex quark chemical potential, the determinant
satisfies 
\be
\det(\gamma_\mu D_\mu + m-\gamma_0\mu)={\det}^*(\gamma_\mu D_\mu+m+\gamma_0\mu^*)\;, 
\label{detpos}
\ee
which is complex unless ${\rm Re} \,\mu=0$. 
However, a complex fermion determinant cannot be interpreted as a probability measure for importance
sampling and the evaluation of the path integral by Monte Carlo methods is spoiled. On the other 
hand, simulations at imaginary $\mu$ or finite isospin chemical potential, $\mu_u=-\mu_d$, feature
a positive determinant and can be simulated in the same way as at $\mu=0$.

It should be stressed that this is merely an algorithmic problem, the partition function and physics
remain completely well defined. The same problem appears also in condensed matter physics, whenever
fermionic systems with a net density of particles vs.~anti-particles or particles vs.~holes are considered.  
Existing solutions to sign problems with cluster 
algorithms \cite{cluster}, world line representations \cite{world} or complex Langevin simulations \cite{langevin} could not yet be adapted to full dynamical QCD. Available methods for QCD are
reweighting techniques \cite{rew}, Taylor expansions about $\mu/T=0$ \cite{tay1,tay2},
simulations at imaginary chemical potentials
followed by analytic continuation \cite{imag1,imag2} and use of the canonical ensemble \cite{can}. All of these introduce (partly related) additional systematic errors and
are valid only for $\mu/T\lsim 1$. For a pedagogical introduction, see \cite{oplh}. 
Because of technical difficulties and computational effort associated with these calculations, most
are restricted to comparatively coarse and small lattices. Recent reviews are given 
in \cite{murev,forcrev,fodrev,grev}. 
Here we merely collect some results for the effect of 
finite baryon density on the equation of state.

First qualitative results for the dependence of the equation of state on quark chemical potential were
produced using reweighting techniques in \cite{bmw_mu1}, for coarse lattices with $N_\tau=4$ employing standard staggered fermions corresponding to $m_\pi\approx 400$ MeV. 
Similarly, the Taylor expansion approach with cumulant approximations to the fermion determinant
was tested on $N_\tau=4$ lattices with two flavours of Wilson fermions with $m_\pi\sim 500$ MeV
\cite{whot_mu}.
In the meantime, there
are more realistic calculations, all of them based on the Taylor expansion approach which we thus
review here, after some more general considerations.

\subsection{Qualitative results from Taylor expansion and imaginary $\mu$}\label{sec:qualmu}

The study of the finite density modifications of the equation of state based on a Taylor expansion is straightforward. Essentially, corrections
to the zero density equation of state are calculated order by order in sufficiently small chemical potentials. 
Let us begin with some qualitative considerations for the theory with $N_f=2$ \cite{im_mu}.  
The pressure is written as a Taylor series 
\be
\frac{\Delta p}{T^4} \equiv \frac{p(\mu_u,\mu_d)- p(0,0)}{T^4} = 
\sum_{n,m=1}\frac1{n!m!}f_{nm}\left(\frac{\mu_u}{T}\right)^n \left(\frac{\mu_d}{T}\right)^m,
\ee
where $f_{nm}$ are the Taylor expansion coefficients, which vanish for odd $(n+m)$ because of 
CP-symmetry, and the zero density contribution has been subtracted. 
When the quark masses are equal, one has $f_{nm}=f_{mn}$.
The standard procedure is to calculate the coefficients at $\mu=0$. However, in \cite{im_mu}
it was remarked that the $f_{nm}$'s are also related to derivatives $\chi_{ij}$ of the pressure measured
at {\it non-zero} chemical potential by 
\be
T^{i+j-4} \chi_{ij}=\frac{\partial^{i+j} (p(\mu_u,\mu_d)/T^4)}
{\partial(\mu_u/T)^i \partial(\mu_d/T)^j}
=\sum_{n=i,m=j}\frac1{(n-i)!(m-j)!}f_{nm}\left(\frac{\mu_u}{T}\right)^{n-i} \left(\frac{\mu_d}{T}\right)^{m-j}\;.
\label{Deri}
\ee
At zero density $\chi_{ij}=f_{ij} T^{4-i-j}$, while  
at non-zero densities $\chi_{ij}$ includes higher order $f_{nm}$ terms,
and does not vanish for odd $(i+j)$.
The proposal of \cite{im_mu} is to evaluate the $\chi_{ij}$ at imaginary or isospin chemical 
potential, where simulations can be carried out, by fitting all of them simultaneously to the polynomial expansion \eq(\ref{Deri}). In practice, imaginary chemical potential $\mu=i\mu_i$ was used and eight different $\chi_{ij}$ evaluated up to $i+j=4$. The data were then fitted to the corresponding polynomial expansions \eq(\ref{Deri}) truncated to
a given order $(n+m)$, which gives the corresponding $f_{nm}$.
Knowledge of the coefficients $f_{nm}$  then
allows to reconstruct the thermodynamic potentials. 

In the confined phase the equation of state is well described by the hadron resonance
gas. In this case the expansion in chemical potentials reads \cite{allton}
\ba
\label{HRG}
\frac{\Delta p(\mu_u,\mu_d)}{T^4} & = &G[\cosh(\frac{2\mu_{Is}}{T})-1] +R[\cosh(\frac{3\mu_q}{T})\cosh(\frac{\mu_{Is}}{T})-1] \\ 
&+&W[\cosh(\frac{3\mu_q}{T})\left(\cosh(\frac{\mu_{Is}}{T})+\cosh(\frac{3\mu_{Is}}{T})\right)-2], \nonumber
\ea
where $G$, $R$ and $W$ are constants related to the hadron spectrum which can be translated again
to the $f_{nm}$.
Quark and isospin chemical potentials $\mu_q$ and $\mu_{Is}$ are defined as
$\mu_q=(\mu_u+\mu_d)/2$ and $\mu_{Is}=(\mu_u-\mu_d)/2$ respectively. 

To test those propositions,
simulations have been performed for $N_f=2$ flavours of standard staggered quarks
on exploratory $8^3\times 4$ lattices, with a bare quark mass $am=0.05$, with a range of imaginary
chemical potentials $a\mu_i=0.0,\ldots 0.24$ and a temperature range $T/T_c=0.83,\ldots 2.0$ \cite{im_mu}.
\fig\ref{fig:Taylor} shows examples for two Taylor coefficients extracted by fitting a sixth order polynomial.
They are found to agree well with coefficients computed directly as derivatives, 
but the results from fitted polynomials have smaller errors. The main finding of the investigation concerns
the possibility to capture the physics accurately with truncated polynomials.
In \fig\ref{fig:Taylor} (right) the $\chi^2/\text{dof}$ of the fits is shown as a function of temperature.
A fourth order polynomial does not allow for good fits anywhere in the temperature range, and even
a sixth order polynomial has difficulties around $T_c$. The conclusion is that for a reliable description
one needs eighth order polynomials {\it at least}, which is the best currently available information
on coarse lattices within the Taylor series approach \cite{gg}.
\begin{figure}
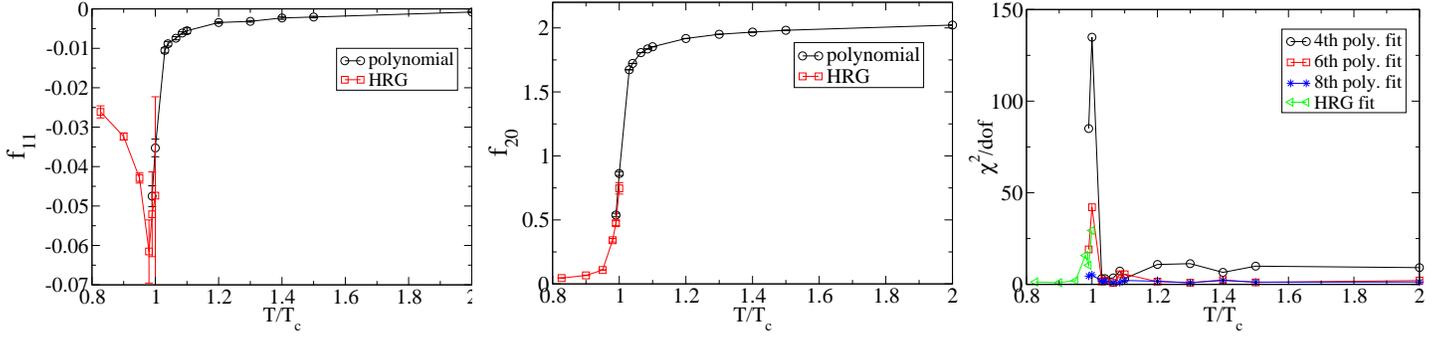

\vspace{5mm}
\hspace{-5mm}
\includegraphics[height=4.4cm]{plots/c11all2-2.eps}
\includegraphics[height=4.4cm]{plots/c20all2-2.eps}
\includegraphics[height=4.4cm]{plots/chisq_poly.eps}
\caption[]{Left, Middle:
Taylor coefficients $f_{11}$ and $f_{20}$.
Right: $\chi^2/dof$ of various polynomials as a function of $T/T_c$ with fitting range $a\mu_I$ is 0.0-0.24. 
From \cite{im_mu}.}
\label{fig:Taylor}
\end{figure}

\subsection{Equation of state to order $\mu^6$ with almost physical quarks}\label{sec:mual}

The Taylor expansion approach is followed by the MILC  and hotQCD collaborations for $N_f=2+1$
flavours, with chemical potentials for the light and strange quarks, $\mu_{l,s}/T$, respectively \cite{lev}.
The pressure and trace anomaly are now sums in these variables,
\ba
{p\over T^4}&=&{\ln  Z \over T^3V}=
\sum_{n,m=0}^\infty c_{nm}(T) \left({\bar{\mu}_l\over T}\right)^n
\left({\bar{\mu}_s\over T}\right)^m,
\label{p}\\
{I\over T^4}&=&-{N_\tau^3\over N_s^3}{d\ln Z \over d\ln a}=\sum_{n,m}^\infty b_{nm}(T)\left({\bar{\mu}_l\over T}\right)^n
\left({\bar{\mu}_s\over T}\right)^m.
\label{I}
\ea
The bar over the chemical potential indicates that they are measured in units of MeV rather than in units
of temperature. The partition function is an even function of chemical potential due to its $CP$ symmetry, i.e.~$n+m$ is even.
The coefficients $c_{nm}, b_{nm}$ are the appropriate derivatives evaluated at zero density.
Since all $\mu$-dependence resides in the determinants, what is required are expressions like
\ba
\frac{\partial\ln\det M}{\partial \mu_i}&=&\Tr\left(M^{-1}\frac{\partial M}{\partial \mu_i}\right) \nn\\
\frac{\partial^2\ln\det M}{\partial \mu_i^2}&=&\Tr\left(M^{-1}\frac{\partial^2M}{\partial \mu_i^2}\right)
-\Tr \left(M^{-1}\frac{\partial M}{\partial \mu_i}M^{-1}\frac{\partial M}{\partial \mu_i}\right)\nn\\
&\mbox{etc.}&
\ea
These are traces of local operators which can be evaluated using random noise vectors.
Explicit expressions for the coefficients can be found in \cite{milc_mu1}.

\begin{figure}[t]
\includegraphics[width=\textwidth]{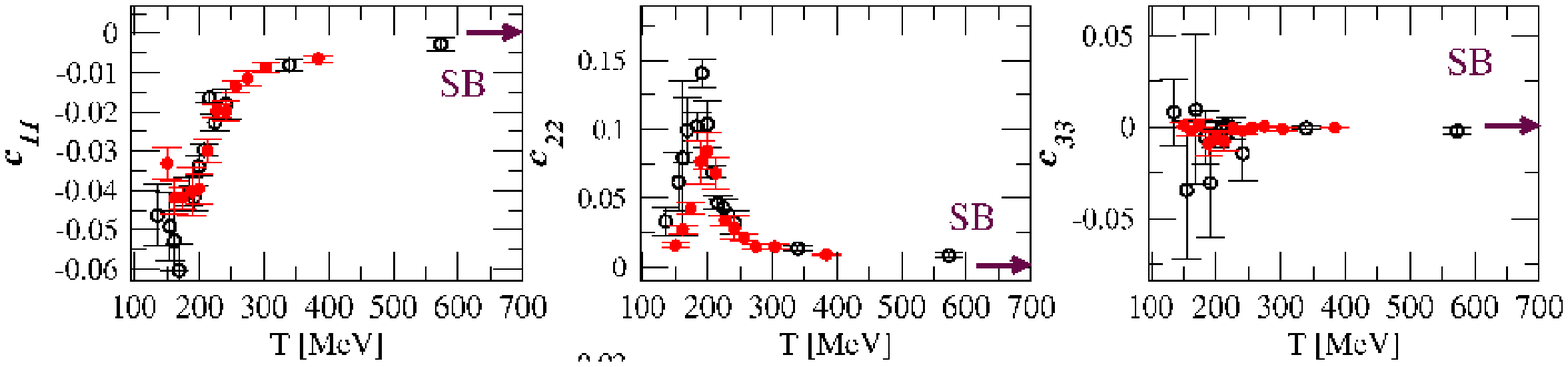}
\includegraphics[width=0.5\textwidth]{plots/dP}
\includegraphics[width=0.5\textwidth]{plots/PO6Isentr}
\caption[]{Top row: Some coefficients $c_{nm}$  in the Taylor expansion of 
the pressure 
as a function of temperature. Arrows indicate the respective (massless)
Stefan-Boltzmann limits. Bottom row:
Change in the pressure due to the non-zero chemical potentials (left) and pressure along
different isentropes (right). 
From \cite{lev}.}
\label{fig:muco}
\end{figure}
The coefficients have been computed through $\mu^6$ 
using the configurations generated with the asqtad staggered 
fermion action for physical strange quark mass and $m_l/m_s=0.1$, $m_\pi\approx 220$ MeV, i.e.~those entering the zero density calculations covered in 
Sec.~\ref{sec:almost}. Examples for some of the coefficients are shown in \fig\ref{fig:muco} (top)
for lattices with $N_\tau=4,6$ \cite{lev}. Given previous experience at zero density, it is not surprising that
in some coefficients there are differences between the $N_\tau=4$ to $N_\tau=6$ results, indicating 
the sizeable discretisation errors present at $N_\tau=4$. 

Once the coefficients are at hand, the change $\Delta X=X(\mu_{l,s})-X(0)$ in the thermodynamic potentials
can be computed. Note that since the 
$c_{n1}(T)$ terms are non-zero, a non-vanishing strange quark density $n_s$ is 
induced even for $\mu_s = 0$. Conversely, in order to have zero strange quark density, one needs to tune
the strange quark chemical potential $\mu_s(T,\mu_l)$.
The corresponding results for the pressure are shown in \fig\ref{fig:muco} (bottom left).
As in the coefficients, cut-off effects are clearly visible in the equation of state.

In heavy ion collisions, once the plasma has thermalised, it expands to cool at constant entropy. 
To get closer to the experimental situation, it is then interesting to consider the isentropic
equation of state, for which the ratio of entropy to baryon number is held fixed. Isentropes in heavy 
ion collisions are trajectories in $(\mu_l,\mu_s,T)$-space with $n_s=0$, which are characterised
by different values for $s/n_B=C$. For AGS, SPS, and RHIC, $s/n_B\approx30, 45$, and $300$, respectively \cite{isent}. For a two flavour theory, they were first calculated on the lattice in 
\cite{isent0}.
The pressure evaluated along those isentropes is shown 
in \fig\ref{fig:muco} (bottom right).
Interestingly, the discretisation effects appear smaller in these quantities. The authors explain that this
is due to the fact that the zero density contribution, i.e.~the first term in the Taylor series, is largest 
for these quantities and there the difference between $N_\tau=4,6$ was observed to be small \cite{milc_07}.  However, it should be kept in mind that these
lattices are too coarse for scaling and this can {\it not} be taken as an indication of the proximity to the 
continuum, as we have seen in Sec.~\ref{sec:comp}.

\subsection{Equation of state to order $\mu^2$ for physical quark masses}\label{sec:muphys}

Recently, the finite density corrections to the equation of state have also been evaluated by the 
BMW collaboration \cite{bmw_mu}, 
making use of their $\mu=0$ configurations generated with the stout link improved
staggered action at physical quark masses for various lattice spacings. The results discussed here
are the finite density extension of those covered in Sec.~\ref{sec:bmw}. Specifically, they look at
the leading order correction, which is the second derivative with respect to chemical potentials,
\be
\chi_2^{ij} \equiv \frac{T}{V} \frac{1}{T^2}\left.\frac{\partial ^2 \log Z}{\partial \mu_i \partial \mu_j}\right|_{\mu_i=\mu_j=0}.
\label{eq:chi2def}
\ee
The trace anomaly to quadratic order is then
\be
\frac{I(T,\mu)}{T^4} = T \frac{\partial }{\partial T} \frac{p(T,\mu)}{T^4} \,+\, \frac{\mu^2}{T^2}\chi_2 =
\frac{I(T,0)}{T^4} \,+\, \frac{\mu^2}{2T} \frac{\partial \chi_2}{\partial T}.
\label{eq:imu}
\ee
Chemical potentials were fixed such as to keep strange quark number density to zero, $n_s=0$,
\be
\mu_{L}/3\equiv \mu_u=\mu_d,\quad\quad 
\mu_s = -2\frac{\chi_2^{us}}{\chi_2^{ss}} \mu_u\;.
\label{eq:muL}
\ee
Alternatively, one may assign the same chemical potential for baryon number to all
quark flavours, $\mu_B=\mu_u=\mu_d=\mu_s$.

The required derivatives
were calculated in the temperature range $125$ MeV $<$ T $< 400$ MeV. 
We recall that configurations on $N_\tau=6,8,10,12$ were available for this purpose, which have been
supplemented by $N_\tau=16$ \cite{bmw_fluc}. Hence all intermediate quantities can be extrapolated
to the continuum.
For lower temperatures
the partition function for the hadron resonance gas was used in order to obtain the appropriate 
derivatives. The following figures are therefore to be interpreted as continuum results at the physical
point.
 
\fig\ref{fig:deltap} (left) shows the difference in the pressure due to finite chemical potential for the light quarks.
Again perfect agreement with the continuum hadron resonance gas is observed roughly up to the 
transition temperature. The change in the trace anomaly due to chemical potential $\mu_L=400$ MeV
is shown in \fig\ref{fig:deltap} (right). In a similar fashion, results for the pressure, the energy and densities
as well as the speed of sound are compared for $\mu=0,400$ MeV in \cite{bmw_mu}.
\begin{figure}[t]
\centering
\includegraphics*[width=8.5cm]{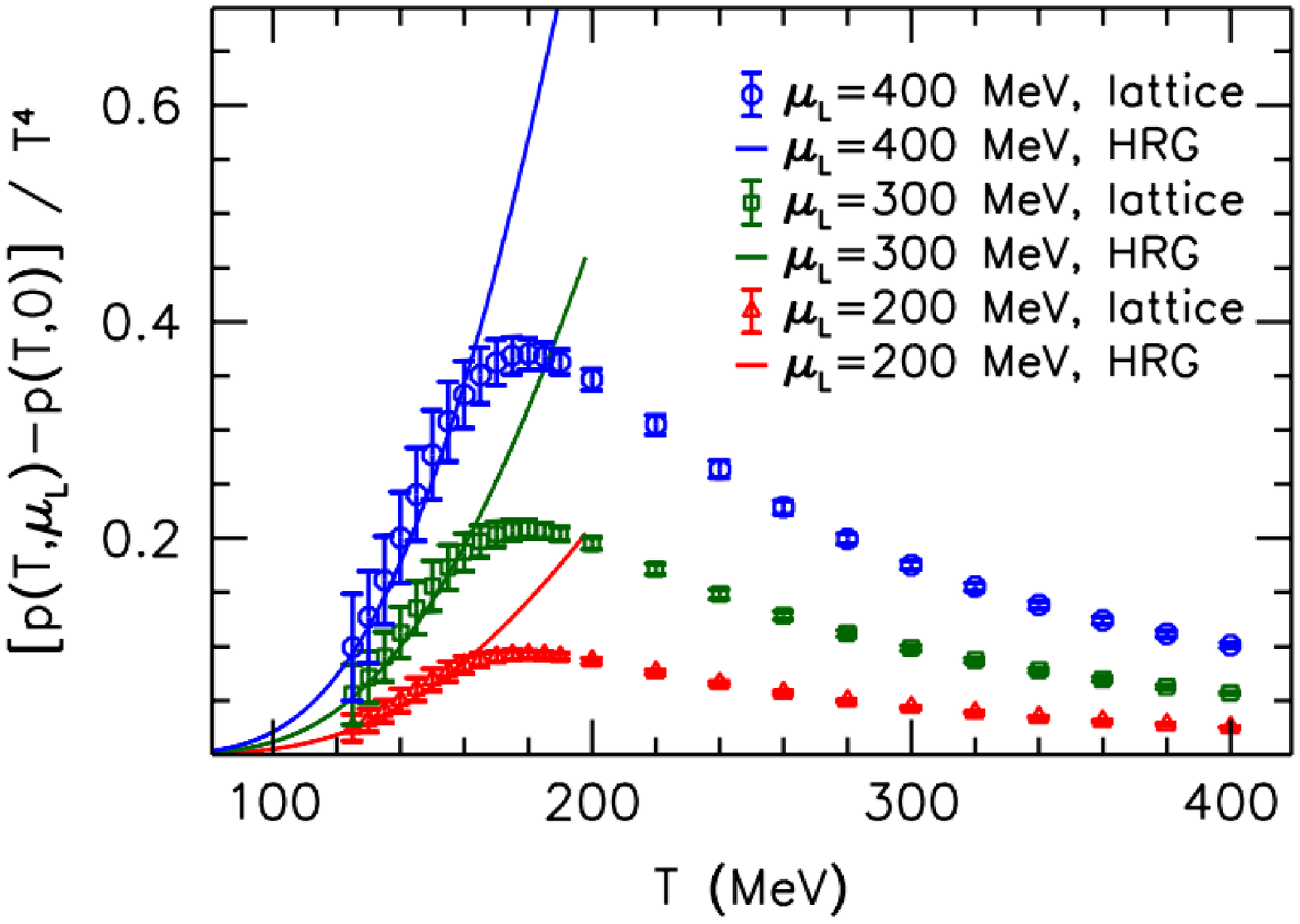}\hspace*{.5cm}
\includegraphics*[width=8.5cm]{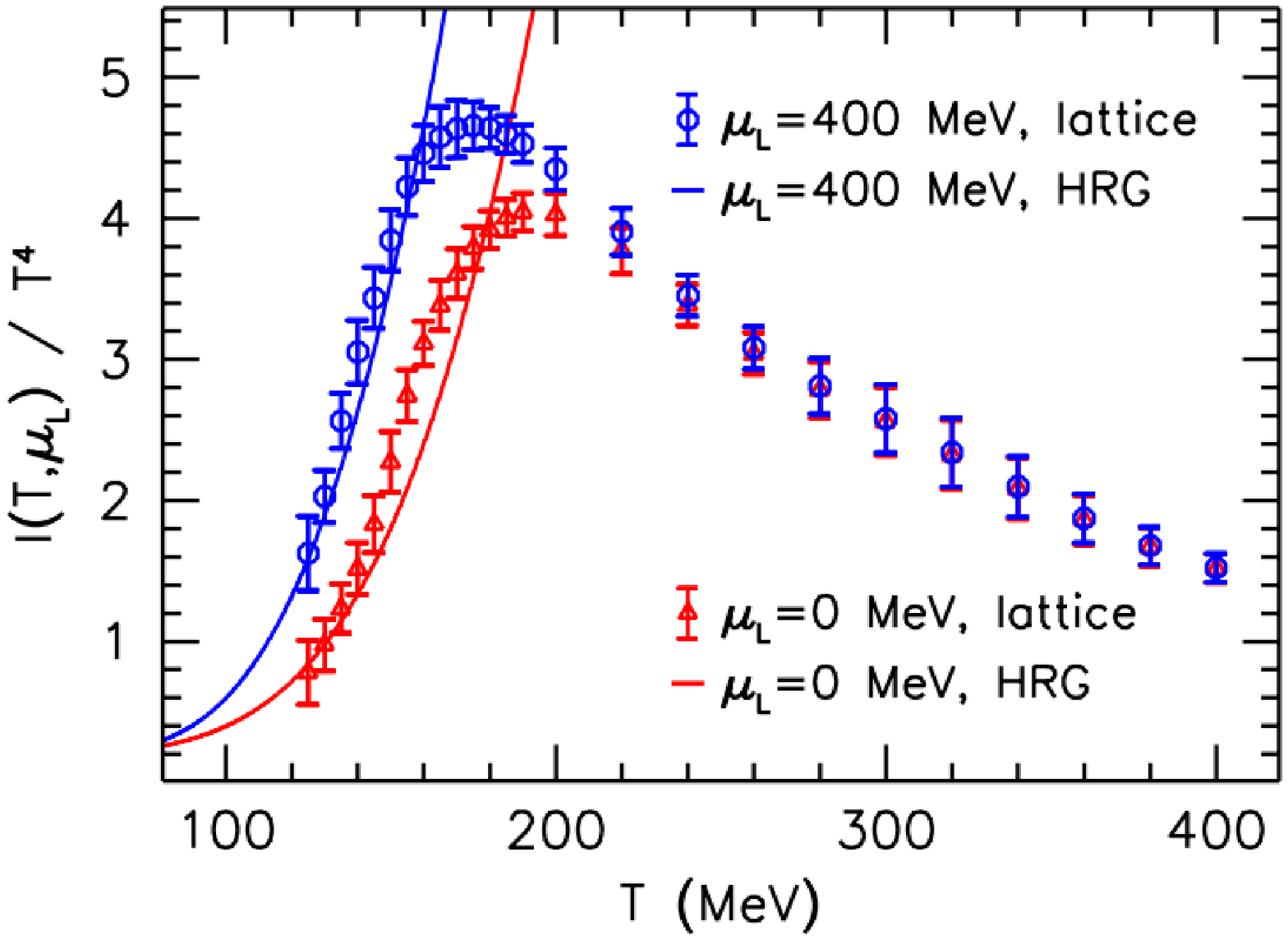}
\vspace*{-0.15cm}
\caption[]{Left: Difference between the pressure at $\mu>0$ and $\mu=0$. Right: Trace anomaly
for non-zero $\mu_L$. From \cite{bmw_mu}.}
\label{fig:deltap}
\end{figure}

In a second step, the isentropic trajectories corresponding to constant entropy over quark number, $S/N$,
are located, \fig\ref{fig:sn} (left). Note that there are significant deviations from leading order 
perturbative predictions. Furthermore, in the low temperature region $T\lsim 130$ MeV, the isentropes
require chemical potentials too large for the Taylor expansion to be valid, so these results are to be
taken with care. The thermodynamic functions evaluated along the isentropes are shown
in Figs.~\ref{fig:sn}, \ref{fig:psn}.
As in the case of zero density, the authors of \cite{bmw_mu} provide convenient parametrisations
that reproduce the data over the whole range $0<T<400$ MeV,
\ba
\frac{I(T)}{T^4}\!&=&
e^{-h_1/t - h_2/t^2}\!\cdot \!\left[ h_0 + \frac{f_0\!\cdot\! [\tanh(f_1\cdot t+f_2)+1]}{1+g_1\cdot t+g_2\cdot t^2} \right]\!,\nn\\
\chi_2(T)&=& e^{-h_3/t - h_4/t^2}\cdot f_3\cdot [\tanh(f_4\cdot t+f_5)+1],
\label{eq:pari}
\ea
with the appropriately fitted constants given in Table \ref{tab:pars}. The trace anomaly for small
chemical potentials follows then from \eq(\ref{eq:imu}), and the pressure by $T$-integration.

\begin{figure}[t]
\includegraphics*[width=8.5cm]{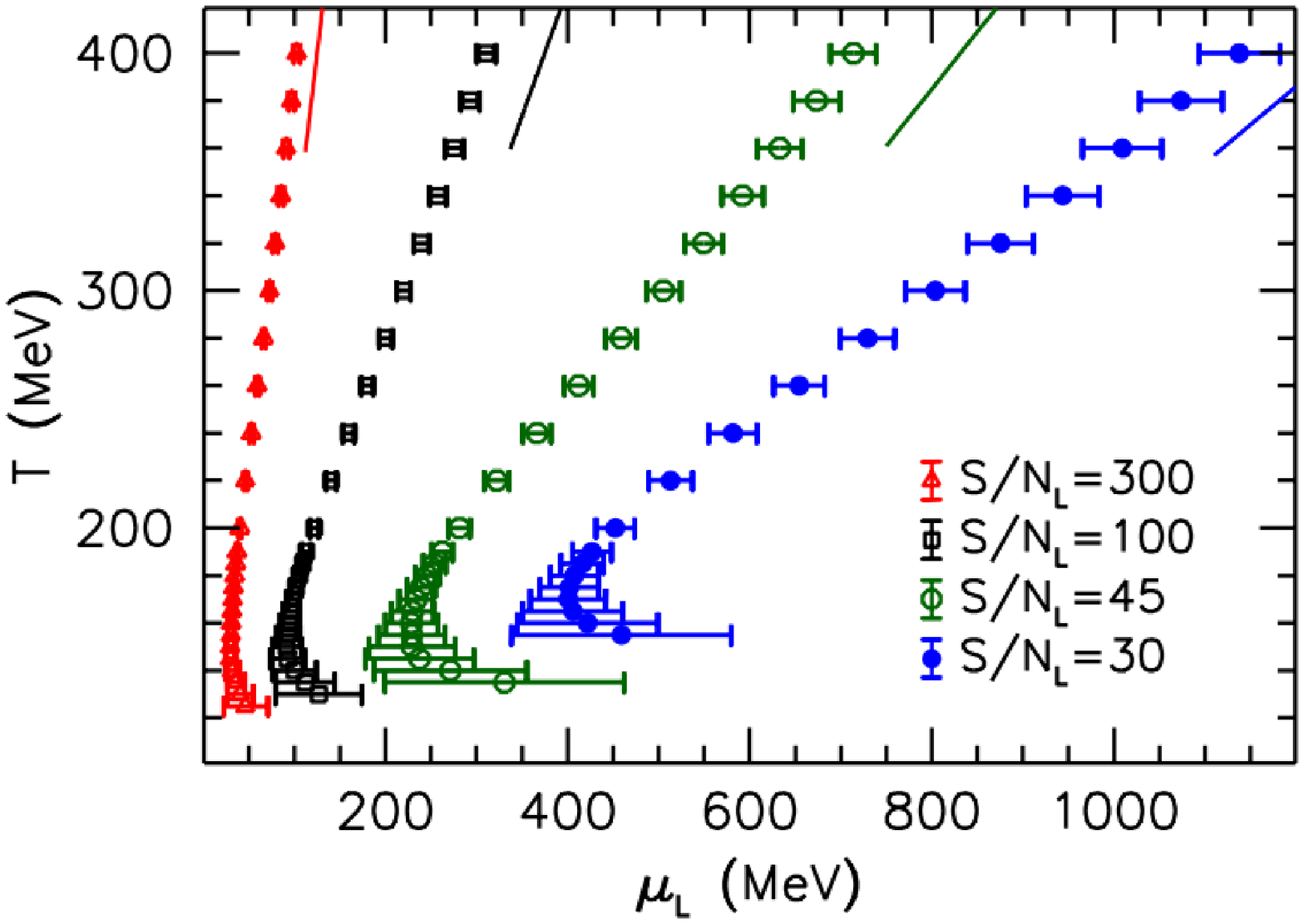}\hspace*{.5cm}
\includegraphics*[width=8.5cm]{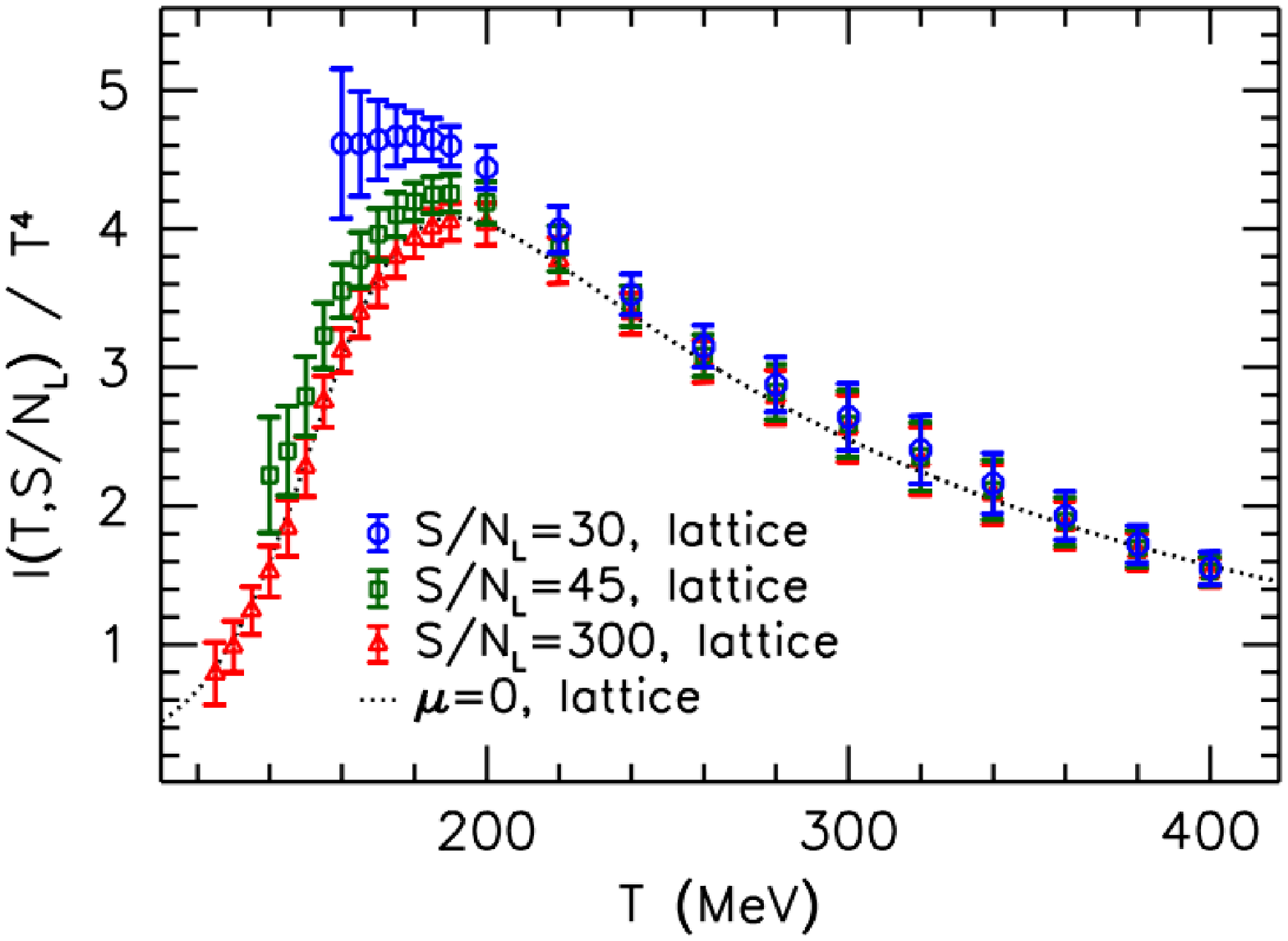}
\vspace*{-0.15cm}
\caption[]{Left: Trajectories of constant $S/N_L$ on the phase diagram. The lines in the upper part of the plot represent the leading order perturbative expressions. Right: Trace anomaly
along the isentropes. From \cite{bmw_mu}.}
\label{fig:sn}
\end{figure}
\begin{figure}[t!]
\includegraphics*[width=8.5cm]{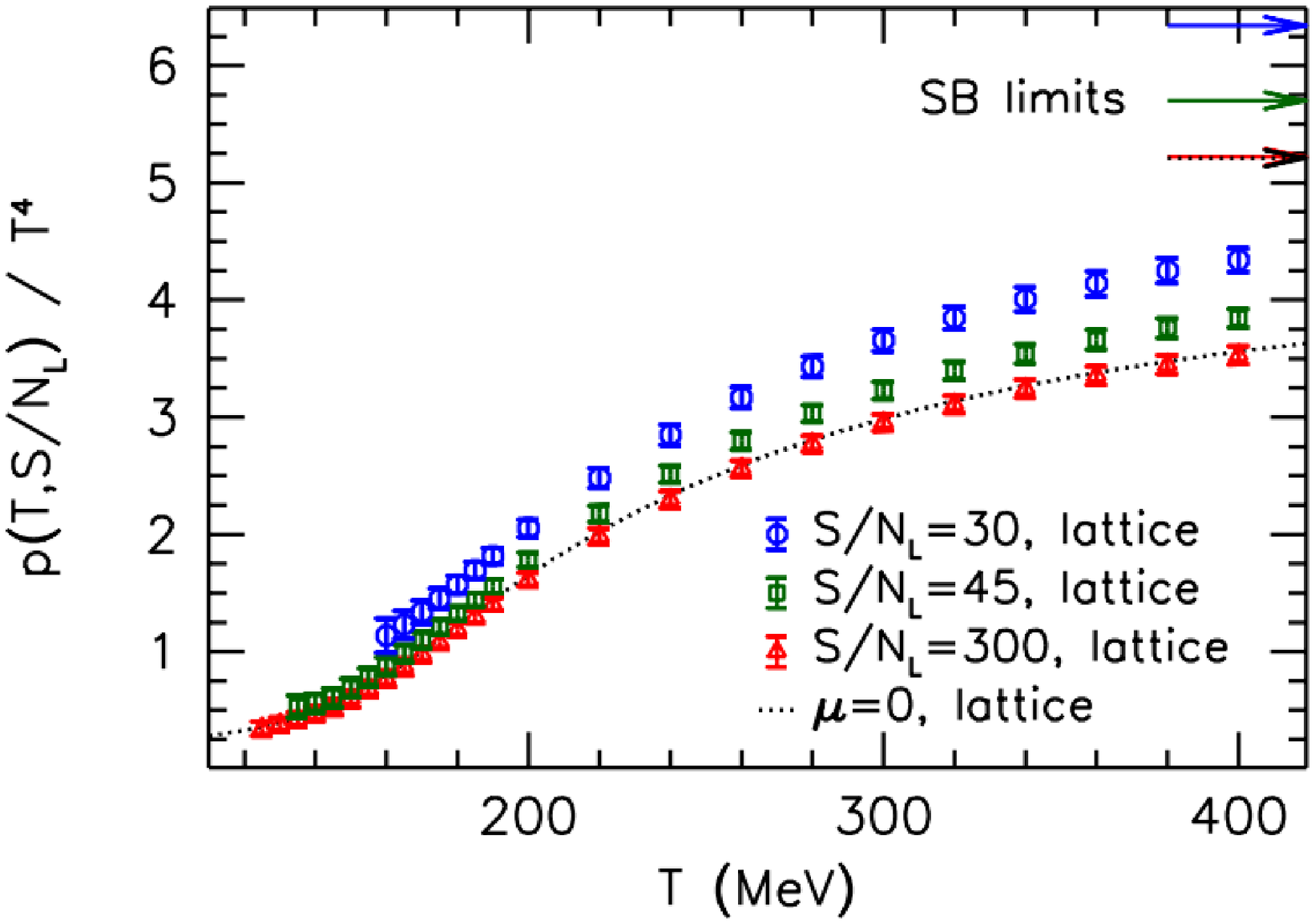}\hspace*{.5cm}
\includegraphics*[width=8.5cm]{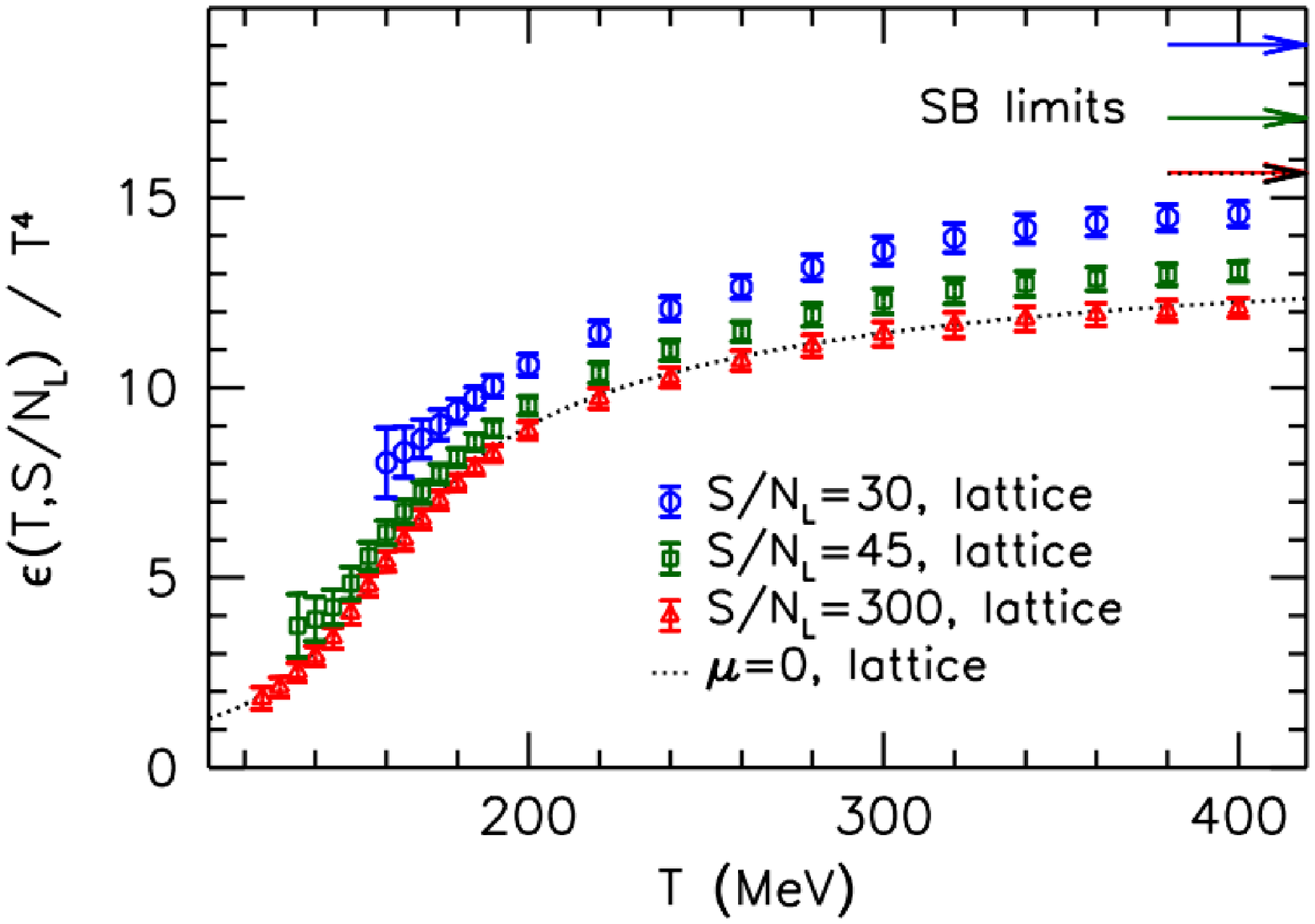}
\vspace*{-0.15cm}
\caption[]{Pressure (left) and energy density (right) along the isentropes. From \cite{bmw_mu}.}
\label{fig:psn}
\end{figure}
\begin{table}[ht!]
\centering
\begin{tabular}{|c|c|c|c|c|c|c|c|}
\hline
$h_0$ & $h_1$ & $h_2$ & $f_0$ & $f_1$ & $f_2$ & $g_1$ & $g_2$ \\
\hline
0.1396 & -0.1800 & 0.0350 & 2.76 & 6.79 & -5.29 & -0.47 & 1.04\\
\hline
\end{tabular}\\[0.1cm]
\begin{tabular}{|c|c|c|c|c|c|}
\hline
$\mu$ & $h_3$ & $h_4$ & $f_3$ & $f_4$ & $f_5$ \\
\hline
$\mu_L$ & 
-0.3364 & 0.3902 & 0.0940 & 6.8312 & -5.0907 \\ \hline
$\mu_B$ &
-0.5022 & 0.5950 & 0.1359 & 6.3290 & -4.8303 \\
\hline
\end{tabular}
\caption[]{\label{tab:pars}
Parameters for the functions~(\ref{eq:pari}).}
\end{table}

Finally, a comment on the reliability is in order. The results discussed in this section are valid in the 
continuum and for physical quark masses, but only up to order $\mu^2$.  In order to check the systematics,
also the subleading coefficient $\chi_4$ was checked in \cite{bmw_mu}. The statement there
is that the $(\mu_B/T)^4$ contribution to the pressure makes up $\lsim 10\%$ of the leading order
contribution for $\mu_B/T\leq 2$ and $\lsim 35\%$ up to $\mu_B/T\leq 3$. Hence, good accuracy is only
achieved at high temperatures around and beyond the transition. 
We should also keep in mind the finding of 
Sec.~\ref{sec:qualmu}, according to which a good description of the functional $\mu$-behaviour takes
several more but the leading coefficient. Future calculations of higher order contributions are
thus necessary before the $\mu$-dependence can be regarded as controlled.

\section{Summary and Outlook}

\begin{table}[tb]
\hspace*{-0.8cm}
\begin{tabular}{|c|c|c|c|c|c|c|}
\hline
&&& Pure gauge &&&\\
\hline
\hline
Gauge action  & & & Smallest $a$ & Cont.~limit for & Reference & Section \\[2mm]
\hline
Wilson & & & 0.125 fm & $T\approx220-1350$ MeV & \cite{boyd} & \ref{sec:ymlow}\\
\hline
trS, tad, RG & & & 0.25 fm &  & \cite{ym_imp} & \ref{sec:ymlow}\\
\hline
Wilson & & & 0.068 fm &  & \cite{whot_fix} & \ref{sec:ymlow}\\
\hline
trS & & & 0.125 fm & $T\approx 270-13500$ MeV& \cite{bmw_hight2}& \ref{sec:hight}\\
\hline
\hline
&&& $N_f=2+1$ &&&\\
\hline
\hline
Gauge action & Fermion action & $m_\pi$ & Smallest $a$ & Cont.~limit for & Reference & Section\\
\hline
p4, asqtad & p4, asqtad & 220 MeV & 0.125 fm & & \cite{hotp4asq} & \ref{sec:almost}\\
\hline
p4 & p4 & 156 MeV & 0.125 fm &  & \cite{cheng_09} & \ref{sec:almost} \\
\hline
trS & stout staggered &  135 MeV & 0.083 fm & $T\approx 100-500$ MeV  & 
\cite{bmw_eos2,bmw_charm} & \ref{sec:bmw} \\
\hline
RG & Wilson clover & 620 MeV & 0.07 fm & & \cite{whot} & \ref{sec:reswil} \\
\hline
HISQ/tr & HISQ/tr & 160 MeV & 0.125 fm & & \cite{hot_hisq} & \ref{sec:compstag}\\
\hline
\hline
&&& $N_f=2+1+1$ &&&\\
\hline
\hline
Gauge action & Fermion action & $m_\pi$ & Smallest $a$ & charm & Reference & Section\\
\hline
asqtad & asqtad & 220 MeV &  0.17 fm & quenched& \cite{lev} & \ref{sec:charm} \\
\hline
trS & stout staggered &  135 MeV & 0.1 fm & dynamical & \cite{bmw_charm} & \ref{sec:charm}\\
\hline
\hline
&&& $N_f=2+1, \mu\neq0$ &&&\\
\hline
\hline
Gauge action & Fermion action & $m_\pi$ & Smallest $a$ & order in $\mu$ & Reference & Section\\
\hline
asqtad & asqtad & 220 MeV &  0.17 fm & $\mu^6$ & \cite{lev} & \ref{sec:mual} \\
\hline
trS & stout staggered &  135 MeV & 0.063 fm & $\mu^2$, cont. limit & \cite{bmw_mu} & \ref{sec:muphys}\\
\hline
\end{tabular}
\caption[]{
Some phenomenologically relevant simulations of the equation of state covered 
in this review. trS=tree-level Symanzik, tad=tadpole, RG=fixed point or Iwasaki, p4, asqtad all refer
to improvement schemes explained in Sec.~\ref{sec:impr}. The smallest lattice spacing $a$ is measured 
at $T\approx 200$ MeV. All tabulated $N_f=2+1$ simulations work
with the physical strange quark mass.
}
\label{tab:summary}
\end{table}

After many years of intense studies and improvements on the theoretical, computational and algorithmic
level, we are converging towards a fully non-perturbative and quantitative understanding of the QCD equation of state at finite temperatures and zero density. 
Lattice studies using the staggered fermions with stout link improvement give
accurate predictions for physical masses and temperatures $125<T<1000$ MeV, a complete continuum
extrapolation based on four lattice spacings in this temperature range is around the corner.
We thus have a fully non-perturbative prediction for QCD in the continuum. An independent check
of these results with a different staggered discretisation is also on the way. A last source of potential
systematic error concerns the question whether rooted staggered fermions have a fundamental flaw
in their continuum approach. This will be answered by the corresponding Wilson simulations as discussed
in this review. Extending those comparisons down to the physical light quark masses will require 
a few more years. Furthermore, people have started thermodynamical simulations using dynamical overlap fermions which have the correct (lattice) chiral symmetry \cite{over1,over2}. However, it will take
much longer for those to reach the same level of maturity. 

At low temperatures the lattice results for the equation of state match smoothly to the hadron resonance gas predictions, which in itself is a non-trivial finding supported
by first principles calculations based on strong coupling expansions. Together with the fact that for
temperatures up to $1$ GeV one observes significant deviations $\sim 10\%$  from the ideal gas predictions, this is testament to the strongly  coupled nature of the plasma around $T_c$, with significant
contributions from strongly interacting soft modes up to much higher temperatures. 
Within pure gauge theory contact to the perturbative regime could be established and there is no 
fundamental problem extending this to the dynamical case. 
Therefore, we may conclude that a complete description of 
the continuum equation of state at zero density and all temperatures is within reach.

Finally, some interesting extensions have also been achieved, though not yet at the same level of
refinement. We have learned that the inclusion of the charm quark has a significant effect at high
temperatures relevant for the LHC. A continuum limit for these predictions is also in reach within a few 
years. Moreover, significant progress has been made in the treatment of small baryon densities.
Leading order $\mu^2$ modifications on the equation of state have been computed, and the technologies
to extend these to higher orders using Taylor expansion and simulations at imaginary chemical potentials
or employing reweighting
are available. 
A quick reference to all these developments is provided in Table \ref{tab:summary}.
We may thus conclude that the quark hadron transition at zero and low baryon density,
together with the associated equation of state, constitute the first fully non-perturbative, first principles
predictions from the lattice for physical QCD under extreme conditions.  

\section*{Acknowledgements}
I thank Christopher Pinke as well as both referees for their careful reading and 
constructive comments.
This work is supported by the German BMBF, contract number 06MS9150, and by the Helmholtz International Center for FAIR within the LOEWE program launched by the State of Hesse.

\end{document}